\documentstyle[aps,floats]{revtex}

\def\hybrid{\topmargin 0pt      \oddsidemargin 0pt
	\headheight 0pt \headsep 0pt \textheight 9in \textwidth 6.25in
\marginparwidth .875in \parskip 5pt plus 1pt \jot = 1.5ex}

 \hybrid

\begin{document}

\title{\Large Let's Baxterise\footnote{"Dedicated to R.J.Baxter on the
occasion of his 60th birthday"}}

\author{ 
S. Boukraa \dag ,  J-M.Maillard\footnote{e-mail
: maillard@lpthe.jussieu.fr} \dag \dag  }

\date{\today}

\maketitle

\centerline{\dag Universit\'e de Blida, Institut d'A\'eronautique,
Route de Soumaa, BP 270, Blida Algeria}

\centerline{\dag \dag LPTHE, Tour 16, 1er \'etage, 4 Place Jussieu, 
75252 Paris Cedex, France}

\begin{abstract}

We recall the concept of Baxterisation of an R-matrix, or
of a monodromy matrix, which corresponds to build, from
one point in the $\, R$-matrix parameter
space, the algebraic variety where the spectral parameter(s) live. 
We show that the Baxterisation, which amounts to studying
the iteration of a birational transformation,
 is a ``win-win'' strategy: it enables to 
discard efficiently the non-integrable situations, focusing 
directly on the two interesting cases where the algebraic varieties
are of the so-called ``general type'' (finite order iteration)
or are Abelian varieties (infinite order iteration).
We emphasize the heuristic example
of the sixteen vertex model and 
provide a complete description of the
finite order iterations situations for the Baxter model.
We show that the  Baxterisation procedure can be introduced in much larger 
frameworks where the existence of some underlying Yang-Baxter
structure is not  used: we   Baxterise
  L-operators, local quantum Lax matrices,
and  quantum Hamiltonians.

\end{abstract}

\vskip .1cm 

{PAR--LPTHE 2000-- 13}

\vskip .1cm 
Submitted to J. Stat. Phys.
\vskip .3cm 

 \noindent
\vskip .2cm 

\vskip .8cm

\vskip .8cm

\section{Introduction}

The Yang-Baxter equations are known to be 
a sufficient condition~\footnote{
And to some extend, necessary condition~\cite{LoMa86}.}
for the commutation of transfer matrices. 
Moreover, it has been shown that the commutation of transfer matrices
necessarily yields a parameterization of the R-matrices
in term of {\em algebraic varieties}~\cite{Ma86}, and that the 
set of {\em inversion relations}, combined together
with the geometric symmetries of the lattice,
yield a set of  {\em birational} symmetries 
of the parameter space of the model,
which are  {\em discrete symmetries} of the 
Yang-Baxter equations~\cite{bmv2,bmv2b}. Generically 
this set of birational symmetries is an infinite set.
Combining these facts, one gets the following
result: the Yang-Baxter integrability
is {\em necessarily} parameterized in term of 
 {\em algebraic varieties} having a set of 
{\em discrete birational symmetries}. An algebraic variety
with an infinite set of (birational) symmetries
cannot be an algebraic variety of the 
so-called ``general type''~\cite{Ma86}. 
In the following  this set of discrete birational
symmetries will be mainly seen as generated by the iteration of a 
 birational transformation, thus canonically associating a
 {\em discrete dynamical system}. This birational transformation can be
of finite order, yielding a finite set of discrete symmetries :
many Yang-Baxter integrable models correspond to this  {\em finite order}
situation~\cite{MaVi96} (see below : RSOS models~\cite{Andrews}, 
integrable chiral Potts model~\cite{HaMa88},
free-fermion models, tetrahedron relations~\cite{Za81,Ba83,Ba86}, ...).
However, such birational transformations are generically (seen as a
 discrete dynamical system) {\em infinite order} transformations.
The iteration of one point under such an
infinite order  birational transformation,
yields an infinite number of points which can actually 
``densify'' an algebraic variety (elliptic curve, Abelian surface,
... see~\cite{BoMaRo95,BoMa95}). One can thus 
deduce the algebraic variety 
from the iteration of one point by this  birational transformation.
This iteration procedure actually {\em solves}~\cite{bmv2,bmv2b} 
{\em the so-called ``Baxterisation'' problem} introduced
by V. Jones in a framework of knot theory. The Baxterisation
problem corresponds to actually find, from one (isolated) $\, R$-matrix, 
satisfying the Yang-Baxter equation, a whole family of  $\, R$-matrices
depending on one, or several, ``spectral parameter(s)''
satisfying the Yang-Baxter equation. In other words, the  Baxterisation
problem corresponds to actually build, from
one point in the $\, R$-matrix parameter
space (one  Yang-Baxter integrable $\, R$-matrix),
the algebraic variety where the 
spectral parameter(s) live. All this is also true for higher 
dimensional generalizations (tetrahedron relations~\cite{Ba83}, ...)
of the Yang-Baxter equations, and for Baxterisation
problems in dimensions greater than two~\cite{BeMaVi91e}.

This paper will be illustrated by many examples of Baxterisation,
in particular, the  heuristic example
of the Baxterisation of the sixteen vertex model which is,  generically, non
Yang-Baxter integrable, with a detailed analysis of the
finite order situations of the Baxter model.  One will finally show that the 
 Baxterisation procedure can be introduced in much larger 
frameworks where the existence of some underlying Yang-Baxter
structure is not clear, and not used. For instance, 
we will give several examples of  Baxterisation of quantum Hamiltonians.
We will also give
several examples of Baxterisation
of differential operators,
starting with the simple example of the  Baxterisation
of the Toda L-operator, and then giving examples of   Baxterisation
of other simple local quantum Lax matrices.

\section{The discrete symmetry group associated with
the Baxterisation procedure}

Let us consider a quite
 general vertex model where one direction, denoted as
direction $(1)$, is singled out.
Pictorially this can be represented as follows:
\begin{equation}
\label{fig2}
\setlength{\unitlength}{0.005in}%
\begin{picture}(120,131)(20,700)
\thicklines
\put( 80,820){\line( 0,-1){120}}
\put( 20,760){\line( 1, 0){120}}
\put(125,769){\makebox(0,0)[lb]{\raisebox{0pt}[0pt][0pt]{\elvrm $k$}}}
\put( 85,710){\makebox(0,0)[lb]{\raisebox{0pt}[0pt][0pt]{\elvrm $J$}}}
\put( 85,810){\makebox(0,0)[lb]{\raisebox{0pt}[0pt][0pt]{\elvrm $L$}}}
\put( 20,769){\makebox(0,0)[lb]{\raisebox{0pt}[0pt][0pt]{\elvrm $i$}}}
\put(160,755){\makebox(0,0)[lb]{\raisebox{0pt}[0pt][0pt]{\elvrm $(1)$}}}
\end{picture}
\end{equation}
where $i$ and $k$ (corresponding to direction $(1)$) can take $q$ values,
while  $J$ and $L$, in the other direction, take $m$ values.

One can define a ``partial'' transposition on direction $(1)$ denoted $t_1$.
The action of $t_1$ on the $R$-matrix is given by \cite{BoMaRo95,BoMa95}:
\begin{eqnarray}
(t_1R)^{iJ}_{kL}\, \, = \, \,   R^{kJ}_{iL}\, 
\end{eqnarray}
The $R$-matrix is a $\, (q\,m) \times (q\,m)\, $ matrix which can be seen as
$q^2$ blocks which are $m \times m$ matrices:
\begin{eqnarray}
\label{3t1bloc}
R= \;\;\; 
\pmatrix{ 
A[1,1] & A[1,2] & A[1,3] & \cdots & A[1,q]  \cr 
A[2,1] & A[2,2] & A[2,3] & \cdots & A[2,q] \cr 
A[3,1] & A[3,2] & A[3,3] & \cdots & A[3,q]\cr 
\vdots & \vdots & \vdots & \ddots & \vdots\cr
A[q,1] & A[q,2] & A[q,3] & \cdots & A[q,q] \cr 
}
\end{eqnarray}
where $A[1,1]$, $A[1,2]$, ..., $A[q,q]$ are $m \times m$ matrices.
With these notations the partial transposition $t_1$ amounts to permuting 
matrices  $A[\alpha,\beta]$ and   $A[\beta,\alpha]$,
for all these block matrices. We use the 
same notations as in~\cite{BoMaRo93a,BoMaRo93b,BoMaRo93c}, that is, we
introduce the two following transformations on matrix $\, R$,
 the matrix inverse $\,
{\widehat I}$
and  the homogeneous matrix inverse $I$:
\begin{eqnarray}
{\widehat I}: R \, \longrightarrow \, \, R^{-1} 
\, , \qquad \qquad 
I: R \, \longrightarrow \, \, \, \det(R) \cdot R^{-1}
\end{eqnarray}
and we introduce the (generically) infinite order homogeneous, and
inhomogeneous, transformations :
\begin{eqnarray}
\label{KK}
K  \, = \,\, \, t_1 \cdot I\, , \qquad \quad 
{\widehat K} \, = \,\, \, t_1 \cdot {\widehat I} \, , \qquad 
\end{eqnarray}
The {\em homogeneous} inverse $\, I$ is a 
{\em polynomial} transformation on each 
of the entries $\, m_{ij}$ of $\, R$,
which associates to each $m_{ij}$ its corresponding cofactor.
Transformation $\widehat{I}$ is an {\em involution}, whereas
 $ \, \,I^2\, =\, ( \det(R) )^{q\, m\, -\, 2} \cdot {\cal I}$,
where ${\cal I}$ denotes the identity transformation.
Transformation $t_1$
is also an involution.
The transformation ${\widehat K}$ is clearly
a {\em birational transformation} on the entries  $m_{ij}$ since
 its inverse transformation is $\, \widehat{I} \cdot t \, $
which is obviously also a rational transformation. Transformation $\, K$ is a
 {\em homogeneous polynomial transformation} on the entries  $\, m_{ij}$
of the $\, R$-matrix.

For such vertex models of lattice statistical mechanics,
transformations $\, {\widehat I}$ and $\, t_1\, $  come from
the {\em inversion relations}~\cite{St79,Ba82}, and 
the geometrical symmetries of the lattice,
in the framework of integrability and {\em beyond} integrability.
These involutions generate a {\em discrete} group of ({\em birational}) automorphisms 
of the Yang-Baxter equations~\cite{bmv2,bmv2b} and their higher dimensional
generalizations~\cite{BeMaVi91e}. They also  generate
 a discrete group of (birational) 
automorphisms of the algebraic varieties canonically associated with
the Yang-Baxter equations (or their higher dimensional
generalizations)~\cite{Ma86}. In the generic case
where the birational transformation $\, {\widehat K} \, = \,\, \, t_1
\cdot {\widehat I}\, $ is an infinite order transformation, this set
of birational automorphisms corresponds, essentially, to the iteration of
$\, {\widehat K}$. When the infinite order 
transformation $\, {\widehat K} \,$ densifies
the algebraic varieties~\cite{BoMa95}, one can deduce the equations 
of these algebraic varieties exactly  and {\em thus solve}
 the so-called "Baxterisation problem". 

{\bf Terminology :} In the following, we will call ``$R$-matrix'' the matrix
 we iterate with (\ref{KK}), even if this ``$R$-matrix''
does not satisfy a Yang-Baxter relation.
From now, on we will call ``{\em Baxterisation}'' the
procedure which amounts, for a given $\, R$-matrix, to iterating
transformation (\ref{KK}), and finding the algebraic variety 
which contains all the points of this iteration. A good example 
corresponds, for instance, to get, from isolated $\, R$-matrices 
satisfying the tetrahedron relations~\cite{Hi93b}, $\, R$-matrices 
also solutions of 
the  tetrahedron relations 
{\em but now depending on spectral parameters}.

\section{Baxterisation : a win-win strategy}

When one ``Baxterises'' an R-matrix 
one can get three different kinds 
of situations : either 
the orbits  of (\ref{KK}) are {\em chaotic}, and therefore 
one {\em cannot} expect any ``nice'' parametrization
of the lattice model, 
or the orbits correspond to algebraic varieties,
and one can actually introduce some ``well-suited'' 
 parametrization of the model associated 
to these algebraic varieties, which will be extremely precious
for any further analysis of the model
(analysis of the Yang-Baxter equations, or higher dimensional
generalizations, calculations 
of partition function per site, ...), or, finally, 
 these orbits are finite orbits. When the 
iteration of (\ref{KK}) is {\em infinite} and yields 
algebraic varieties, one can show that 
the algebraic varieties are not of the so-called ``general-type''~\cite{Ma86}
(using the algebraic geometry terminology). For instance, for
 algebraic surfaces, one can actually classify 
the various  surfaces that are not of the ``general type'': 
product of elliptic curves, Enriques surfaces, Kummer surfaces,
Abelian surfaces, ... 
When the  orbits are {\em finite}
one is back to   algebraic varieties
of the  ``general-type'', which is a very large set of different 
 situations, {\em quite hard to classify}~\cite{Ma86}.
 However, writing for some integer $\, N$, the (projective) condition :
\begin{eqnarray}
\label{finitecond}
K^N(R) \, = \, \, \zeta \cdot R
\end{eqnarray}
{\em actually gives the equations} of the algebraic varieties 
corresponding to this very condition (\ref{finitecond}). For 
instance, one can find the 
algebraic subvariety 
of the chiral Potts model on which the higher genus 
Au-Yang-Baxter-Perk solutions~\cite{AuYang} live, as a finite order
condition~\cite{HaMa88}. 

{\bf To sum up :} one has the following situations :

$\bullet $ Either the group is  {\em infinite}, and one gets the ``precious''
parametrization of the model (even if it is a 
parametrization in terms of theta functions of $g$ variables, ...).

$\bullet $ The group is {\em finite}, and one actually gets 
the equations of the algebraic varieties corresponding 
to this situation, writing condition 
(\ref{finitecond}). This  finite order situation 
is a ``gold mine'' as far as 
integrable situations are concerned~\cite{MaVi96,HaMa88}.
Let us recall, for instance, the example of the tetrahedron 
solutions~\cite{MaVi96,Ba83}.

$\bullet $  The group is  {\em infinite}, but the orbits, corresponding
to the iteration of (\ref{KK}), are {\em chaotic} : one does not have
algebraic varieties~\footnote{One may have algebraic {\em subvarieties} 
compatible with this situation, the subvarieties corresponding to the
two previous situations.} and therefore one does not have
any Yang-Baxter structure~\footnote{One can, however, try to
characterize 
the ``complexity'' of this chaotic situation, calculating the
 {\em topological entropy}, or the
 Arnold complexity, of the birational mapping (\ref{KK})
one iterates~\cite{zeta,topo,growth}.}. This situation corresponds to
an  {\em exponential growth} of the iteration 
calculations~\cite{zeta,topo,growth}.

In every case the Baxterisation procedure helps to avoid the
 points of the parameter space,
 corresponding to exponential growth of the calculations,
where no Yang-Baxter
relation can be expected, and enables to get analytically the 
 algebraic varieties corresponding to  {\em finite order} 
iterations (see section (\ref{finiord}) below), or 
 enables to actually build the algebraic varieties from the iteration
 of an {\em infinite order} birational transformation. 

The Baxterisation procedure amounts to studying the
iteration of a (resp. several) birational (resp. polynomial) 
mapping which  actually corresponds to
{\em discrete symmetries} of the parameter space
of the models. It thus provides a natural link between 
lattice statistical mechanics (field theory, ...) 
and the theory of {\em discrete dynamical systems}. Furthermore,
it also provides links with many other domains of mathematical physics,
or mathematics. For instance, as far as effective algebraic geometry, or
even arithmetic,
is concerned, the Baxterisation provides many
 natural, and simple, examples of 
 {\em Abelian varieties} with an {\em infinite set of rational
points}~\cite{BoMa95}.

Let us first recall some well-known Yang-Baxter integrable 
situations corresponding to {\em finite order} conditions.

\subsection{Finite order conditions : free-fermion conditions}
\label{freeferm}
 The matrix $R$ of the asymmetric eight-vertex model is of
the form :
\begin{equation}
\label{8v}
R = \pmatrix {  a & 0 & 0 & d' \cr
                0 & b & c' & 0 \cr
                0 & c & b' & 0 \cr
                d & 0 & 0 & a' }
\end{equation}
The free-fermion condition is :
\begin{equation}
\label{ff8}
a a' - d d' + b b' - c c' \, =\, \, 0 \nonumber 
\end{equation}
A matrix of the form (\ref{8v}) may be brought, by similarity
transformations, to a block-diagonal
form:
\begin{eqnarray*}
R = \pmatrix{ R_1 & 0 \cr 0 & R_2 }, \qquad\mbox{ with }
\qquad  R_1 = \pmatrix{ a & d' \cr d & a' } \quad \, \, \, \mbox{and}\, \, \,
\quad R_2 = \pmatrix{ b & c' \cr c & b' }.
\end{eqnarray*}
  If one denotes by $\, \delta_1=a a' - d d'$, and  by $\delta_2 =  b b' - c c'$, the
determinants of the two blocks, then the homogeneous inverse $I$
(polynomial transformation) just reads :
\vskip  .3cm
\begin{eqnarray}
&&a \rightarrow a' \cdot \delta_2, \quad \, \, \,
a' \rightarrow a \cdot \delta_2, \quad \, \, \,
d \rightarrow -d \cdot \delta_2, \quad \, \, \,
d' \rightarrow -d' \cdot \delta_2, \quad \, \, \,  \\
&&b \rightarrow b' \cdot \delta_1, \quad \, \, \,
b' \rightarrow b \cdot \delta_1, \quad \, \, \,
c \rightarrow -c \cdot \delta_1, \quad \, \, \,
c' \rightarrow -c' \cdot \delta_1 \nonumber
\end{eqnarray}
and transformation $\, t_1$ is given by :
$t_1 : \, c \leftrightarrow d'\,\,   {\rm and:} \, \,  d \leftrightarrow c'$.
The condition (\ref{ff8})  is {\em left invariant} by $\, t_1$,
$\, I$, and thus $\, K \, = \, t_1 \cdot I$.
It is straightforward to see that condition (\ref{ff8}) is
$\; \delta_1 = - \delta_2 \;$  and has the effect of {\em linearizing}  $I$
into :
\begin{eqnarray}
a \rightarrow a' , \quad \, \, \,
a' \rightarrow a , \quad \, \, \,
d \rightarrow -d , \quad \, \, \,
d' \rightarrow -d' , \quad \, \, \,
b \rightarrow -b', \quad \, \, \,
b' \rightarrow - b , \quad \, \, \,
c \rightarrow c, \quad \, \, \,
c' \rightarrow c'
\end{eqnarray}
The group, generated by $\, I$ and $\, t_1$, is then 
realized by permutations of the entries, mixed
with
changes of signs, and its orbits are 
thus {\em finite}. The situation depicted here, namely 
an {\em inversion relation} that reduces, on some algebraic subvariety, to
 {\em permutation of the entries up to signs},
 also occurs for {\em free-fermion} two-dimensional vertex models on a
{\em triangular} lattice (see Sacco and Wu in~\cite{SaWu75}), 
or for the  three-dimensional vertex
corresponding to the Zamolodchikov-Baxter solution
of the tetrahedron equations~\cite{SeMaSt95,Ko94,Hi94}. Details
 can be found in~\cite{MaVi96}.

The finite order situations are  extremely favorable 
for Yang-Baxter integrability (chiral Potts 
model~\cite{HaMa88}, RSOS models~\cite{Andrews}, 
tetrahedron solution~\cite{MaVi96}, 
free-fermion solutions, free para-fermions, ...). 
If one tries to find new  Yang-Baxter integrable models, one should
certainly first try to write, systematically, all the algebraic
subvarieties
corresponding to these (projective) finite order conditions. 
However, in this finite order case,  one may say that 
the ``Baxterisation procedure''
does not work, or ``works to well'' : it is too degenerate. 
As a consequence, many remarkable results, and structures,
occur (polynomial representation of  the natural integers
 together with their multiplication, ...). 
These results, and structures, will be sketched
 in section (\ref{finiord}).

\subsection{The group is infinite : 
iterations associated with the sixteen-vertex model}
\label{sixteen}
Let us now consider, with the example of
 the {\em sixteen vertex model}~\cite{LiWu72}, 
 the situation~\footnote{Generic from the point of
view of the discrete dynamical systems.} where $\, K$, or
 $\, {\widehat K}$, is {\em infinite order}, thus yielding a
{\em non-trivial} Baxterisation.

In the case of $4 \times 4$ matrices (see Fig.\ref{fig2} but with $\,
q \, = \, m \, = \, 2$), a particular permutation of the 
entries of the matrix, $t_1$, has been introduced in the 
framework of the symmetries of the 
{\em sixteen-vertex model} \cite{prl2}. The action of 
two partial transpositions 
$\, t_1$, and $\, t_2$, on the $R$-matrix is given by~\cite{prl2}:
\begin{eqnarray}
(t_1 R)^{ij}_{kl}\, = \,   R^{kj}_{il},\qquad \quad
(t_2 R)^{ij}_{kl} \, = \,   R^{il}_{kj},\qquad \quad
t\, = \, t_1 \cdot t_2
\end{eqnarray}
Transformation $\, t$ is nothing but the matrix transposition :  $\, t$
commutes with the matrix inversion 
$\, {\widehat I}$. If one denotes $\, m_{ij}\, $
the entries of the $\, R$-matrix, this permutation corresponds to:
\begin{eqnarray}
\label{deft1}
t_1: \,\,\, \,\,\, m_{13} \leftrightarrow m_{31} ,
 \;\;\;\,\, m_{14} \leftrightarrow m_{32} , \;\;\;\,
\, m_{23} \leftrightarrow m_{41} , \;\;\;\,\, m_{24} \leftrightarrow m_{42}
\end{eqnarray}
which amounts to permuting the two $2 \times 2$ 
(off-diagonal) sub-matrices of the $4 \times 4$ $R$-matrix.
This transposition $\, t_1$ corresponds to a partial transposition
of one direction (say the horizontal one denoted by ``1'', the other
 transposition $\, t_2$ corresponding to the other direction denoted by ``2'') 
of a two-dimensional vertex model~\cite{bmv2,bmv2b,prl2} (see Fig.\ref{fig2}).

Remarkably, the symmetry group, generated by the matrix inverse
 ${\widehat I}$  and
 transformation $\, t_1$, or by the infinite generator 
$\, {\widehat K}\,  = \, t_1 \cdot {\widehat I}$,  has been shown 
to yield {\em elliptic curves}~\cite{bmv2b,prl2}
which {\em foliate the whole parameter space}
 of the sixteen vertex model. One should not
confuse the integrability of the {\em symmetries} of the parameter space
of the sixteen vertex model (namely the mappings considered here)
and the Yang-Baxter integrability~\cite{prl2} : the {\em sixteen vertex
model is not generically
Yang-Baxter integrable}~\footnote{Using this elliptic parameterization~\cite{prl2},
one could imagine that the ``inversion trick''~\cite{St79,Ba82}, 
together with some ``well-suited''
analytical assumptions,  could allow to actually get the exact
expression of the partition function per site of the sixteen vertex
model beyond the Yang-Baxter integrability. The model would be
``calculable''{\em without being Yang-Baxter integrable} : this remains an
open question.}.

The integrability of the birational mapping $\,  {\widehat K}$,
or, equivalently, of the homogeneous (bi-)polynomial 
transformation $\, K$, is closely related to the occurrence 
of  {\em remarkable factorization schemes}~\cite{BoMaRo95,BoMa95}. In order to see this, 
let us consider a $\, 4 \times 4$ matrix $\, M_0 \, = \, R$,
and the successive matrices obtained by iteration of transformation
 $\, K\, = \, t_1 \cdot I$, where $\, t_1$ is
 defined by (\ref{deft1}). Similarly to the factorizations 
described in~\cite{BoMaRo95,BoMa95}, one has, for arbitrary $\, n$, the 
following factorizations for the iterations of $\, K$ :
\begin{eqnarray}
\label{K16}
 M_{n+2} \, = \, \, \, \,
{{K (M_{n+1})}\over{f_n^2}}, \quad \, \,\, f_{n+2} \, =  \, \, \, \,
{{\det(M_{n+1})}\over{f_n^3}},  \quad \, \, \,  \,
 \widehat{K}_{t_1}
(M_{n+2})\,\, =  \,  {{K (M_{n+2})} \over
{\det(M_{n+2})}}\,\, = \,\,\,\, \,  {{M_{n+3}} \over {f_{n+1} f_{n+3}}}
\end{eqnarray} 
where the $\, f_n$'s are  homogeneous
polynomials in the entries of the initial matrix $\, M_0$ and the 
$\, M_n$'s are ``reduced matrices'' with homogeneous
polynomial entries~\cite{BoMaRo95,BoMa95}.

Let us denote by $\, \alpha_n$ the degree of 
the determinant of matrix $M_n$, 
and by $\, \beta_n$ the degree of 
 polynomial $f_n$, and
let us introduce  $\alpha(x)$, $\beta(x)$
 which are the generating functions of these
 $\alpha_n$'s, $\beta_n$'s :
\begin{eqnarray}
\label{g}
\alpha(x)\, = \,  \sum^\infty_{n=0} \alpha_n \cdot x^n, \;\; \, \,
\qquad 
\beta(x) \, = \,  \sum^\infty_{n=0} \beta_n \cdot x^n, \;\; \, \,
\end{eqnarray}
From these factorizations, one can easily get linear relations
 on the exponents $\alpha_n$, $\beta_n$ and exact expressions for their generating
 functions and for the  $\alpha_n$'s and $\beta_n$'s:
\begin{eqnarray}
\label{abmn16}
\alpha (x) = \,\, {{4 (1+3x^2)}\over{(1-x)^3}}, \, \, \,  \quad
\beta(x)=  \, \,{{4x}\over{(1-x)^3}}, \, \, \, \quad
\alpha_n = \,\,  4 \; (2\;n^2+1) , \, \, \, \, \, \quad
\beta_n = \,\,  2n \; (n+1) \, 
\end{eqnarray}
One has a whole hierarchy of recursions 
{\em integrable}, or compatible with 
integrability~\cite{BoMaRo95}. For instance, one has :
\begin{eqnarray}
\label{recufnt1}
{{f_n \, f_{n+3}^2\, - \, f_{n+4 }\, f_{n+1}^2\,} \over {
f_{n-1} \, f_{n+3}\,\, f_{n+4} \, - \, f_n \, f_{n+1 }\, f_{n+5}\,}}
\, \, = \, \, \, \, {{f_{n+1} \, f_{n+4}^2\,
 - \, f_{n+5 }\, f_{n+2}^2\,} \over {
f_{n} \, f_{n+4}\,\, f_{n+5} \, - \, f_{n+1} \, f_{n+2 }\, f_{n+6}\,}}
\end{eqnarray}
In an equivalent way, introducing the 
variable $\, x_n \, = \, \det({\widehat K}^n(R)) \cdot
\det({\widehat K}^{n+1}(R))$, 
one gets a {\em hierarchy} of recursions on the $\, x_n\, $'s,
(see~\cite{BoMaRo95}), the simplest recursion  reading :
\begin{equation}
\label{label16}
{{{x_{n+2}}-1} \over {{x_{n+1}}\,{x_{n+2}}\,{x_{n+3}}-1}}\, \,\, =\,\,
\,\,\, \,\, {{{x_{n+1}}-1} \over {{x_{n}}\,{x_{n+1}}\,{x_{n+2}}-1}} \cdot
{{x_{n}}\,{x_{n+1}\,x_{n+2}^2}}
\end{equation}
Equation (\ref{label16})  is equivalent to (\ref{recufnt1})
since $\, x_n \, = \, (f_n^3 \cdot f_{n+2})/(f_{n+1}^3 \cdot f_{n-1})$.
It can be seen that these recursions (\ref{recufnt1}) and (\ref{label16})
are  {\em integrable} ones~\cite{BoMaRo95}.
For this, one can introduce~\cite{BoMaRo93a} a new 
(homogeneous) variable :
\begin{equation}
q_n\,  = \, \, {{f_{n+1} \cdot f_{n-1}}\over{f_n^2}} \quad \quad \quad
{\rm then:} \quad \quad \quad x_n \, = \, \, {{q_{n+1}}\over{q_n}}
\end{equation}
and end up, after some simplifications, with the following
biquadratic relation between $\, q_n$ and $\, q_{n+1}$ :
\begin{eqnarray}
\label{formt1}
q_n^2\cdot q_{n+1}^2\, +\, \mu \cdot q_n \cdot q_{n+1} \, +\,
 \rho \cdot (q_n+q_{n+1}) \, -\, \lambda 
\,\, \,\, \, = \,\, \, \,\, \, 0 \, 
\end{eqnarray}
which is clearly an integrable recursion~\cite{BoMaRo95}.
In terms of the $f_n$'s, the three parameters $\, \rho\, $,
$\, \lambda\, $ and $\, \mu\, $ read :
\begin{eqnarray}
\rho \, = \, \, 
{\frac {{f_{1}}^2 \, f_4\, -{f_{2}}^2 \, f_{3}}{
f_{1}^3 f_{3}-\,  f_{2}^3 }}\, , \qquad \quad
\lambda \, = \, \, \,{\frac {\, f_{2}\, f_{4}-\,  f_{1}\,{
f_{3}}^2}{ f_{1}^3\,  f_{3}-\, f_{2}^3}}
 , \qquad \mu \, = \, \,{\frac {{ f_{2}}^{5}\,
-{ f_{3}}^{2}{ f_{1}}^{3}-{ f_{1}}^{5} f_{4}
+ f_{3}\,{ f_{2}}^{3}}{ f_{2}\, f_{1}\cdot  (f_{1}^3\,  f_{3}- f_{2}^3 )}}
  \nonumber
\end{eqnarray}
It may also be interesting to introduce :
\begin{eqnarray}
\label{dzeta}
\kappa \, = \, \, {{4 \cdot \lambda \, + \, \mu^2} \over {\rho}}
\end{eqnarray}

\vskip .2cm

\section{Baxterisation of monodromy matrices : $2 \, m \times 2 \, m$ matrices}
\label{frame}

Let us now recall the more general vertex model (see Fig.~\ref{fig2}
above), where the  singled out direction $(1)$ corresponds to a
{\em two-dimensional}
``auxiliary space'' (that is $\, q
\, = \, 2$). The action
 of $\, t_1$, the ``partial'' transposition
 on direction $(1)$, is given by~\cite{prl2}:
\begin{eqnarray}
(t_1R)^{iJ}_{kL}\, \, = \, \,   R^{kJ}_{iL}\,  \quad  \quad \quad
{\rm that} \,\,\, {\rm  is:} \quad \qquad
\label{t1bloc}
t_1: \;\;\; 
\pmatrix{ 
A & B  \cr 
C & D  \cr 
}
\longrightarrow
\pmatrix{ 
A & C  \cr 
B & D  \cr 
}
\end{eqnarray}
where $A$, $B$, $C$ and $D$ are  $m \times m$ matrices. It is a
 straight calculation to see that the matrix inversion reads :
\begin{eqnarray}
\label{inversbloc}
{\widehat I} : \;\;\; 
\pmatrix{ 
A & B  \cr 
C & D  \cr 
}
\longrightarrow \, \, \, 
\pmatrix{ 
\Bigl( A\, - \, B \cdot D^{-1} \cdot C
\Bigr)^{-1} & \Bigl(C  - \, D \cdot B^{-1} \cdot A \Bigr)^{-1}\cr 
\Bigl(B  - \, A \cdot C^{-1} \cdot D \Bigr)^{-1} 
& \Bigl(D  - \, C \cdot A^{-1} \cdot B\Bigr)^{-1}  \cr 
}
\end{eqnarray}

This general framework enables {\em to take into account the analysis of} $N$-site
 monodromy matrices~\cite{BoMaRo95} (take $m= 2^N$)
 of  two-dimensional models, as well as the analysis of 
$d$-dimensional $2^d$-state {\em vertex models} (take $m= 2^{d-1}$).  
Let us just give here a pictorial representation of the two sites ($N=2$)
 monodromy matrix of a two-dimensional model and of a three-dimensional vertex model:
\vskip .2cm
\vskip .4cm
\vskip .4cm

\begin{equation}
\label{fig111}
\setlength{\unitlength}{0.005in}%
\begin{picture}(120,131)(20,700)
\thicklines
\put( -320,820){\line( 0,-1){120}}
\put( -380,760){\line( 1, 0){200}}
\put( -230,820){\line( 0, -1){120}}
\put(-180,769){\makebox(0,0)[lb]{\raisebox{0pt}[0pt][0pt]{\elvrm $k$}}}
\put( -315,710){\makebox(0,0)[lb]{\raisebox{0pt}[0pt][0pt]{\elvrm $j_1$}}}
\put( -315,810){\makebox(0,0)[lb]{\raisebox{0pt}[0pt][0pt]{\elvrm $l_1$}}}
\put( -225,710){\makebox(0,0)[lb]{\raisebox{0pt}[0pt][0pt]{\elvrm $j_2$}}}
\put( -225,810){\makebox(0,0)[lb]{\raisebox{0pt}[0pt][0pt]{\elvrm $l_2$}}}
\put( -380,769){\makebox(0,0)[lb]{\raisebox{0pt}[0pt][0pt]{\elvrm $i$}}}
\setlength{\unitlength}{0.0075in}%
\put( 135,465){\line( 1, 1){ 80}}
\put( 180,570){\line( 0,-1){120}}
\put( 120,510){\line( 1, 0){120}}
\put(215,545){\makebox(0,0)[lb]{\raisebox{0pt}[0pt][0pt]{\elvrm $j'$}}}
\put( 125,455){\makebox(0,0)[lb]{\raisebox{0pt}[0pt][0pt]{\elvrm $j$}}}
\put(225,515){\makebox(0,0)[lb]{\raisebox{0pt}[0pt][0pt]{\elvrm $i'$}}}
\put( 185,460){\makebox(0,0)[lb]{\raisebox{0pt}[0pt][0pt]{\elvrm $k$}}}
\put( 185,560){\makebox(0,0)[lb]{\raisebox{0pt}[0pt][0pt]{\elvrm $k'$}}}
\put( 120,515){\makebox(0,0)[lb]{\raisebox{0pt}[0pt][0pt]{\elvrm $i$}}}
\put( 175,520){\makebox(0,0)[rb]{\raisebox{0pt}[0pt][0pt]{\elvrm $R$}}}
\end{picture}
\end{equation}
\hfill\null\break\smallskip
Denoting $\, s\, =2 \, m$ the size of the matrices, the analysis of the  corresponding 
factorizations yields
 for arbitrary  $n$, ``string-like''
 factorizations~\cite{BoMaRo95,BoMa95}. For
 arbitrary $\, m$ (equal to $2^{d-1}$ or not), the analysis 
of the factorizations of the iterations of transformation $\, K$ yields:
\begin{eqnarray}
\label{factot2q}
&&M_1 \, = \, K(M_0),\;\; \, 
f_1 \, = \, \det(M_0),\;\;\, 
f_2 \, = \, {{\det(M_1)}\over {f_1^{s-4}}},\;\;\, 
M_2 \, = \,  {{K(M_1)}\over {f_1^{s-5}}},\;\;\, 
f_3 \, = \, {{\det(M_2)}\over {f_1^7 \cdot f_2^{s-4}}},\;\; 
M_3 \, = \, {{K(M_2)}\over{f_1^5 \cdot f_2^{s-5}}}, \nonumber \\
&& f_4 \, = \, {{\det(M_3)}\over{f_1^{2\,(s-4)} \cdot f_2^7 \cdot
f_3^{s-4}}},\;\;\, \,\, \,
M_4 \, = \, {{K(M_3)}\over{f_1^{2\,(s-5)} \cdot
 f_2^5 \cdot f_3^{s-5}}},\;\;\,\, \, \, \, f_5 \, = \, {{\det(M_4)}\over{f_1^8 \cdot 
f_2^{2\,(s-4)} \cdot f_3^7 \cdot f_4^{s-4}}} \, \, \, \, \, \cdots \nonumber
\end{eqnarray}
and, for arbitrary  $n$, the following ``string-like'' factorizations :
\begin{eqnarray}
\label{factt2q}
K(M_n)\,\,   &=& \,\,  \, M_{n+1} \cdot  f_{n}^{s-5}  \cdot f_{n-1}^5 
 \cdot f_{n-2}^{2(s-5)} \cdot f_{n-3}^6 \cdot f_{n-4}^{2(s-5)}
 \cdot f_{n-5}^6 \, \cdots \, , \nonumber \\
det(M_n)\,\,   &=& \,\,  \, f_{n+1}  \cdot f_{n}^{s-4} 
 \cdot f_{n-1}^7  \cdot f_{n-2}^{2(s-4)} 
 \cdot f_{n-3}^8  \cdot f_{n-4}^{2(s-4)} \cdot f_{n-5}^8  
\cdot f_{n-6}^{2(s-4)} \, \cdots
\end{eqnarray}
One easily gets from (\ref{factt2q}) :
\begin{eqnarray}
\label{gener2q}
&&\alpha(x) \,\,  = \, \,\, \; {{s} \over {1+x}} \; \,\, 
+ \,\, \, s^2 \; {{\;x\;(1+x^2)}\over{(1+x) \; (1-x)^4}}, \;
\qquad \quad \beta(x)  \,\, =  \,\,\, {\frac {s\,x}{\left (1-x\right
)^{3}}},
 \; \nonumber \\
&&\alpha_n =\, {{s}\over{3}}\;(2n+1)\,(2n^2+2n+3)\;, \; \qquad 
 \quad \beta_n  =\, {{s} \over {2}} \,n\,(n+1) \quad
\end{eqnarray}
The $\alpha_n$'s and $\beta_n$'s are, respectively, {\em cubic and 
quadratic functions} of $\, n$. 

\subsection{Towards Bethe Ansatz : the propagation property}
This {\em polynomial growth} of the calculation can be understood as
follows. One of the ``keys'' to the Bethe Ansatz is the existence 
(see equations (B.10),  (B.11a) in~\cite{Ba73}) of
 vectors which are {\em pure tensor products} (of the form
 $v \otimes w$) and which $\, R$ maps onto pure tensor product $v'
\otimes w'$. This key property~\footnote{Which is ``almost''
a sufficient condition for the Yang-Baxter equation. In the case of
the Baxter model this non trivial relation corresponds to
some intertwining relation of the product of two theta functions, 
which is nothing but the
{\em quadratic Frobenius relations} on theta
functions~\cite{frobenius,Ga83}.}
was called {\em propagation property} by R. J. Baxter, and 
corresponds to the existence of a {\em Zamolodchikov algebra}~\cite{ZaZa79}
for the Baxter model~\footnote{The existence of a Zamolodchikov algebra
is, at first sight, a {\em sufficient} condition for the Yang-Baxter
equations to be verified. Theta functions of $\, g$ variables
{\em do satisfy  quadratic Frobenius relations}~\cite{Ga83}, 
 consequently {\em yielding 
 a Zamolodchikov algebra parameterized in term of  theta functions of
several variables}. However this Zamolodchikov algebra is apparently
{\em not sufficient} for Yang-Baxter
equations to be satisfied~\cite{Sch82}. For the 
corresponding vertex models ($\, g$ replicas of the Baxter model
coupled together) it may be possible that the
associated $\, R$-matrices, which are parametrised in term of 
theta functions of several variables (and thus correspond to a
``nice'' Baxterisation),  do not satisfy the Yang-Baxter
equations, but could be such that the partition function per
site could be calculated exactly using the ``inversion trick'' : this
remains an open question.}. 
This ``propagation'' equation  reads here :
\begin{eqnarray}
\label{pt}
 R \; ( u \otimes V ) \,\, \,  = \, \, \mu \cdot u' \otimes V' \quad 
\qquad {\rm with:} \qquad \, \, 
u=\pmatrix{ 1 \cr p \cr}, \quad  \quad
u'=\pmatrix{ 1 \cr p' \cr}, \quad 
\end{eqnarray}
vectors V and V' having $\, m\, $ coordinates.
One can rewrite (\ref{pt}) under the form:
\begin{equation}
\pmatrix{                                A  & B  \cr
                                         C  & D  \cr           }
\pmatrix{ V \cr p \cdot V \cr}
\, \, =\, \,\, \, 
\mu \cdot \pmatrix{ V' \cr p' \cdot V' \cr}
\end{equation}

Actually, for all the vertex models for which transposition
 $\, t_1\, $ can be represented as (\ref{t1bloc}) (namely 
monodromy matrices,  or
 $d$-dimensional vertex models, with ``arrows''
taking {\em two} colors, ... ), one can associate 
 an {\em algebraic curve} of equation  :
\begin{eqnarray}
\label{curve}
\det(A \, p'\, -C \, -\,  D \, p +\, p \, p' \, B) 
\,  \,\, \, \,= \, \, \,\, \, 0
\end{eqnarray}
which form is invariant by  $\, t_1\, $,  $\, {\widehat I}\, $ and thus 
by $\,{\widehat  K}\, $ or $\, {\widehat K}^2$. 
As a byproduct this provides 
a {\em canonical Jacobian variety for such vertex models}, namely the 
 Jacobian variety 
associated with curve (\ref{curve}). This procedure, which 
associates with an $R$-matrix the 
algebraic curve (\ref{curve}), originates from 
a key ``propagation''  relation (\ref{pt}), closely related to 
the action of the birational transformations $\, {\widehat K}$~\cite{prl2}.

\subsection{Continuous symmetries generalizing the gauge symmetries }
\label{genergauge}
The birational transformation
$\, {\widehat K}\, = \, t_1 \cdot {\widehat I}\,$ can be
represented as follows :
\begin{eqnarray}
\label{inversblocK}
{\widehat K}\, = \, t_1 \cdot {\widehat I}\, : \;\;\; 
\pmatrix{ 
A & B  \cr 
C & D  \cr 
}\, \,
\longrightarrow \, \, \, \, \,
\pmatrix{ 
A' & B'  \cr 
C' & D'  \cr 
}
\, = \, \, \, \,
\pmatrix{ 
\Bigl( A\, - \, B \cdot D^{-1} \cdot C
\Bigr)^{-1} &
\Bigl(B  - \, A \cdot C^{-1} \cdot D \Bigr)^{-1} \cr 
 \Bigl(C  - \, D \cdot B^{-1} \cdot A \Bigr)^{-1}
& \Bigl(D  - \, C \cdot A^{-1} \cdot B\Bigr)^{-1}  \cr 
}
\end{eqnarray}
Let us introduce the following $\, SL(m) \times SL(m)\, $
transformation $\,  {\cal G}^{(m)}\, $ :
\begin{eqnarray}
 {\cal G}^{(m)}\, : \, \, \, 
\pmatrix{ 
A & B  \cr 
C & D  \cr 
}
\longrightarrow \, \, \,\, 
\pmatrix{ 
G^{(m)}_L \cdot A \cdot G^{(m)}_R & \quad G^{(m)}_L \cdot B \cdot  G^{(m)}_R \cr 
G^{(m)}_L \cdot C \cdot G^{(m)}_R & \quad G^{(m)}_L \cdot D \cdot  G^{(m)}_R  \cr 
} 
\nonumber
\end{eqnarray}
where $\, G^{(m)}_L \,$ and  $\, G^{(m)}_R \,$ are two $\, SL(m)$
matrices. It is straightforward to see that :
\begin{eqnarray}
{\widehat K}  \pmatrix{ 
G^{(m)}_L \cdot A \cdot G^{(m)}_R & G^{(m)}_L \cdot B \cdot  G^{(m)}_R \cr 
G^{(m)}_L \cdot C \cdot G^{(m)}_R & G^{(m)}_L \cdot D \cdot  G^{(m)}_R  \cr 
}
 \, = \, \, 
\pmatrix{ 
(G^{(m)}_R)^{-1} \cdot A' \cdot (G^{(m)}_L)^{-1} 
&  \quad (G^{(m)}_R)^{-1} \cdot B' \cdot  (G^{(m)}_L)^{-1} \cr 
(G^{(m)}_R)^{-1} \cdot C' \cdot (G^{(m)}_L)^{-1} 
& \quad  (G^{(m)}_R)^{-1} \cdot D' \cdot  (G^{(m)}_L)^{-1}  \cr 
}
\end{eqnarray}
and, in a second step, that :
\begin{eqnarray}
\label{slmslm}
 {\cal G}^{(m)} ({\widehat K}^2 (R)) \, = \, \,\, 
{\widehat K}^2 ( {\cal G}^{(m)}(R))
\end{eqnarray}
In fact, such a result is {\em not specific of
 a two-dimensional auxiliary
space}. Recalling the  $\, (q\, m) \times  (q\, m)\, $ $\, R$-matrix
(\ref{3t1bloc}), and the associated partial transposition $\, t_1$,
and introducing a  $\, SL(m) \times SL(m)\, $
transformation $\,  {\cal G}^{(m)}\, $ which transforms 
each $\, m \times m\, $ block $\, A[\alpha, \, \beta]\, $
into $\, G_L \cdot  A[\alpha, \, \beta] \cdot  G_R  $, 
one recovers, again, relation (\ref{slmslm}).
Of course there is nothing specific with $\, t_1$, and one 
finds the same results for the partial 
transposition $\, t_2\, $ corresponding to the vertical line,
with this time, a commutation between $\, {\widehat K}^2\, $
and a  $\, SL(q) \times SL(q)\, $
transformation $\,  {\cal G}^{(q)}$. Since $\, t_2\, = \, t_1 \cdot
 t$,  and since $\, t$ commutes with $\, {\widehat I}$,
one immediately gets~\footnote{ Similarly for a 
three-dimensional $\, 2^3 \times 2^3\, $ $\, R$-matrix
one gets easily $\, 
\widehat{K}^2(  g_{1L}^{-1} \otimes g_{2L}^{-1} \otimes 
g_{3L}^{-1} \cdot  R_{3D}  \cdot g_{1R} \otimes g_{2R}  \otimes g_{3R} )\,
 \, = \, \, \,\,
 g_{1L}^{-1} \otimes g_{2L}^{-1} \otimes g_{3L}^{-1} \cdot {\widehat K}^2( R_{3D} )
 \cdot g_{1R} \otimes g_{2R} \otimes g_{3R}$.} with obvious notations the
 following $\,  SL(q) \times SL(m) \times  SL(q) \times
SL(m)\,$ symmetry for  $\, {\widehat K}^2\, $ :
\begin{eqnarray}
\label{35}
{\widehat K}^2 \Bigl(
G^{(q)}_L  \otimes G^{(m)}_L \cdot R \cdot   G^{(q)}_R  \otimes  \,G^{(m)}_R 
\Bigr)\, = \, \, \, 
G^{(q)}_L  \otimes G^{(m)}_L \cdot
\Bigl( {\widehat K}^2 (R)\Bigr)
 \cdot  G^{(q)}_R  \, \otimes G^{(m)}_R 
\end{eqnarray}
This symmetry drastically generalizes the
well-known {\em gauge symmetries}~\cite{GaHi75} 
which correspond to $\, G^{(q)}_R \, = \, (G^{(q)}_L)^{-1}\, $ 
and  $\, G^{(m)}_R \, = \, (G^{(m)}_L)^{-1}$. Let us note 
that this kind of symmetries, generalizing the gauge transformations,
have actually been used by R. J. Baxter to map  the Baxter
model onto
an {\em inhomogeneous} six vertex model, in order
to build the
Bethe Ansatz of the Baxter model~\cite{Ba73}.

\section{ Sixteen vertex model and Baxter model : 
revisiting the elliptic curves}
\label{revisit}

Considering the (non-generically Yang-Baxter integrable)
sixteen vertex model,
 one finds that a  {\em canonical} parameterization
in terms of {\em  elliptic curves}
occurs in the sixteen homogeneous parameter space of the
model~\cite{prl2}.
This canonical parameterization is obtained~\footnote{In the 
Yang-Baxter integrable subcase,
the Baxter model, this elliptic parameterization, deduced from the
Baxterisation procedure,
is actually the  elliptic parameterization 
introduced by R. J. Baxter to solve the Baxter
 model~\cite{Ba81}.} from the {\em Baxterisation} 
procedure~\cite{prl2}. In fact several elliptic curves (associated
with different ``spaces'') occur :
one corresponding to the factorization analysis of 
section (\ref{sixteen}), another from the propagation property (\ref{pt}),
and another one from the iteration 
of the birational transformation $\, {\widehat K}^2$ 
in the sixteen homogeneous parameter space of the model~\cite{prl2}.
Let us analyse the relations between these various elliptic curves
and show that they {\em actually identify}.

Let us recall the results and notations concerning the sixteen vertex
model~\cite{prl2}.
Let us use the following notation for  $R$ :
\begin{eqnarray}
\label{Ra1a2}
R=
\pmatrix{	a_1 &	a_2 & 	b_1 & 	b_2 \cr
		a_3 & 	a_4 &	b_3 &	b_4 \cr
		c_1 & 	c_2 & 	d_1 & 	d_2 \cr
		c_3 & 	c_4 & 	d_3 & 	d_4 \cr	}
\end{eqnarray}

\subsection{ Propagation property for the sixteen vertex model and the
Baxter model}
\label{preb}
Considering the sixteen vertex model (\ref{Ra1a2}),
the propagation relation (\ref{pt}), for $\,  m\, = \, 2$,
becomes~\footnote{
See the propagation property for the Baxter model, namely (B.10),  (B.11a) in~\cite{Ba73}.}

$\,  R \; ( v_n \otimes w_n )\,\,   =\, \,\,\,  \mu \cdot
v_{n+1}\otimes w_{n+1} \,\,\, $ where :
\begin{eqnarray}
\label{propag}
v_n=\pmatrix{ 1 \cr p_n \cr}, \quad \quad w_n=\pmatrix{ 1 \cr \tilde{p}_n \cr}, \quad \quad
v_{n+1}=\pmatrix{ 1 \cr p_{n+1} \cr}, \quad \quad w_{n+1}=\pmatrix{ 1 \cr \tilde{p}_{n+1} \cr}
\end{eqnarray}
and yields, by eliminations of $\, p_n, \,   p_{n+1}$ (resp.
 $\, {\tilde p}_n, \,   {\tilde p}_{n+1}$),   the two
 biquadratic relations~\cite{prl2} :
\begin{eqnarray} 
\label{biq1}
&&l_4\,  +\,  l_{11}  \cdot p_n \, - l_{12}
  \cdot p_{n+1} +\, l_2  \cdot p_n^2 \, + l_1  \cdot p_{n+1}^2 \, 
-\,  (l_9 + l_{18}) \cdot p_n \cdot  p_{n+1}  \\
&& \qquad \qquad  \qquad  \qquad   \qquad  -\,  l_{13} \cdot  p_n^2  \cdot  p_{n+1} \, 
+ \, l_{10} \cdot  p_n  \cdot  p_{n+1}^2\,  +\,  l_3  \cdot p_n^2  \cdot  p_{n+1}^2
\, \, \, =\,\, \, \, 0 \nonumber
\end{eqnarray}
\begin{eqnarray}
\label{biq2}
&&l_7 \, + \, l_{16} \cdot  \tilde{p}_n\,  - \, l_{15} \cdot
 \tilde{p}_{n+1} +\, l_8 \cdot \tilde{p}_n^2
 +\,  l_5  \cdot \tilde{p}_{n+1}^2 -\,  (l_{9} - l_{18})  \cdot
 \tilde{p}_n  \cdot \tilde{p}_{n+1} \\
&& \qquad \qquad  \qquad  \qquad   \qquad 
 -\,  l_{17}  \cdot  \tilde{p}_n^2  \cdot \tilde{p}_{n+1}\,  + l_{14}
 \cdot \tilde{p}_n
 \tilde{p}_{n+1}^2 
+\, l_6\,  \tilde{p}_n^2 \, \tilde{p}_{n+1}^2\, \, \, =\,\, \, \, 0 \nonumber
\end{eqnarray}
where the $\, l_i$'s are quadratic expressions of the 
entries (\ref{Ra1a2}) of the $\, R$-matrix~\cite{prl2}.
These two biquadratics can be seen as :
\begin{eqnarray}
[p'^2, p', 1] \cdot R_3^{(1)} \cdot 
\left [\begin {array}{c} p^2\\
\noalign{\medskip}p\\
\noalign{\medskip}1
\end {array}\right ] \, = \, \, 0
\quad \qquad \mbox{and :}\quad
\qquad
[q'^2, q', 1] \cdot R_3^{(2)} \cdot
\left [\begin {array}{c} q^2\\
\noalign{\medskip}q\\
\noalign{\medskip}1
\end {array}\right ] \, = \, \, 0 \nonumber
\end{eqnarray}
where the two $3 \times 3$ matrices read :
\begin{eqnarray}
\label{3x3}
R_3^{(1)} \, = \, \, \, \,
 \left [\begin {array}{ccc} 
l_1 &l_{10}&l_3\\
-l_{12}&-(l_9+l_{18})& -l_{13}\\
l_4&l_{11}& l_2
\end {array}\right ]
\qquad\quad \mbox{and :}\qquad\quad
R_3^{(2)} \, = \, \, \, \,
 \left [\begin {array}{ccc} 
l_5&l_{14}&l_6\\
-l_{15}&-(l_9-l_{18})& -l_{17}\\
l_7 &l_{16}& l_8
\end {array}\right ]
 \end{eqnarray}
In the Baxter limit 
these two matrices reduce to a 
{\em only one} $\, 3 \times 3$ matrix :
\begin{eqnarray}
\label{form}
R_3^{bax} \, \, \, = \, \, \, \, \,
 \left [\begin {array}{ccc} 
J_x+J_y &0&J_x-J_y \\
0&4\, J_z& 0\\
J_x-J_y&0& J_x+J_y 
\end {array}\right ]
\end{eqnarray}
where $\, J_x\, $ , $J_y$, and $J_z$
are the three well-known quadratic expressions
 of the $\,XYZ\, $ Hamiltonian :
\begin{eqnarray}
J_x \, = \, \,  \,a \cdot b\, + \, c \cdot d \, , \, \qquad \quad 
J_y \, = \, \,  \,a \cdot b\, - \, c \cdot d \, , \, \quad 
 \qquad J_z \, = \, \, {{a^2+b^2-c^2-d^2} \over {2}} , \nonumber
\end{eqnarray}

Some  $\, {\widehat K}^2$-invariants can be 
deduced from $\, SL(3)\, $ invariants of the two $3 \times 3$ 
matrices (\ref{3x3}), namely a quadratic expression in the $l_i$'s  
(for $R_3^{(1)}$ for instance) $\, 
l_1 \cdot l_2 \, + \, l_3 \cdot l_4  \, - \, l_{10} \cdot l_{11} 
 \, - \, l_{12} \cdot l_{13} + (l_9 \, + \, l_{18})^2$, 
a cubic (in the  $l_i$'s ) which is nothing but their {\em determinant},
and a quartic one. Eighteen (algebraically related) {\em quadratic} polynomials
($p_1,\dots,p_{18}$) which are linear combinaisons
of the $\, l_{i}$'s, and 
transform very simply  under $\, t_1$ and $\,
I$, have been found~\cite{prl2}. Introducing the ratio
of these covariants $\, p_i$'s, one gets {\em invariants}
of $\, {\widehat K}^2\, $ thus giving the equations of the
elliptic curves : the elliptic curves are given by the intersection
of fourteen quadrics~\cite{prl2}.

\subsection{Reduction of a sixteen vertex model to a
 $\, K^2$-effective Baxter model}

From (\ref{35}) for $\, q \, = \, m \, = \,  2$, one deduces a
 $\, sl_2 \times sl_2 \times sl_2 \times sl_2$ symmetry
on the sixteen vertex model~\cite{prl2}.
Furthermore, the $\, R$-matrix of the sixteen vertex 
{\em can actually be decomposed}~\footnote{Finding, 
for a given  $\, R$-matrix
of the sixteen vertex 
model, the elements of this decomposition, namely $\, R_{Baxter} \, $ 
and  $g_{1R}$, $g_{2R}$,  $g_{1L}$, $g_{2L}$, is an extremely difficult
process that will not be detailed here. Conversely, one can show easily
that the matrices of the form (\ref{decompoBMV}) {\em span the whole space}
of $\, 4 \times 4$ matrices.} as :
\begin{eqnarray}
\label{decompoBMV}
R_{sixteen} \,\,\,\, = \, \,\, \,\,\, \,\,\,
 g_{1L}^{-1} \otimes g_{2L}^{-1}\cdot R_{Baxter} \cdot g_{1R} \otimes g_{2R}
\end{eqnarray}
where $\, R_{Baxter} \, $ denotes the $\, R$-matrix of an
 ``effective''
 Baxter model
and  $\, g_{1R}$, $\, g_{2R}$,  $\, g_{1L}$, $\, g_{2L}$ are $\, 2 \times 2$
matrices. The sixteen homogeneous parameters of 
the sixteen vertex are thus decomposed into four  homogeneous parameters of 
an ``effective'' Baxter model and four times three parameters (four homogeneous
parameters) of the 
various $2 \times 2$ matrices :  $\, g_{1R}$, $\, g_{2R}$,  $\,
g_{1L}$, $\, g_{2L}$.
Using this very decomposition  (\ref{decompoBMV}),
and the previous symmetry relation (\ref{35}) for $\, q\, = \, m\,
=2$, one actually gets :
\begin{eqnarray}
\label{decompofinqq}
\widehat{K}^2(  R_{sixteen})\, \, = \, \, \,\, \, \, \,\,
 g_{1L}^{-1} \otimes g_{2L}^{-1}\cdot \widehat{K}^2( R_{Baxter})
 \cdot g_{1R} \otimes g_{2R}
\end{eqnarray}
The matrices  $\,\,  g_{1R}$, $\,\,  g_{2R}$,  $\,\,  g_{1L}$, $\,\,  g_{2L}$
of the decomposition (\ref{decompoBMV}) 
can thus be seen as {\em constants of motion}
of the iteration of $\widehat{K}^2$.

If  $R_{Baxter}$ belongs to a ``special''  manifold, or algebraic variety,
$R_{sixteen}$, given by (\ref{decompoBMV}), will
{\em also belong} to  a ``special'' manifold, or algebraic variety : for instance, if 
$\, R_{Baxter}$ belongs to a 
{\em finite order} algebraic variety for the iteration
of $\widehat{K}^2$, namely $\widehat{K}^{2 \, N}(R_{Baxter})\, = \, $
$\,\eta \cdot  R_{Baxter}$, then $\, R_{sixteen}$ will
also belong to
a finite order algebraic variety for the  iteration
of $\widehat{K}^2$ : $\widehat{K}^{2 \, N}(R_{sixteen})\, = \, $
$\eta \cdot R_{sixteen}$.
If $\, R_{Baxter}$ belongs to a {\em critical variety} then 
$R_{sixteen}$ given by (\ref{decompoBMV}) should
also belong to  a critical variety.
This last result does not come from the fact that 
 $g_{1L}$, $g_{2L}$, 
 $g_{1R}$, $g_{2R}$ are symmetries of the partition function (they are
not, except in the gauge case : $g_{1L}$
 $\, = \, g_{1R}$ with  $\, g_{2L}$
 $\, = \, g_{2R}$) : they are symmetries of $\, \widehat{K}^2$
which is a {\em symmetry of the critical manifolds}.
Therefore they are {\em symmetries of the critical manifolds
even if they are not symmetries of the partition function}.

A decomposition, like (\ref{decompoBMV}),
is closely associated to
the {\em parametrization of the sixteen vertex model}
 in {\em elliptic curves}~\cite{prl2} :
given $R_{Baxter}$, $g_{1L}$, $g_{2L}$, 
 $g_{1R}$ and $g_{2R}$, one can easily deduce 
$R_{sixteen}$. Conversely, given  $R_{sixteen}$,
it is extremely difficult to get $R_{Baxter}$, $g_{1L}$, $g_{2L}$, 
 $g_{1R}$ and $g_{2R}$, however, and remarkably,
it is quite simple to get $\, R_{Baxter}$. Since  $\, g_{1L}$,
 $\,  \, g_{1R}$,   $g_{2L}$, 
 $\, g_{2R}$ are $\, {\widehat K}^2$-invariants,
one can try to relate, directly, the ``${\widehat K}^2$-effective''
covariants $\, J_x$, $\, J_y$ and $\, J_z$ with the 
 $\, {\widehat K}^2$-invariants 
related to the recursion on the  $\, x_n$'s or the $\,q_n$'s,
namely $\,\rho$, $\, \mu$, $\, \lambda$, or
 $\, \kappa$ (see  (\ref{formt1}), (\ref{dzeta})). In terms
 of these well-suited algebraic covariants, the previous
parameters read :
\begin{eqnarray}
\label{44}
&&\rho \, = \, \, \, 4 \, J_z^2 \, J_x^2 \, J_y^2\,, \qquad \quad
 \mu  \, = \, \, -\, 2 \cdot
(J_z^{2}J_x^{2}+J_z^{2}J_y^{2}+J_x^{2}J_y^{2} ) \, , \qquad 
\kappa  \, = \, \, 4 \cdot  (J_z^2\, +\, J_x^2\, +\, J_y^2)\nonumber \\
&&\lambda  \, = \, \,-\,  (J_z^2 \, J_x^{2}\, +J_z^{2}J_y^{2}\, +J_x^{2}
J_y^2 )^2\, +\, 4 \cdot  (J_z^2\, +J_x^2\, +\, J_y^2)
 \cdot J_z^{2}J_x^{2}J_y^{2}\,  
\end{eqnarray}

One immediately recognizes some symmetric
polynomials of $\,J_x^2$, $\,J_y^2$, $\,J_z^2$. Therefore
 it is easy to see
that the $J_x$, $\, J_y$, $\, J_z$ can be
straightforwardly obtained from
a cubic polynomial
$\, P(u)\, $ :
\begin{eqnarray}
\label{45}
P(u) \, = \, \, 4 \cdot u^3 \, -\kappa \cdot u^2\, -2 \cdot \mu \cdot
u\, - \rho \, = \, \,
4 \cdot (u-J_x^2) \cdot  (u-J_y^2) \cdot  (u-J_z^2) 
\end{eqnarray}
This is remarkable, because trying to get  $\, J_x$, $\, J_y$, $\, J_z$,
by brute-force eliminations
from (\ref{decompoBMV}), yields huge 
calculations. In fact, one only gets, from (\ref{45}), 
the {\em squares} of the  $\, J_x$, $\, J_y$, $\, J_z$,
but  the critical manifold, as well as the
finite order conditions (see below),
only depend on   $J^2_x$, $\, J^2_y$, $\, J^2_z$.

\subsection{Biquadratic (19) versus biquadratic (35)}
\label{biqbiq}
The Baxterisation process is associated with the iteration of 
$\, {\widehat K}$,
or, rather, $\, {\widehat K}^2$.
Since, as far as $\, {\widehat K}^2\, $ is concerned,  one 
can reduce a sixteen vertex model to 
an ``effective'' Baxter model, one can 
try to revisit, directly, the relation 
between  the biquadratic (\ref{formt1}) and the
 ``propagation curve'' (\ref{biq1})
for the Baxter model. For the Baxter model relation (\ref{formt1}) becomes the biquadratic :
\begin{eqnarray}
\label{qnqnp1t1}
&&\,q_n^{2}\,  q_{n+1}^{2}\,\,  -\, 2 \cdot
(J_z^{2}J_x^{2}+J_z^{2}J_y^{2}+J_x^{2}J_y^{2} ) \cdot q_n \cdot q_{n+1} \, +\, 4
 J_z^2 \, J_x^2 \, J_y^2\,\cdot 
 ( q_n +q_{n+1} ) \nonumber \\
&&\qquad \qquad \qquad  +\,  (J_z^2 \, J_x^{2}\, +J_z^{2}J_y^{2}\, +J_x^{2}
J_y^2 )^2\, -\, 4 \cdot  (J_z^2\, +J_x^2\, +\, J_y^2)
 \cdot J_z^{2}J_x^{2}J_y^{2}\,\, \, \,\, = \, \,\, \,\, \, \, 0
\end{eqnarray}
which should be compared with the ``propagation'' biquadratics
(\ref{biq1})
of the Baxter model~\cite{Ba73,Ba81} :
\begin{eqnarray}
\label{vacuumJ1} 
&&\Gamma_1(p_n , \, p_{n+1})\, \, \, = \, \, \, \, \, \,
(J_x \, - \, J_y) \cdot (p_n^{2}\, p_{n+1}^{2}\, +1 )\, \, 
-\, (J_x \, + \, J_y) \cdot  (p_n^2+p_{n+1}^2 )\, +\,4 \cdot  J_z
\cdot p_n \cdot p_{n+1}\,  \nonumber \\
&&\, \, \, = \, \, \, \, \, \,
(p_n^{2}\,-1) \cdot (p_{n+1}^2 \,-1)\cdot J_x - (p_n^{2}\,+1) \cdot
(p_{n+1}^2 \,+1)\cdot J_y
+\, \,4 \cdot  J_z
\cdot p_n \cdot p_{n+1}\,
\, \, \, = \, \, \, \, \, \, 0
\end{eqnarray} 

It is known that simple ``propagation'' curves, like (\ref{vacuumJ1}), have
the following elliptic parameterization~\cite{Ba73,Ba81}:
\begin{eqnarray}
p_n \, = \, sn(u_n, k)\, , \quad \quad p_{n+1} \, = \, sn(u_{n+1},
k)\,\quad \quad {\rm where :} \quad \quad  u_{n+1}\, = \, u_n \, \pm \, \lambda
\end{eqnarray}
where $\, sn(u, k)\, $ denotes the elliptic sinus of modulus $\, k$
and $\, \lambda$ denotes some ``shift''.
The modulus~\cite{BoMa92b} $\, k\, $ is equivalent to the 
following modulus which has a very simple 
expression in terms of  $\, J^2_x$, $\, J^2_y$ and $\, J^2_z$:
\begin{eqnarray}
\label{Mod}
M_{od} \, \, = \, \, \,
   {{J_x^2 \, + \, J_z^2 - 2 \cdot J_y^2 } \over {J_x^2 \, - \, J_z^2}}\, \,  \, 
\end{eqnarray}

At first sight it seems that one has two different 
elliptic curves (biquadratics), namely
 (\ref{qnqnp1t1}) which is symmetric under permutations of $\, J_x^2$, $\, J_y^2$
and $\, J_z^2$,
and  (\ref{vacuumJ1}) which {\em breaks this symmetry}.
Let us also consider the same ``propagation
curve'' (\ref{vacuumJ1}), but now between $\, p_{n+1}$ and $\, p_{n+2}$,
and let us eliminate $p_{n+1}$ between these two algebraic curves. One
gets, after the factorization of $\,(p_n \, - \,  p_{n+1})^2\, $  :
\begin{eqnarray}
\label{vacuumJ2} 
&&\Gamma_2(p_n , \, p_{n+2})\,
 \, \, \, \, = \, \, \, \, \, \, \\
&& \quad \, \, = \, \, 
 2\, J_z^{2}\cdot  (J_y^{2}-J_x^2 )\cdot (p_n^{2}\, p_{n+2}^{2}\, +1 )\, \, 
+\,2\, J_x^{2}J_y^{2}\cdot  (p_n^2+p_{n+2}^2 )\,  
+\, 4 \cdot  \,(J_x^{2}J_y^{2}\, -J_z^{2}J_x^{2}\, -J_z^{2}J_y^{2}) \cdot p_n
\cdot p_{n+2}   \nonumber \\
&& \quad  \, = \, \, (p_n^{2}\,-1) \cdot (p_{n+2}^2 \,-1)\cdot J^{(2)}_x 
- (p_n^{2}\,+1) \cdot
(p_{n+2}^2 \,+1)\cdot J^{(2)}_y\,  +\, \,4 \cdot  J^{(2)}_z
\cdot p_n \cdot p_{n+2}\,\,\, \, \, = \, \, \, \, \, \, 0\nonumber 
\end{eqnarray}
where $\,  J^{(2)}_x $,  $\,  J^{(2)}_y $ and $\,  J^{(2)}_z $
are given below.
One remarks that (\ref{vacuumJ2}) is actually of the {\em same form}
as (\ref{vacuumJ1}).
The two biquadratic curves  (\ref{qnqnp1t1})
and (\ref{vacuumJ2}) are (birationally) equivalent, their
shift $\, \lambda\, $ and modular invariant~\cite{BoMa92b} being equal. 
Actually one can find directly the homographic
 transformation 
\begin{eqnarray}
\label{homog11}
 \, q_n \, = \, \, {{\alpha \cdot p_n \, +
\, \beta} \over {\gamma \cdot p_n \, + \, \delta}}\, , \qquad \qquad 
 \, \, q_{n+1} \, = \, \, {{\alpha \cdot p_{n+2} \, +
\, \beta} \over {\gamma \cdot p_{n+2} \, + \, \delta}}
\end{eqnarray}
which maps (\ref{qnqnp1t1})
onto (\ref{vacuumJ2}), the parameters 
$\,\alpha$, ... $\delta\, $ of the homographic
transformation
(\ref{homog11}) being quite involved. The calculations are quite
tedious and will be given elsewhere.

The previously described elimination of $\, p_{n+1}$, changing  
$\, \Gamma_1\, $ into
$\, \Gamma_2$, amounts to eliminating  $\, u_{n+1}$ between 
 $\, u_n \, \longrightarrow \,  u_{n+1}\, = \,
u_n \, \pm \lambda \,  $ and  $\, u_{n+1} \, \longrightarrow \,  u_{n+2}\, = \,
u_{n+1} \, \pm \lambda \,  $ thus getting
 $\, u_n \, \longrightarrow \,  u_{n+2}\, = \,
u_n \, \pm 2 \cdot \lambda$,  together with two times $\,u_n \,
 \longrightarrow \,  u_{n+2}\, = \,u_n$. Considering the coefficients 
of the biquadratic (\ref{vacuumJ2})
one thus gets, very simply, a  {\em polynomial representation} of the shift 
doubling $\, \lambda   \longrightarrow \, 2 \cdot \lambda \,  $ :
\begin{eqnarray}
\label{doubling}
&&J_x \qquad \longrightarrow \qquad
 J^{(2)}_x \, \, = \, \,  \,\,\,
-J_z^2 \, J_x^2+J_z^2 \, J_y^2-J_x^2 \, J_y^2
 \nonumber \\
&&J_y \qquad \longrightarrow \qquad
 J^{(2)}_y \,\, = \, \,\,\,\,
J_z^2 \, J_x^2-J_z^2 \, J_y^2-J_x^2 \, J_y^2
\nonumber \\
&&J_z \qquad \longrightarrow \qquad
 J^{(2)}_z \,\, = \, \,\,\,\,
-J_z^2 \, J_x^2-J_z^2 \, J_y^2+J_x^2 \, J_y^2
\end{eqnarray}
The modulus (\ref{Mod}) is (as it should) 
{\em invariant} by
 (\ref{doubling}), which represents the
shift doubling transformation.

Of course there is nothing specific with the 
shift doubling : similar calculations can be 
performed to get  {\em polynomial representations} of 
$\, \lambda   \longrightarrow \, M \cdot \lambda$,
for {\em any integer} $\, M$. A general calculation,
corresponding to eliminations between 
two biquadratics  $\, \Gamma_1\, $ of same modulus 
(\ref{Mod}), but different shifts $\, \lambda  \, $
and  $\, \lambda'$, will be given elsewhere.

{\bf Terminology problem :} From the point of view of discrete
dynamical systems a mapping like (\ref{doubling}) (and the mappings
given below (\ref{tripling}), (\ref{mult5}), ...) 
could, at first sight, be called ``integrable'' : the iteration of 
this (two-dimensional) mapping ``densifies''
 algebraic curves (elliptic curves) foliating the
whole two-dimensional space,
namely $\, M_{od} \, \, = \, \, \, constant$, exactly as an integrable
mapping does~\cite{BoMaRo93a,BoMaRo93c}. One can even write 
{\em explicit analytical expressions} for the $\, N$-th iterate,
{\em for any} $\, N$. However, this mapping is {\em not reversible},
the growth of the calculations~\cite{zeta,topo,growth}
 is exponential ($\, 2^N$ {\em exponential
growth}, $\, \ln(2)$ topological entropy, ...). In fact this very example 
of ``calculable'' chaos is
the exact equivalent of the situation encountered with the logistic
map $\, x \, \rightarrow \, \alpha \cdot x \cdot (1-x)\, $ for 
$\, \alpha \, = \, 4\, $ : one does not have a representation of 
a translation $\, \theta \, \rightarrow \, \theta \,+ \,  N \cdot
\lambda$, but 
 a representation of the iteration of a multiplication 
by $\, 2\, $ : $\, \theta \, \rightarrow \, 2^N \cdot \theta $.

\section{Polynomial representations of the multiplication of the shift
by an integer and associated finite order conditions}
\label{shift}

The multiplication of the shift by three can be obtained
using the previous elimination procedure, namely eliminating $y$ between
$\, \Gamma_2(x , \, y)\,$
and $\, \Gamma_1(y , \, z)\,$
(or equivalently eliminating $y$ between
$\, \Gamma_1(x , \, y)\,$
and $\, \Gamma_2(y , \, z)$), 
thus yielding a resultant which factorizes
into two biquadratics of the same form as the two previous ones,
namely $\,  \Gamma_1(x , \, z)\,$
and $\,  \Gamma_3(x , \, z)\,$ :
\begin{eqnarray}
\label{vacuumJ33} 
&&\Gamma_3(x , \, z)\, \, \, = \, \, \, \, \, \,
(x^2\,-1) \cdot (z^2 \,-1)\cdot J^{(3)}_x - (x^{2}\,+1) \cdot
(z^2 \,+1)\cdot J^{(3)}_y
+\, \,4 \cdot  J^{(3)}_z
\cdot x \cdot z\,
\, \, \, = \, \, \, \, \, \, 0
\end{eqnarray}
where $\,  J^{(3)}_x $,  $\,  J^{(3)}_y $ and $\,  J^{(3)}_z $
are given below.

The multiplication of the shift by three
has the following  {\em polynomial representation} $\, (J_x, \, J_y, \, J_z) \,
\rightarrow \,  (J^{(3)}_x, \, J^{(3)}_y, \, J^{(3)}_z) $ :
\begin{eqnarray}
\label{tripling}
&&J^{(3)}_x \, = \, \,J_x \cdot (-2 \, J_z^2 \, J_y^2 \, J_x^4
-3 \, J_y^4 \, J_z^4+2 \, J_y^2 \, J_z^4 \, J_x^2+J_y^4 \, J_x^4
+2 \, J_y^4 \, J_z^2 \, J_x^2+J_z^4 \, J_x^4)
 \nonumber \\
&&J^{(3)}_y \, = \, \,J_y \cdot (-3 \, J_z^4 \, J_x^4
+2 \, J_z^2 \, J_y^2 \, J_x^4+J_y^4 \, J_x^4-2 \, J_y^4 \, J_z^2 \,
J_x^2
+J_y^4 \, J_z^4+2 \, J_y^2 \, J_z^4 \, J_x^2)
 \nonumber \\
&&J^{(3)}_z \, = \, \,J_z \cdot (J_y^4 \, J_z^4+2 \, J_y^4 \, J_z^2 \,
J_x^2
-3 \, J_y^4 \, J_x^4-2 \, J_y^2 \, J_z^4 \, J_x^2+2 \, J_z^2 \, J_y^2 \, J_x^4+J_z^4 \, J_x^4)
\end{eqnarray}
The modulus  (\ref{Mod}) is (as it should) invariant by
the polynomial representation (\ref{tripling}) of the
multiplication of the shift
by three (\ref{tripling}).

The multiplication of the shift by four
has the following polynomial  representation $\, (J_x, \, J_y, \, J_z) \,
\rightarrow \,  (J^{(4)}_x, \, J^{(4)}_y, \, J^{(4)}_z) $ :
\begin{eqnarray}
\label{mult4}
&&J^{(4)}_x \, = \, \,-4\, J_x^6\, J_y^6\, J_z^4-6\, J_z^8\, J_y^4\,
J_x^4
+4\, J_z^8\, J_y^6\, J_x^2+4\, J_z^8\, J_y^2\, J_x^6+4\, J_z^6\,
J_y^6\, J_x^4
-4\, J_z^2\, J_x^8\, J_y^6+10\, J_z^4\, J_x^8\, J_y^4\nonumber \\
&&\quad -4\, J_z^6\, J_x^8\, J_y^2-J_x^8\, J_y^8-J_z^8\, J_y^8-J_z^8\, J_x^8
+4\, J_z^2\, J_y^8\, J_x^6-4\, J_z^6\, J_x^6\, J_y^4+4\, J_z^6\,
J_y^8\, J_x^2
-6\, J_z^4\, J_y^8\, J_x^4 \nonumber \\
&&J^{(4)}_y \, = \, \, -J_x^8\, J_y^8-J_z^8\, J_y^8-J_z^8\, J_x^8
-4\, J_x^6\, J_y^6\, J_z^4-6\, J_z^8\, J_y^4\, J_x^4
+4\, J_z^8\, J_y^6\, J_x^2+4\, J_z^8\, J_y^2\, J_x^6-4\, J_z^2\, J_y^8\, J_x^6\nonumber \\
&&\quad +10\, J_z^4\, J_y^8\, J_x^4-4\, J_z^6\, J_y^6\, J_x^4
-4\, J_z^6\, J_y^8\, J_x^2+4\, J_z^2\, J_x^8\, J_y^6
-6\, J_z^4\, J_x^8\, J_y^4+4\, J_z^6\, J_x^8\, J_y^2+4\, J_z^6\, J_x^6\, J_y^4\nonumber \\
&&J^{(4)}_z \, = \, \,-J_x^8\, J_y^8-J_z^8\, J_y^8-J_z^8\, J_x^8
-6\, J_z^4\, J_x^8\, J_y^4+4\, J_z^6\, J_x^8\, J_y^2
+4\, J_z^2\, J_y^8\, J_x^6-4\, J_z^6\, J_y^6\, J_x^4
+10\, J_z^8\, J_y^4\, J_x^4\nonumber \\
&&\quad -4\, J_z^8\, J_y^6\, J_x^2
-4\, J_z^8\, J_y^2\, J_x^6\, 
+4\, J_x^6\, J_y^6\, J_z^4-4\, J_z^6\, J_x^6\, J_y^4
+4\, J_z^2\, J_x^8\, J_y^6+4\, J_z^6\, J_y^8\, J_x^2-6\, J_z^4\, J_y^8\, J_x^4
\end{eqnarray}
which can be obtained, either
by the  elimination of $\, y\, $ between
$\, \Gamma_2(x , \, y)\,$
and $\, \Gamma_2(y , \, z)\,$ (and extracting a
$\, (x-z)^2\, $ factor in the resultant),
or, equivalently,  by the  elimination of $\, y\, $ between
$\, \Gamma_1(x , \, y)\,$
and $\, \Gamma_3(y , \, z)$,
or the  elimination of $\, y\, $ between
$\, \Gamma_3(x , \, y)\,$
and $\, \Gamma_1(y , \, z)$ (and extracting a $\, \Gamma_2$ factor in
the resultant).
Again, one gets $ \, \Gamma_4(x , \, z)\, $ :
\begin{eqnarray}
\label{vacuumJ44} 
&&\Gamma_4(x , \, z)\, \, \, = \, \, \, \, \, \,
(x^2\,-1) \cdot (z^2 \,-1)\cdot J^{(4)}_x - (x^{2}\,+1) \cdot
(z^2 \,+1)\cdot J^{(4)}_y
+\, \,4 \cdot  J^{(4)}_z
\cdot x \cdot z\,
\, \, \, = \, \, \, \, \, \, 0
\end{eqnarray}
where $\,  J^{(4)}_x $,  $\,  J^{(4)}_y $ and $\,  J^{(4)}_z $
are given above.  It can easily be verified that (\ref{mult4})
can be obtained directly combining (\ref{doubling}) with
itself. 

The multiplication of the shift by five
has the following polynomial representation $\, (J_x, \, J_y, \, J_z) \,
\rightarrow \,  (J^{(5)}_x, \, J^{(5)}_y, \, J^{(5)}_z) $ :
\begin{eqnarray}
\label{mult5}
&&J^{(5)}_x \, = \, \,J_x \cdot P^{(5)}_x(J_x, \, J_y, \, J_z) 
 \nonumber \\
&&J^{(5)}_y \, = \, \,J_y \cdot P^{(5)}_y(J_x, \, J_y, \, J_z) 
\, = \, \,J_y \cdot P^{(5)}_x(J_y, \, J_z, \, J_x) 
 \nonumber \\
&&J^{(5)}_z \, = \, \,J_z \cdot P^{(5)}_z(J_x, \, J_y, \, J_z) 
\, = \, \,J_z \cdot P^{(5)}_x(J_z, \, J_x, \, J_y) 
\end{eqnarray}
where :
\begin{eqnarray}
&&P_5(J_x, \, J_y, \, J_z) \,\, \,  = \, \, \, \, 
J_z^{12} \, J_x^{12}+5 \, J_z^{12} \, J_y^{12}+6 \, J_y^2 \, J_z^{12}
\, J_x^{10}
+14 \, J_y^4 \, J_z^{10} \, J_x^{10}-29 \, J_y^4 \, J_z^{12} \, J_x^8
-6 \, J_y^2 \, J_z^{10} \, J_x^{12}
\nonumber \\
&&\quad -20 \, J_y^8 \, J_z^6 \, J_x^{10}+15 \, J_y^8 \, J_z^4 \, J_x^{12}
-36 \, J_y^8 \,J_z^{10} \, J_x^6
+50 \, J_y^8 \, J_z^8 \, J_x^8-9
 \, J_y^8 \, J_z^{12} \, J_x^4+15 \, J_y^4 \, J_z^8 \, J_x^{12}\nonumber \\
&&\quad -20 \, J_y^6 \, J_z^8 \, J_x^{10}+4 \, J_y^6 \, J_z^{10} \, J_x^8
+36 \, J_y^6 \, J_z^{12} \, 
J_x^6-20 \, J_y^6 \, J_z^6 \, J_x^{12}-10 \, J_y^{10} \, J_z^{12} \,J_x^2
+34 \, J_y^{10} \, J_z^{10} \, J_x^4\nonumber \\
&&\quad +14 \, J_y^{10} \, J_z^4 \, J_x^{10}+4 \, J_y^{10} \, J_z^6 \, J_x^8
-6 \, J_y^{10} \, J_z^2 \, J_x^{12}
-36 \, J_y^{10} \, J_z^8 \, J_x^6\nonumber \\
&&\quad -9 \, J_y^{12} \, J_z^8 \, J_x^4-29 \, J_y^{12} \, J_z^4 \, J_x^8
-10 \, J_y^{12} \, J_z^{10} \, J_x^2+J_y^{12} \, J_x^{12}
+36 \, J_y^{12} \, J_z^6 \, J_x^6+6 \, J_y^{12} \, J_z^2 \, J_x^{10}
\end{eqnarray}
The modulus  (\ref{Mod}) is, again, invariant by
 the polynomial representation of the multiplication of the
shift by five, namely (\ref{mult5}). One remarks that $P^{(5)}_x(J_x, \, J_y,
 \, J_z)$ singles out $J_x$ and is invariant under 
 the permutation  $\, J_y \, \leftrightarrow \, J_z\, $ and, similarly,
 $P^{(5)}_y(J_x, \, J_y, \, J_z)$ singles out $J_y$ and is invariant under 
the permutation  $\, J_x \, \leftrightarrow \, J_z\, $ and 
 $P^{(5)}_z(J_x, \, J_y, \, J_z)$ singles out $J_z$ and is invariant under 
the permutation  $\, J_x \, \leftrightarrow \, J_y$. One has  
similar results for the polynomial
representation
of the  multiplication of the shift by $\, M\, = \, 6, \, 7, \, 9, \,
11, \, \cdots $ The explicit expressions of these  polynomial
representations are given in Appendix A.

 Let us denote by $\, \Gamma_N\, $ a biquadratic corresponding to $\,\, 
 u \, \, \rightarrow \,\,  u \, \pm  N \cdot \lambda\, $ :
\begin{eqnarray}
\label{vacuumJ44N} 
&&\Gamma_N(x , \, z)\, \, \, = \, \, \, \, \, \,
(x^2\,-1) \cdot (z^2 \,-1)\cdot J^{(N)}_x - (x^{2}\,+1) \cdot
(z^2 \,+1)\cdot J^{(N)}_y
+\, \,4 \cdot  J^{(N)}_z
\cdot x \cdot z\,
\, \, \, = \, \, \, \, \, \, 0
\end{eqnarray}
In general, it should be noticed that the 
elimination of $\, y \, $ between
$\, \Gamma_M(x , \, y)\,$
and $\, \Gamma_{M'}(y , \, z)$, yields 
a resultant which is factorized 
into $\, \Gamma_{(M+M')}(x , \, z)\,$ 
and $\, \Gamma_{(M-M')}(x , \, z)\,$ (for $\, M \ge M'$).
When seeking for a new $\, \Gamma_N(x , \, z)\,$
there may be many $\, (M, \, M')\, $
enabling to get  $\, \Gamma_N(x , \, z)\,$ (that is 
such that $\, N \, = \, M+M'$). One can verify 
that all these calculations give, as it should, the {\em same} result 
(in agreement with a polynomial
representation of $\, u \, \rightarrow \, u \, \pm M \cdot \lambda \pm
M' \cdot \lambda\, $ giving $\,  
u \, \pm (M+M')\cdot \lambda \quad {\rm or :} \quad  u \,\pm (M-M')
\cdot \lambda$. Let us denote $\, T_N\, $ these  homogeneous polynomial 
representations of the multiplication of the shift
by the natural integer $\, N$. 
In the same spirit one can verify, 
for $\, N \, = \, M \cdot M'\, $ 
($N, \, M, \, M'$ natural integers),
that :
\begin{eqnarray}
&&T_N (J_x, \, J_y, \, J_z)\, = \, \, \, \, (T_M)^{M'} (J_x, \, J_y, \, J_z)\, = \, \, \,
 T_M(T_M(T_M( \cdots T_M(J_x, \, J_y, \, J_z)\cdots )))\, \, = \, \,
 \, \nonumber \\
&&\, = \, \, \, \, (T_{M'})^M (J_x, \, J_y, \, J_z)\, = \, \, \,
 T_{M'}(T_{M'}(T_{M'}( \cdots T_{M'}(J_x, \, J_y, \, J_z)\cdots )))
\end{eqnarray}
 One can, for instance, easily
 verify  that 
 $\, T_2\, $ and $\, T_3\, $ 
{\em commute}, as well as  $\, T_2\, $ and $\, T_5$. Similarly one can verify,
in a brute-force way, that
  $\, T_3\, $ and $\, T_5\, $ commute. This commutation relations
are true for   $\, T_N\, $ and $\, T_M$, for any $\, N\, $ and $\, M$.
One thus has a {\em polynomial representation of the natural integers
 together with their multiplication}.

In general, the multiplication of the shift by a {\em prime} number $\,
N \ne 2\, $
has the following {\em polynomial representation} $\, (J_x, \, J_y, \, J_z) \,
\rightarrow \,  (J^{(N)}_x, \, J^{(N)}_y, \, J^{(N)}_z) $ :
\begin{eqnarray}
\label{mult11}
&&J^{(N)}_x \, = \, \,J_x \cdot P^{(N)}_x(J_x, \, J_y, \, J_z) 
 \nonumber \\
&&J^{(N)}_y \, = \, \,J_y \cdot P^{(N)}_y(J_x, \, J_y, \, J_z) 
\, = \, \,J_y \cdot P^{(11)}_x(J_y, \, J_z, \, J_x) 
 \nonumber \\
&&J^{(N)}_z \, = \, \,J_z \cdot P^{(N)}_z(J_x, \, J_y, \, J_z) 
\, = \, \,J_z \cdot P^{(N)}_x(J_z, \, J_x, \, J_y) 
\end{eqnarray}
One remarks that the $\,  P^{(N)}_x(J_x, \, J_y, \, J_z)$'s 
(and thus $\,  P^{(N)}_y(J_x, \, J_y, \, J_z)$, $\,  P^{(N)}_z(J_x, \, J_y, \,
J_z)$) are polynomials of $\, J_x^2$,  $\, J_y^2$  and $\, J_z^2$. One
verifies  easily that the  homogeneous
 polynomial transformations $\, T_M\, $
are all of degree $\, M^2$, in $\, J_x, \, J_y, \, J_z$, 
for $\, M\, = \, 2, \, 3, \, 4, \, 5,\, 6,  \,
 7, \,9, \,  11$.

\subsection{Finite order conditions and associated algebraic varieties}
\label{finiord}
Let us show that one can deduce the (projective) finite order
conditions $\, K^{M}(R) \, = \, \zeta \cdot R$, 
from the previous polynomial representations.

Since one knows that the (projective) 
{\em finite order conditions} of $\, {\widehat K}^2$
often play a 
singled-out role for integrability (see the previous section
(\ref{freeferm})), and, in
 particular, since one knows~\cite{MaVi96} that the
free-fermion conditions
of the Baxter model correspond to 
$\, K^{4}(R) \, = \, \zeta \cdot R$, one can, as an exercise, try to
systematically write, for the
sixteen vertex model, the  (projective) finite order  conditions 
$\, K^{2\, N}(R) \, = \, \zeta \cdot R$, 
with $\, N\, $ natural integer, 
the corresponding algebraic varieties being ``good candidates'' for 
new free-(para?)-fermions, or new equivalent of the integrable
chiral Potts model~\cite{HaMa88,AuYang}. Let us 
first give these finite order
conditions for the 
Baxter model where these algebraic varieties are naturally associated 
with the set of RSOS models~\cite{Andrews}.

Let us recall~\cite{MaVi96} that the free-fermion condition 
 for the 
$\, XYZ\, $ Hamiltonian,
$\, J_z \, = \, 0$, corresponds to 
the finite order (projective) condition 
$\, K^{4}(R) \, = \, \zeta \cdot R$. Recalling 
the  polynomial representation
(\ref{doubling}) of the shift doubling, 
one can easily get convinced that 
$\, J_z^{(2)}\, = \, \, 0\, $ should correspond~\footnote{The idea
 here is the following : $\, K$, or 
$\, \widehat{K}$, corresponding, with
some well-suited spectral
parameter, to $ \, \theta \, \, \longrightarrow \, \, \, \theta \,+ \, \,  \eta
$,  $\, K^2$, or $\widehat{K}^2$, must correspond to  
$ \, \theta \, \, \longrightarrow \, \, \, \theta \,+ \, \,  2 \cdot \eta\,
$. A finite order condition of order $\, M$ 
corresponds to a commensuration of $\,
\eta\, $ with a period of the elliptic curves : $\, \eta \, = \, {\cal
P}/M$. Changing $\, K$ into $K^2$ amounts to changing 
$\,\eta\, $ into $\, 2 \cdot \eta$, or equivalently, changing the
order $\, M$ into $\, 2\, M$. } to
$\, K^{8}(R) \, = \, \zeta \cdot R$. This can be verified 
by a straight calculation. More generally,
a polynomial condition $\, C_M(J_x, \, J_y, \, J_z)\, = \,\,  0 $, 
corresponding to $\, K^{M}(R) \, = \, \zeta \cdot R$, 
has to be compatible with the  polynomial 
representations of $\, \lambda \, \longrightarrow 
\, N \cdot \lambda $. This  compatibility
is, in fact, an efficient way to get these finite
 order conditions. Explicit expressions of finite order 
conditions, as well as their compatibility
with the shift doubling and, more generally, 
 the polynomial representations of $\, \lambda \, \longrightarrow 
\, M \cdot \lambda$,  are given in Appendix B. For a prime integer
$\, N\,\ne 2 $
the algebraic varieties $\, P^{(N)}_x(J_x,\, J_y, \, J_z)\, = \, 0 $, 
 $\, P^{(N)}_y(J_x,\, J_y, \, J_z)\, = \, 0$,
and  $\, P^{(N)}_z(J_x,\, J_y, \, J_z)\, = \, 0 $ give 
order $\, 4 \, N $ conditions :
\begin{eqnarray}
K^{4 \, N}(R) \, = \, \zeta \cdot R
\end{eqnarray}
Since the $\, P^{(N)}$'s (and the  $\, J^{(N)}$'s for $\, N$ even)
are functions of  $\, J_x^2$,  $\, J_y^2$ and $\, J_z^2$,
the order $\, 4 \, N$ conditions,  $\, K^{4 \, N}(R) \, = \, \lambda \cdot R$,
are also  functions of the square $\, J_x^2$,  $\, J_y^2$ and $\,
J_z^2$. One can easily get infinite families
of finite order conditions. For instance, iterating the shift doubling
 (\ref{doubling}) (resp. (\ref{tripling})), and using this transformation
on $\, J_z\, = \, 0$,  one easily gets 
an {\em infinite number} of algebraic varieties
corresponding to the finite order conditions of order $\, 2^N$
(resp. $\, 3^N$). Combining  (\ref{doubling}) and (\ref{tripling}),
one gets straightforwardly the finite order conditions of order $\,
2^N \times 3^M$.
More details will be given elsewhere.

{\bf Remark : Finite order conditions for the sixteen vertex model. }

Using the previous results, the decomposition
(\ref{decompoBMV}) of the sixteen vertex model
into a ``$ {\widehat K}^2$-effective'' Baxter model, and relation
(\ref{decompofinqq}), one can obtain 
the finite order conditions, 
$\, K^{4 \, N}(R) \, = \, \, \zeta \cdot R$,
for the sixteen vertex model and find that they are actually
{\em codimension-one} algebraic varieties. Recalling
(\ref{decompofinqq}), 
one easily deduces that these 
 finite order conditions are actually  given in terms of 
the finite order conditions of the ``${\widehat K}^2$-effective'' Baxter
model. These  finite order conditions are  simply expressed in
terms of the associated ``${\widehat K}^2$-effective'' variables $\, J_x$, $\, J_y$
and $\, J_z$ which can be obtained 
from relations (\ref{44}), (\ref{45}). 

From this example one sees that the, at first sight,
``hardly Baxterisable'' case of {\em finite order} iteration
provides, to some extend, more  results, and structures, than a ``standard'' 
{\em infinite order Baxterisation}: one gets, for instance, a
{\em polynomial representation of the natural integers
 together with their multiplication}, this 
polynomial representation  leaving invariant
the {\em modular invariant}~\cite{BoMa92b} of the elliptic curves,
and giving codimension-one 
algebraic varieties
compatible with this structure  ...

\section{Let us Baxterise quantum Hamiltonians }
\label{baxHamil}
Let us now consider a typical problem for any theoretician who
wants to provide some ``interesting contribution'' in High-Tc.
Let us consider a strongly correlated
 quantum Hamiltonian which looks like a t-J model~\cite{tJ,tJ2,tJ3}, or a
 Hubbard model~\cite{Hubbard,Hubbard2},
or some coupled $\, XYZ\, $  quantum chains : how is it possible 
to see if this 
  quantum Hamiltonian is integrable ? More generally, let us consider
 a quantum Hamiltonian. How to see if it is possible to solve 
this  quantum Hamiltonian ? The Bethe Ansatz only works if one has
some ``conservation operator somewhere''~\footnote{ The typical 
example is the operator $\, \sum_{n} \sigma^{z}_n \, $ 
for the $ \, XXZ\, $ Hamiltonian~\cite{LiWu72,Ka74}.
} that enables to see this model
as some ``avatar'' of the six-vertex model (XXZ chain). Beyond this restricted 
``six-vertex'' framework, and if one cannot, or
 does not know how to, associate a (commuting) family 
of transfer matrices, commuting with this Hamiltonian,
one has very few tools left to solve this quantum 
Hamiltonian~\footnote{The random matrix theory analysis
 of the level spacing distribution 
remains a possible tool to simply ``detect'' integrability~\cite{random}.
However, besides the technical difficulties
associated with the unfolding procedure, the 
calculations become very large
for the $\, 16 \times 16$ (Hubbard) $\, R$-matrices, or
 for two coupled spin chains.}.
For a Yang-Baxter integrable model depending 
on one spectral parameter, the associated integrable quantum Hamiltonian
can be seen as the derivative of the (logarithm) of the transfer
matrix at a singled-out value of the spectral parameter : therefore
the  integrable quantum Hamiltonian {\em does not depend 
on the spectral parameter}. If the Bethe Ansatz is too complicated, or
does not exist (higher genus curves, ...), how can we make the spectral
parameter(s) ``emerge'' so that the integrability structure becomes
crystal clear ? This is clearly another type of Baxterisation
problem. Let us sketch how this can be done.

\subsection{Let us Baxterise the $\, XYZ\, $  quantum Hamiltonian }
\label{letsbaxHamil}
Let us recall  the  $\, XYZ$
 Hamiltonian~\cite{Ba73,Ba72b}, $\, H_{XYZ} \, =
\, \sum_n H_{n,n+1}$, and let us represent $\, H_{n,n+1} \, = \,
 J_x \cdot \sigma_n^x \cdot \sigma_{n+1}^x \, + \,  J_y
 \cdot\sigma_n^y \cdot \sigma_{n+1}^y \, + \, 
 J_z \cdot \sigma_n^z \cdot \sigma_{n+1}^z \, $
and  $\, P$, the matrix of permutation of the vertical
 space $n$ and $n+1$,
as  $4 \times 4$ matrices :
\begin{eqnarray}
H_{n,n+1} \, \, = \, \,
\left [\begin {array}{cccc} 
{\it Jz}&0&0&{\it Jx}-{\it Jy}\\
\noalign{\medskip}0&-{\it Jz}&{\it Jx}+{\it Jy}&0\\
\noalign{\medskip}0&{\it Jx}+{\it Jy}&-{\it Jz}&0\\
\noalign{\medskip}{\it Jx}-{\it Jy}&0&0&{\it Jz}
\end {array}\right ]
\, , \qquad 
P \, = \, \, \left [\begin {array}{cccc}
 1&0&0&0\\
\noalign{\medskip}0&0&1&0\\
\noalign{\medskip}0&1&0&0\\
\noalign{\medskip}0&0&0&1
\end {array}
\right ]
\end{eqnarray}

Recalling Baxter's notations, the $R$-matrix, 
 reads at order one in some small expansion parameter $\,\epsilon\, $ :
\begin{eqnarray}
&&R \, = \, \, P \, + \, \, 
\epsilon \cdot P \cdot H_{n,n+1} \, + \, \cdots \, \, = \, \, \,
\left [\begin {array}{cccc}
 1&0&0&0\\
\noalign{\medskip}0&0&1&0\\
\noalign{\medskip}0&1&0&0\\
\noalign{\medskip}0&0&0&1
\end {array}
\right ]\, + \, \, 
\epsilon \cdot \left [\begin {array}{cccc}
 {\it Jz}&0&0&{\it Jx}-{\it Jy}\\
\noalign{\medskip}0&{\it Jx}+{\it Jy}&-{\it Jz}&0\\
\noalign{\medskip}0&-{\it Jz}&{\it Jx}+{\it Jy}&0\\
\noalign{\medskip}{\it Jx}-{\it Jy}&0&0&{\it Jz}
\end {array}\right ]\, + \, \, \cdots \nonumber \\
&& \qquad \, = \, \,
\left [\begin {array}{cccc}
 1+\epsilon\, {\it Jz}&0&0&\epsilon \cdot \left ({\it Jx}-{\it Jy}
\right )\\
\noalign{\medskip}0&\epsilon \cdot \left ({\it Jx}+{\it Jy}\right )&1
-\epsilon \cdot {\it Jz}&0\\
\noalign{\medskip}0&1-\epsilon \cdot {\it Jz}&\epsilon\, \left ({\it Jx}+{\it Jy}
\right )&0\\
\noalign{\medskip}\epsilon \cdot \left ({\it Jx}-{\it Jy}\right )&0&0&1+
\epsilon \cdot {\it Jz}
\end {array}\right ]
\, + \, \, \cdots \, \, \, = \, \, \, \, \,
\left [\begin {array}{cccc} 
a&0&0&d\\
\noalign{\medskip}0&b&c&0\\
\noalign{\medskip}0&c&b&0\\
\noalign{\medskip}d&0&0&a\end {array}
\right ]
\end{eqnarray}
One has the correspondence :
\begin{eqnarray}
a \, = \, \, 1 \, + \epsilon \cdot J_z \,+ \cdots  \,,  \quad 
b \, = \, \,  \,\epsilon \cdot (J_x\, + \, J_y)\,+ \cdots  \,,  \quad 
c \, = \, \, 1 \, - \epsilon \cdot J_z\,+ \cdots  \,, \quad 
d \, = \, \, \,\epsilon \cdot (J_x\, - \, J_y)\,+ \cdots \nonumber 
\end{eqnarray}

One verifies, immediately, that the
algebraic $\, K$-covariants of the
Baxter model read :
\begin{eqnarray}
{{a^2+b^2-c^2-d^2} \over {2}}\, = \, \, 
2 \epsilon \cdot J_z\, , \quad \quad
{{a \cdot b \, + \, c \cdot d }} \, = \, \, 
2 \epsilon \cdot J_x\, , \quad \quad
{{a \cdot b \, - \, c \cdot d }} \, = \, \, 
2 \epsilon \cdot J_y
\end{eqnarray}
the  $\, {\widehat K}$-invariants being (for instance)
the ratio $\, J_x/J_z$ and $\, J_y/J_z$.
The $\, R$-matrix $\, P$ can be seen to 
belong to {\em all} the elliptic curves of this foliation 
of the Baxter model (the invariants $\, (a^2+b^2-c^2-d^2)/a/b$
or $\, a\, b/c/d$ are of the form $\, 0/0$).
Matrix $\, P$ is really the equivalent of the {\em base point} of an elliptic 
foliation~\cite{BoHaMa97}. 

Having the $\, XYZ$ Hamiltonian and willing to 
``recover'' the Baxter $\, R$-matrix (and its canonical elliptic
parameterization) is a slight modification of the previous 
Baxterisation process of $\, R$-matrices, or monodromy matrices,
where we were building the algebraic variety 
from one  arbitrary point of the algebraic 
variety. We do not have a point of the
elliptic curve here,
but rather a {\em singular point}, which is the equivalent of the 
 {\em base point} of an elliptic 
foliation, namely point $\, P$,  and a ``vector'' $\, (J_x, \, J_y, \,
J_z)\, $ giving the {\em tangent} to the elliptic curve 
at point $\, P$. It is however
clear that  $\, [P, \, (J_x, \, J_y, \,
J_z)]\, $ is {\em sufficient} to build, in a unique way, the elliptic
curve, and thus the spectral parameter.

{\bf A heuristic remark :} Point  $\, P$ is a {\em singular point} of $\,
K^2$ and the  {\em diagonal} matrices
 are {\em fixed points} of $\, {\widehat K}^2
\, $ :
\begin{eqnarray}
K(P) \, = \, \, \,
\left [\begin {array}{cccc} 
-1&0&0&-1\\
\noalign{\medskip}0&0&0&0\\
\noalign{\medskip}0&0&0&0\\
\noalign{\medskip}-1&0&0&-1\end
{array}\right ]\, , \quad 
K(K(P)) \, = \, \,\,
\left [\begin {array}{cccc} 
0&0&0&0\\
\noalign{\medskip}0&0&0&0\\
\noalign{\medskip}0&0&0&0\\
\noalign{\medskip}0&0&0&0
\end {array}\right ]\, , \quad 
{\widehat K}^2 \left [\begin {array}{cccc}
 A&0&0&0\\
\noalign{\medskip}0&B&0&0\\
\noalign{\medskip}0&0&C&0\\
\noalign{\medskip}0&0&0&D
\end {array}\right ]
\, = \, \, \left [\begin {array}{cccc}
 A&0&0&0\\
\noalign{\medskip}0&B&0&0\\
\noalign{\medskip}0&0&C&0\\
\noalign{\medskip}0&0&0&D
\end {array}\right ] \nonumber
\end{eqnarray}
More generally, considering the most general $\, 4 \times 4\, $
matrices (sixteen vertex model), the singular matrices for $\, K^2$
are the $\, R$-matrices such that $\, K(R)$ is a rank two matrix.
For the Baxter model  singular matrices, such that 
$\, K^2(R)\, $ is the null matrix, are, for instance,
$\, d\, = \, b \, = \, 0$ or $\, a \, = \, c\, = \, 0$.
The {\em base points}
 of the elliptic foliation of the parameter space
of the Baxter model are a subset of these singular subvarieties. The 
 {\em base points} of the elliptic
foliation are the points such that the 
$\, K$-invariants, $\, (a^2+b^2-c^2-d^2)/a/b$
or $\, a\, b/c/d$, are both of the form $\, 0/0$, namely $\, d = b =
0,  c = a $,  or $  \, d = b = 0, \, a = -c$,  or 
$  \, a= d = 0, \,  b = c$,
 or $  \, b = -c, \,  a= d = 0$. This can be easily 
generalized to the sixteen vertex model by imposing that
the $\, p_i$'s $\, K$-covariants~\cite{prl2}, mentioned 
in section (\ref{preb}), are all equal
to zero. 

In general these two sets of subvarieties, namely 
the singular subvarieties and the  subvarieties 
of fixed points of $\, {\widehat K}^2$, play a crucial role 
in the Baxterisation of a Hamiltonian. The subvariety
of fixed points of $\, {\widehat K}^2$ always contains 
the set of all the {\em diagonal} matrices. Not
 surprisingly we will see, in
the next section, that diagonal matrices naturally occur (in a
non-trivial way)  in the
Baxterisation process (see for instance  (\ref{good}) below).

\subsection{Let us Baxterise  the $\, t-J$ quantum Hamiltonian }

Let us recall the  Hamiltonian of the $t-J$ model~\cite{tJ},
and let us consider  an ordering of the $\, 9 \times 9\, $ $\, R$-matrices,
 well-suited for the 
partial transposition $\, t_1$ :
\begin{eqnarray}
\label{lawandorder}
(+,+),\,\, (+,-),\,\, (+,0),\,\, (-,+),\, \,(-,-),\, \,(-,0),\, \,(0,+),\,\,
(0,-),\, \,(0,0)
\end{eqnarray}
With this ordering, the ``local Hamiltonian'' of the $t-J$
model, equivalent of the previous $\, H_{n,n+1} \,$,
 and the permutation matrix $\, P$, read respectively :
\begin{eqnarray}
\label{hnnpluun}
H_{n,n+1} \, = \, \, 
\left [\begin {array}{ccccccccc}
V\,+ {{J}\over {4}}&0&0&0&0&0&0&0&0\\
\noalign{\medskip}0&V-{\frac {J}{4}}&0&{\frac {J}{2}}&0&0&0&0&0\\
\noalign{\medskip}0&0&0&0&0&0&-t&0&0\\
\noalign{\medskip}0&{\frac {J}{2}}&0&V\,-{\frac {J}{4}}&0&0&0&0&0\\
\noalign{\medskip}0&0&0&0&V\,+ {{J}\over {4}}&0&0&0&0\\
\noalign{\medskip}0&0&0&0&0&0&0&-t&0\\
\noalign{\medskip}0&0&-t&0&0&0&0&0&0\\
\noalign{\medskip}0&0&0&0&0&-t&0&0&0\\
\noalign{\medskip}0&0&0&0&0&0&0&0&0
\end {array}\right ], \quad 
P \, = \, \, \left [\begin {array}{ccccccccc}
 1&0&0&0&0&0&0&0&0\\
\noalign{\medskip}0&0&0&1&0&0&0&0&0\\
\noalign{\medskip}0&0&0&0&0&0&1&0&0\\
\noalign{\medskip}0&1&0&0&0&0&0&0&0\\
\noalign{\medskip}0&0&0&0&1&0&0&0&0\\
\noalign{\medskip}0&0&0&0&0&0&0&1&0\\
\noalign{\medskip}0&0&1&0&0&0&0&0&0\\
\noalign{\medskip}0&0&0&0&0&1&0&0&0\\
\noalign{\medskip}0&0&0&0&0&0&0&0&1
\end {array}\right ]
\end{eqnarray}
The integrability cases~\cite{Schlotmann} correspond to $\, 
(V, \, J) \, = \, \, (-t/2, \, 2\, t) $ 
or $ \, (t/2, \, -2\, t) $ or $ \, (3\, t/2, \, 2\, t) $ or $ \, (-3\,
t/2, \, -2\, t) $. The $\, R$-matrix reads  at order one 
 in some small expansion parameter $\,\epsilon\, $:
\begin{eqnarray}
&&R \, \, \, \, \,  = \, \, \, \, \,  \, P \, +\, \epsilon \cdot P \cdot H_{n,n+1} 
\, + \, \cdots \, \,  \, \,\, \,  \, \,  = \, \, \,  \, \\
&&\left [\begin {array}{ccccccccc} 
1\, +\epsilon \cdot \left (V+J/4\right )&0&0&0&0&0&0&0&0\\
\noalign{\medskip}0&\epsilon \cdot J/2&0&1+\, \epsilon \cdot \left (V-J/4\right
)&0&0&0&0&0\\
\noalign{\medskip}0&0&-\epsilon \cdot t&0&0&0&1&0&0\\
\noalign{\medskip}0&1
+\epsilon \cdot \left (V-J/4\right )&0&\epsilon \cdot J/2&0&0&0&0&0\\
\noalign{\medskip}0&0&0&0&1+\epsilon \cdot \left (V+J/4\right )&0&0&0&0\\
\noalign{\medskip}0&0&0&0&0&-\epsilon \cdot t&0&1&0\\
\noalign{\medskip}0&0&1&0&0&0&-\epsilon \cdot t&0&0\\
\noalign{\medskip}0&0&0&0&0&1&0&-\epsilon \cdot t&0\\
\noalign{\medskip}0&0&0&0&0&0&0&0&1
\end {array}\right ]\,  \, \, + \, \, \cdots \nonumber
\end{eqnarray}

The (supersymmetric) integrable case 
$\, (V,\, J) \, = \, (-\,t/2, \, 2\, t) \, $ 
 yields the (supersymmetric) ``local Hamiltonian'' 
$\, H_{n,n+1}(J =  2 \, t, V =  -\,t/2) \, \, = \, \, \, \,
t \cdot H_{susy}\,$ :
\begin{eqnarray}
\label{hsusy}
t \cdot H_{susy}\, = \, \, \,  \, \, 
t \cdot \left [\begin {array}{ccccccccc} 
0&0&0&0&0&0&0&0&0\\
\noalign{\medskip}0&-1&0&1&0&0&0&0&0\\
\noalign{\medskip}0&0&0&0&0&0&-1&0&0\\
\noalign{\medskip}0&1&0&-1&0&0&0&0&0\\
\noalign{\medskip}0&0&0&0&0&0&0&0&0\\
\noalign{\medskip}0&0&0&0&0&0&0&-1&0\\
\noalign{\medskip}0&0&-1&0&0&0&0&0&0\\
\noalign{\medskip}0&0&0&0&0&-1&0&0&0\\
\noalign{\medskip}0&0&0&0&0&0&0&0&0
\end {array}\right ]
\end{eqnarray}
Let us ``absorb'' the homogeneity of $\,  H_{susy}\,$ by introducing 
$\, x \, = \, - \epsilon \cdot t$. 
At first sight, one wants to Baxterise:
\begin{eqnarray}
\label{raw}
R_{aw} \,\, = \,\,\, \, P \, - \, x \cdot P \cdot H_{susy} 
\, + \,\, \cdots \, \,\, \,\,  \, \, 
\end{eqnarray}
Let us introduce
the two $\, 9 \times 9\, $ matrices :
\begin{eqnarray}
\label{nn'}
&&{\widehat N} \, = \, 
\left [\begin {array}{ccc}
1\,&0&0\\
\noalign{\medskip}0\,&1&0\\
\noalign{\medskip}0\,&0&0
\end {array}\right ]
\otimes
 \left [\begin {array}{ccc}
1\,&0&0\\
\noalign{\medskip}0\,&1&0\\
\noalign{\medskip}0\,&0&1
\end {array}\right ]
\, + \, \, 
\left [\begin {array}{ccc}
1\,&0&0\\
\noalign{\medskip}0\,&1&0\\
\noalign{\medskip}0\,&0&1
\end {array}\right ]
\otimes
 \left [\begin {array}{ccc}
1\,&0&0\\
\noalign{\medskip}0\,&1&0\\
\noalign{\medskip}0\,&0&0
\end {array}\right ] 
\nonumber \\
&&{\widehat N'} \, = \, 
\left [\begin {array}{ccc}
1\,&0&0\\
\noalign{\medskip}0\,&1&0\\
\noalign{\medskip}0\,&0&0
\end {array}\right ]
\otimes
 \left [\begin {array}{ccc}
0\,&0&0\\
\noalign{\medskip}0\,&0&0\\
\noalign{\medskip}0\,&0&1
\end {array}\right ]
\, + \, \, 
\left [\begin {array}{ccc}
0\,&0&0\\
\noalign{\medskip}0\,&0&0\\
\noalign{\medskip}0\,&0&1
\end {array}\right ]
\otimes
 \left [\begin {array}{ccc}
1\,&0&0\\
\noalign{\medskip}0\,&1&0\\
\noalign{\medskip}0\,&0&0
\end {array}\right ] 
\end{eqnarray}
With  ordering (\ref{lawandorder}) matrix $\,{\widehat N}\,$
 is associated with  the {\em electron-counting}
operator which commutes with the t-J Hamiltonian. Matrix $\,{\widehat N}\,$
commutes with $\, P$ and is such that the matrix $\,
D_{iag}\, = \, \,  \, \, $ $\,  P \cdot H_{susy} -\, P \, +\, P \cdot
\widehat{N}  \,  $ is a {\em diagonal} matrix of successive diagonal entries :
$\, 1, \, 1, \, -1, \, 1, \, 1, \, -1, \, -1, \, -1, \, -1$. Actually,
 as far as the integrability of the model is 
concerned, one can add 
to the Hamiltonian an operator {\em which commutes with 
it}. Obviously one can also 
add to the $\, R$-matrix, a term 
$\,x \cdot  P$ which just amount to a global $\, 1+x\, $
multiplicative 
factor in front of $\, P$. This amounts to introducing, instead of
(\ref{raw}),
the following $\, R$-matrix:
\begin{eqnarray}
\label{good}
&&R_1(x)  \, \, \, \,  =  \, \, \, \, \,  P \, -\, x \cdot  D_{iag}
 \,  \, \,  = \, \,\, \, \,
(1\, +\, x) \cdot P -\, x \cdot  P \cdot H_{susy}\, 
- \, x \cdot P \cdot  {\widehat N}  \,\,  \,\,  = \, \\
&& \qquad  \,   = \, \,\, \, \,
(1\, +\, x) \cdot P \cdot \Bigl({\cal I}_d  \, - \, {{x} \over {1+x}} \cdot
 (H_{susy}\,+ {\widehat N}) \Bigr) \, \, \, = \,  \, \,
\left [\begin {array}{ccccccccc} 
1-x&0&0&0&0&0&0&0&0\\
\noalign{\medskip}0&-x&0&1&0&0&0&0&0\\
\noalign{\medskip}0&0&x&0&0&0&1&0&0\\
\noalign{\medskip}0&1&0&-x&0&0&0&0&0\\
\noalign{\medskip}0&0&0&0&1-x&0&0&0&0\\
\noalign{\medskip}0&0&0&0&0&x&0&1&0\\
\noalign{\medskip}0&0&1&0&0&0&x&0&0\\
\noalign{\medskip}0&0&0&0&0&1&0&x&0\\
\noalign{\medskip}0&0&0&0&0&0&0&0&1+x
\end {array}\right ] \nonumber 
\end{eqnarray}
where $\, {\cal I}_d \, $ denotes the $\, 9 \times 9\, $
identity matrix. Under transformation $\, {\widehat K}^4\, $
the $\, R$-matrix $\, R_1(x)\, \, $
becomes another matrix, of the {\em same form}~\footnote{Note that 
$\, {\widehat K}^2(R_1)\, $ has a form 
similar to $\, R_1(x+1)\, $ up to {\em some change of signs} of some entries
(supersymmetric graduation).}, but where 
$\, x \, \rightarrow \, x\, + \, 2\, $ :
\begin{eqnarray}
 {\widehat K}^4(R_1(x)) \, \,  \,  = \, \,  \,  \, 
{{x-1} \over {x+1}} \cdot R_1(x+2)
\end{eqnarray}
This solves the Baxterisation problem of (\ref{hsusy}), reducing it,
after
addition of ``well-suited'' commuting
 operators, to a {\em simple linear interpolation}. 

$\, \bullet $ Another integrable case~\cite{Schlotmann} for (\ref{hnnpluun}) is
$ \,(V, \, J)=\, (3\, t/2,\, 2\, t) $, for which one has to introduce 
{\em another correction} namely $\, {\widehat N'}\, $ in (\ref{nn'}), 
 yielding the {\em same} $\, R$-matrix (\ref{good}),
namely $\, R_1 \,   =  
\, \, \,  P \, -\, x \cdot  D_{iag}$,  the diagonal matrix $\,
D_{iag}\, $ being, now,
equal to $\,   P \cdot H_{tJ} \, \,-\, P \, \, +\, P \cdot
{\widehat N'}$, where  $\, H_{tJ}\, $ is given by (\ref{hnnpluun})
for $ \,(V, \, J)=\, (3\, t/2,\, 2 \, t)$.

$\, \bullet $  Two other  integrable cases~\cite{Schlotmann}  for (\ref{hnnpluun})
namely $ \,(V, \, J)=\, (-3\, t/2,\, -2\, t) \, $ and 
$ \,(V, \, J)=\, (t/2,\, -2\, t) \, $
 yield the {\em same} $ R$-matrix $\, R_2$.
For  $ \,(V, \, J)=\, (-3\, t/2,\, -2\, t) \, $ and 
 $ \,(V, \, J)=\, (t/2,\, -2\, t) \, $
 one has to introduce the 
 correction $\, {\widehat N}\, $ and  $\, {\widehat N'}\, $
respectively, yielding a very simple  $\, R$-matrix $\, R_2\, = \, 
x \cdot {\cal I}_d \, + \, P $.
Similarly, under transformation $\, {\widehat K}^2$,
matrix $\, R_2(x)\, \, $ becomes 
another matrix of the {\em same form} but where 
 $\, x \, \rightarrow \, x\, - \, 3$ :
\begin{eqnarray}
 {\widehat K}^2(R_2(x)) \, \,  \,  = \, \,  \,  \, 
{{x^2-1} \over {x \cdot (x-3) }} \cdot R_2(x-3)
\end{eqnarray}
thus solving the Baxterisation problem of these last two integrable cases
as a {\em simple linear interpolation} between $\, {\cal I}_d\, $ and $\, P$.

\subsection{Hamiltonian dependence of the Baxterisation procedure}

One certainly wants the Baxterisation procedure of a quantum Hamiltonian to
be ``universal'' : it should be compatible with the addition
of quantum operators which commute with the quantum Hamiltonian,
and, thus, do not modify the integrability of the  Hamiltonian.

 The Baxterisation procedure 
is, of course,  compatible with the
{\em gauge transformations} on the quantum Hamiltonian :
\begin{eqnarray}
\label{gauge}
H_{n,n+1} \qquad \longrightarrow \,\qquad  H_{n,n+1} \, \, + \, \,
(G^{\alpha}_n \otimes
{\cal I}_n\, - \,{\cal I}_{n+1} \otimes
G^{\alpha}_{n+1}) \, 
\end{eqnarray}
where $\, {\cal I}_n\, $ and $\, {\cal I}_{n+1}\, $ denote the
identity operators on site $\, n\, $ and $\, n+1$ respectively,
the ``exponentiation'' of (\ref{gauge})
giving the $\, R \, \rightarrow \, g^{-1} \otimes h^{-1} \cdot R \cdot
g \otimes h\, $ well-known gauge symmetries~\cite{GaHi75}, compatible with
the  Baxterisation: $\, {\widehat K}^2(g^{-1} \otimes h^{-1} \cdot R \cdot
g \, \otimes\, h) \,= \, \, g^{-1} \otimes h^{-1} \cdot {\widehat K}^2(R) \cdot
g \, \otimes \, h$. Furthermore, from the symmetries of the Baxterisation
(see for instance (\ref{35})), one deduces that the  Baxterisation 
of a quantum Hamiltonian is necessarily compatible 
with the following deformations of $\,H_{n,n+1} \, $ 
generalizing~\footnote{ A simple
example corresponds, for instance, to
 adding  operator $\, \sum_{n} \sigma^{z}_n \, $ 
to the $ \, XXZ\, $ Hamiltonian~\cite{LiWu72,Ka74}.
} (\ref{gauge}) :
\begin{eqnarray}
\label{generalizing}
H_{n,n+1} \qquad \longrightarrow \,\qquad  H_{n,n+1} \, \, + \, \,
(G^{\alpha}_n \otimes
{\cal I}_n\, - \,{\cal I}_{n+1} \otimes
G^{\alpha}_{n+1}) \, + \, (\tilde{G}^{\beta}_n \otimes
{\cal I}_n\, + \,{\cal I}_{n+1} \otimes
\tilde{G}^{\beta}_{n+1})
\end{eqnarray}
The Baxterisation should also provide  (at least 
as far as integrability is concerned) compatible
 results when one adds, to the quantum Hamiltonian $\,\sum_n H_{n,n+1}$, 
 operators  commuting 
with it :
\begin{eqnarray}
\label{intdefo}
H_{n,n+1} \qquad \longrightarrow \,\qquad  H_{n,n+1} \, \, + \, \,
O^{\alpha}_{n, \, n+1}\,, \qquad \alpha \, = \, \, 1, \cdots , r
\end{eqnarray}
 where the $\,   \, \sum_n 
O^{\alpha}_{n, \, n+1}$'s commute with   $\, 
 \sum_n H_{n,n+1}$. When  the 
quantum Hamiltonian $\,\sum_n H_{n,n+1}\, $ is integrable
one would like to have a complete description of {\em all} the
 {\em integrable deformations} (\ref{intdefo}), and a complete
description of the algebraic variety associated 
with the Baxterisation of the largest integrable deformation
of  the quantum Hamiltonian (as many $\, O^{\alpha}_{n, \, n+1}$'s as
possible). Are all the integrable deformations  $\, O^{\alpha}_{n, \, n+1}\, $
necessarily of the $\, {\widehat K}^2$-compatible form
(\ref{generalizing}) ? All the  $\, O^{\alpha}_{n, \, n+1}$'s are such
that the $\,   \, \sum_n 
O^{\alpha}_{n, \, n+1}$'s commute with   $\, 
 \sum_n H_{n,n+1}$, but do all the  $\, O^{\alpha}_{n, \, n+1}$'s
commute with  $\, H_{n,n+1}$ and
commute~\footnote{In the case where $\, {\widehat K}^2\, $ is an
infinite order transformation densifying an algebraic variety,
this variety is an Abelian variety and one can deduce, from this
Abelian property, that the $\, P \cdot O^{\alpha}_{n, \, n+1}$'s should
commute.
} all together ? Is there a way to find $\, r$, the largest number of 
 integrable deformations  $\, O^{\alpha}_{n, \, n+1}$, by
simple arguments?
Is it possible to find lower and upper 
bounds~\footnote{For Abelian varieties parameterized by theta functions
of $\, g$ variables, one could imagine that $\, r$ should be related
to $\, g \, + \, N_{g}$, where  $\, N_{g}\, $ denotes the number of
independent gauge deformations (\ref{gauge}).}  for $\, r$ ? 
All these questions will be analysed in details elsewhere. Let us just
give, in the following sections, some partial answers to these questions
based on deformations of 
the two previous ``heuristic'' integrable cases, namely $\, R$-matrices  $\, R_1(x)$
and $\, R_2(x)$.

Let us try to see what happens when one changes the integrable
 Hamiltonian  adding an operator 
which commutes with it. Suppose that, instead
of introducing the $\, R$-matrix $\, R_1(x)$,
which yields a simple linear interpolation for the Baxterisation
process (\ref{good}),  one introduces :
\begin{eqnarray}
\, R(x, \, y) \, \,  = \, \, 
 P \, -\, x \cdot  D_{iag}\, + \, y \cdot  P \cdot  {\widehat N}
 \,  \, \,  = \, \,\, \, \, \,
(1\, +\, x) \cdot P -\, x \cdot  P \cdot H_{susy}\, 
- \, (x\, -\, y) \cdot P \cdot  {\widehat N}  \,  + \, \cdots \, 
\end{eqnarray}
 One can actually easily verify that
the $n$-th $\, \widehat{K}^{2}$-iterate of $\,  R(x, \, y) \,$
are all linear combinaisons of $\, R(x, \, y)$,
 $\, \widehat{K}^{2}(R(x, \, y))$,
 $\, \widehat{K}^{4}(R(x, \, y))$,
  $\, \widehat{K}^{6}(R(x, \, y))$  and $\, \widehat{K}^{8}(R(x, \, y))$
or, equivalently,  $\, R(x, \, y)$,
 $\, \widehat{K}^{2}(R(x, \, y))$,
 $\, \widehat{K}^{4}(R(x, \, y))$, $\, P$ and $\,  P \cdot  {\widehat
N}\, $  :
\begin{eqnarray}
\label{form3}
&&\widehat{K}^{2\, n}(R(x, \, y))\, = \, \, \,  \\
&&\, = \, \, \,A_0^{(n)} \cdot  R(x, \, y) \, + \, A_1^{(n)} \cdot \widehat{K}^{2}(R(x, \, y))
 \, + \, A_2^{(n)} \cdot \widehat{K}^{4}(R(x, \, y))
 \, + \, A_3^{(n)} \cdot P + \, A_4^{(n)} \cdot P \cdot  {\widehat N} \, = \,\nonumber  \\
&&\, = \, B_0^{(n)} \cdot  R(x, \, y) \, + \, B_1^{(n)} \cdot \widehat{K}^{2}(R(x, \, y))
 \, + \, B_2^{(n)} \cdot \widehat{K}^{4}(R(x, \, y))
 \, + \, B_3^{(n)} \cdot \widehat{K}^{6}(R(x, \, y)) 
+ \, B_4^{(n)} \cdot \widehat{K}^{8}(R(x, \, y)) \nonumber 
\end{eqnarray}
In other words, the Baxterisation acts in the 
four-dimensional vector 
space~\footnote{Five homogeneous parameters.
Note that the odd $\, \widehat{K}$-iterate
do not belong to this four-dimensional vector space
but to its transformed under $\, t_1$.
} of the $\, R$-matrices
of the form (\ref{form3}). The Baxterisation of $\, R(x, \, y) \, $ gives  successive 
$\, 9 \times 9 \, $ matrices $\, \widehat{K}^{2\, n}(R(x, \, y))\, $
of the form:
\begin{eqnarray}
\label{form2}
R(A, \, B, \, C, \, D, \, E, \, F)  \,\,\,\,\,\, 
= \,\, \,\left [\begin {array}{ccccccccc}
 A&0&0&0&0&0&0&0&0\\
\noalign{\medskip}0&B&0&C&0&0&0&0&0\\
\noalign{\medskip}0&0&D&0&0&0&E&0&0\\
\noalign{\medskip}0&C&0&B&0&0&0&0&0\\
\noalign{\medskip}0&0&0&0&A&0&0&0&0\\
\noalign{\medskip}0&0&0&0&0&D&0&E&0\\
\noalign{\medskip}0&0&E&0&0&0&D&0&0\\
\noalign{\medskip}0&0&0&0&0&E&0&D&0\\
\noalign{\medskip}0&0&0&0&0&0&0&0&F\end {array}\right ]\, , \qquad
{\rm where :} \quad A \, = \, B\, + \, C
\end{eqnarray}

{\bf Factorization scheme :} Let us Baxterise $\, m_0 \, = \, R(x, \, y)\, $ 
using the {\em homogeneous} transformation $\, K$.
The first iteration reads $\, m_1 \, = \, \, K(m_0)$.
Let us denote $\, G_1\, $ the gcd of  {\em all the entries} of matrix
 $\, m_1$. In order to go on iterating,
it is better to introduce the ``reduced'' matrix,
 $\, M_1 \, = \, \, m_1/G_1$. Let us denote
$\, m_2 \, = \, \, K(M_1)$. Similarly, 
one can introduce $\, G_2\, $ the gcd of all the entries of matrix
 $\, m_2$, and define  the ``reduced'' matrix,
 $\, M_2 \, = \, \, m_2/G_2$, and so on.
Introducing $\, G(w)\, $ the generating function of 
the degree of the successive gcd's $\, G_n$'s, and $\, D(w)\, $
the generating function of 
the degree of the determinants of the successive
 reduced matrices $\, M_n$'s,
one gets for the iteration
of (for instance) $\, R(x, \, 2)\, $ :
\begin{eqnarray}
\label{haha2}
&&G(w) \, = \,
\,4\,w+25\,{w}^{2}+34\,{w}^{3}+144\,{w}^{4}+164\,{w}^{5}\, +
604\, w^6 \, + \,684\,  w^7\, + \,
+\, 2444\, w^8\, +\, 2764\, w^9 \cdots  \\
&&D(w) \, =
\,9+36\,w+63\,{w}^{2}+198\,{w}^{3}+288\,{w}^{4}+828\,{w}^{5}
 \,+ \, 1188 \, w^6 \, + \,3348 \,  w^7\,\,
+4788\, w^8\, \, +\, 13428\, w^9 + \, \cdots \nonumber
\end{eqnarray}
Of course one easily verifies that
$\, \, (1\, -8\, w) \cdot D(w) \,+\, 9 \, G(w)\, -\, 9 \, = \, \, 0 $,
which is simply deduced from :
\begin{eqnarray}
\label{gcd}
M_{n+1} \, = \, \, {{ K(M_n)} \over {G_{n+1}}}
\end{eqnarray}
One easily verifies, up to order nine,
  that (\ref{haha2}) are the expansions of the two
{\em rational}
expressions :
\begin{eqnarray}
G(w) \, = \, \,{\frac {w \cdot  (4+21\,w-7\,{w}^{2}\, +26\,{w}^{3}\, -16\,{w}^4 )}{
(1-w)\, (1+\, 2\,w) \, (1-2\,w )}}
\, , \, \qquad 
D(w) \, = \, \,
9\cdot {\frac { (1+3\,w -2\,{w}^2)\,(1\, +\, w^2)}{
 (1-w)\, (1+\, 2\,w) \, (1-2\,w )}}
\end{eqnarray}
There is nothing specific with $\, y \, =2\, $ : beyond $\, R(x, \, 2)$,
one also gets (generically) 
a $\, 2^n$ {\em exponential growth} of the iteration calculations
for $\, R(x, \, y)$, thus excluding Yang-Baxter integrability.
This may, at first sight, seem in contradiction with the fact that
the (local) Hamiltonian $\, H \, = \, $
$\,  H_{susy}\, + \rho \cdot  {\widehat N}\, $ 
is ``as integrable as'' 
  $\,  H_{susy}\,$ (or   $\,  H_{susy}\, +\,   {\widehat N}$). In fact
the (local)
Hamiltonian $\, H \, = \,\,  H_{susy}\, + \rho \cdot  {\widehat N}\, $ 
should also yield a {\em polynomial growth}
 of the calculation
 corresponding to  some
algebraic subvariety of  (\ref{form2}) such that 
the {\em tangent space} to this algebraic subvariety,
at point $\, R \, = \, \, P$, contains the vector 
 $\, P \cdot H \, = \,\,  P \cdot (H_{susy}\, + \rho \cdot  {\widehat N})$.
In other words, $\, R(x, \, y)\, $ should only be integrable 
in the $\, (x\, , \, y) \, \rightarrow \, (0, \, 0)\, $ limit,
but {\em not for finite values} for $\, (x\, , \, y)$. In our case 
the plane containing point $\, P$ and the two integrable ``vectors''
 $\, P \cdot  H_{susy}\, $ and 
 $\, P \cdot (H_{susy}\, +\,  \rho \cdot  {\widehat N})$,
is not included in the algebraic subvariety of polynomial growth. 
This is a quite general situation :  in general, integrability does not correspond
to {\em linear spaces} but to {\em algebraic subvarieties}
with some ``curvature''.

Deforming (\ref{good}) with  $\,  P \cdot  {\widehat N}$,
one thus finds that
the integrability of (\ref{good})  is ``surrounded'' 
by a (four-dimensional) space corresponding, generically, 
to a $\, 2^n$ {\em exponential growth}. However this 
{\em does not mean} that the {\em whole}
four-dimensional space (\ref{form2}) (except (\ref{good})
of course) corresponds to a $\, 2^n$ exponential growth: it is possible that some
algebraic {\em subvarieties} of  (\ref{form2}) 
could correspond to a {\em polynomial growth} 
of the iteration calculations and, possibly, correspond to a Yang-Baxter
integrability. The question is thus to find such polynomial growth, and
possibly Yang-Baxter
integrable, algebraic subvarieties~\footnote{This should be
particularly difficult when these subvarieties are not codimension-one
subvarieties, but codimension two, three (or more ...)
subvarieties}. The next section gives an explicit example of 
particular Yang-Baxter integrable deformations of the simple linear
interpolation $\, R_2 \, = \, \, {\cal
I}_d\,  \, + \, x \cdot P$, and gives the associated 
integrable algebraic subvariety. Section (\ref{requres}) 
sketches a strategy to find systematically~\footnote{Of course one can
always iterate {\em numerically} a point, very close to point $\, R\, = \,
P$, moving away from this point along directions like 
 (for $\, R_1$)  $\,  P \cdot H_{susy}
$,    $\,  P \cdot  {\widehat N}\, $ or   $\,  P \cdot  {\widehat N'}\, $
and visualize these integrable subvarieties~\cite{BoMa95}.  
} these polynomial growth,
and integrable, subvarieties (or, more generally, the subvarieties
of smaller topological entropy~\cite{zeta,topo}).

\subsection{Beyond 
 simple linear interpolation :
an integrable deformation of  $\, R_2 \, = \, \, 
{\cal I}_d\,  \, + \, x \cdot P\, $}
\label{baxbax}

Matrix  $\, R_2 \, = \, \, {\cal I}_d\,  \, + \, x \cdot P\, $ 
gives a Baxterisation of some integrable t-J 
Hamiltonian as
a simple {\em linear interpolation} between $\, P$
 and the identity matrix $\, {\cal I}_d$.
Let us try to extend this simple linear interpolation,
and ``merge'' it into a larger integrable $\, R$-matrix.
Similarly to what has been previously performed, let us consider
two matrices $\, N_{ew}\, $ and $\,  A_{sym}$,  which respectively
commute, and {\em anticommute}, with  $\, P$  :
\begin{eqnarray}
\label{new} 
&&N_{ew} \,\, = \, 
\left [\begin {array}{ccc}
1\,&0&0\\
\noalign{\medskip}0\,&0&0\\
\noalign{\medskip}0\,&0&0
\end {array}\right ]
\otimes
 \left [\begin {array}{ccc}
1\,&0&0\\
\noalign{\medskip}0\,&0&0\\
\noalign{\medskip}0\,&0&0
\end {array}\right ]
\, + \, \, 
\left [\begin {array}{ccc}
0\,&0&0\\
\noalign{\medskip}0\,&1&0\\
\noalign{\medskip}0\,&0&0
\end {array}\right ]
\otimes
 \left [\begin {array}{ccc}
0\,&0&0\\
\noalign{\medskip}0\,&1&0\\
\noalign{\medskip}0\,&0&0
\end {array}\right ]
\, + \, \, 
\left [\begin {array}{ccc}
0\,&0&0\\
\noalign{\medskip}0\,&0&0\\
\noalign{\medskip}0\,&0&1
\end {array}\right ]
\otimes
 \left [\begin {array}{ccc}
0\,&0&0\\
\noalign{\medskip}0\,&0&0\\
\noalign{\medskip}0\,&0&1
\end {array}\right ]
 \\
&&A_{sym}  \, = \, \,
\left [\begin {array}{ccc}
0\,&0&0\\
\noalign{\medskip}1\,&0&0\\
\noalign{\medskip}0\,&0&0
\end {array}\right ]
\otimes
 \left [\begin {array}{ccc}
0\,&-1&0\\
\noalign{\medskip}0\,&0&0\\
\noalign{\medskip}0\,&0&0
\end {array}\right ]
\, + \, \, 
\left [\begin {array}{ccc}
0\,&0&0\\
\noalign{\medskip}0\,&0&0\\
\noalign{\medskip}1\,&0&0
\end {array}\right ]
\otimes
 \left [\begin {array}{ccc}
0\,&0&1\\
\noalign{\medskip}0\,&0&0\\
\noalign{\medskip}0\,&0&0
\end {array}\right ]
\, + \, \, 
\left [\begin {array}{ccc}
0\,&1&0\\
\noalign{\medskip}0\,&0&0\\
\noalign{\medskip}0\,&0&0
\end {array}\right ]
\otimes
 \left [\begin {array}{ccc}
0\,&0&0\\
\noalign{\medskip}1\,&0&0\\
\noalign{\medskip}0\,&0&0
\end {array}\right ] \nonumber \\
&&\quad 
+\, \left [\begin {array}{ccc}
0\,&0&1\\
\noalign{\medskip}0\,&0&0\\
\noalign{\medskip}0\,&0&0
\end {array}\right ]
\otimes
 \left [\begin {array}{ccc}
0\,&0&0\\
\noalign{\medskip}0\,&0&0\\
\noalign{\medskip}-1\,&0&0
\end {array}\right ]
\, + \, \, 
\left [\begin {array}{ccc}
0\,&0&0\\
\noalign{\medskip}0\,&0&1\\
\noalign{\medskip}0\,&0&0
\end {array}\right ]
\otimes
 \left [\begin {array}{ccc}
0\,&0&0\\
\noalign{\medskip}0\,&0&0\\
\noalign{\medskip}0\,&1&0
\end {array}\right ]
\, + \, \, 
\left [\begin {array}{ccc}
0\,&0&0\\
\noalign{\medskip}0\,&0&0\\
\noalign{\medskip}0\,&1&0
\end {array}\right ]
\otimes
 \left [\begin {array}{ccc}
0\,&0&0\\
\noalign{\medskip}0\,&0&-1\\
\noalign{\medskip}0\,&0&0
\end {array}\right ]
\end{eqnarray}
Do note that $\, P \cdot  A_{sym} \, $ is a {\em diagonal} matrix
with diagonal entries $\, 0, \, -1, \, 1, \, 1, \, 0, \, -1, \, -1, \,
1, \, 0$. 
Let us consider a deformation of 
 $\, R_2 \, = \, \, {\cal I}_d\,  \, + \, x \cdot P\, $ with these two matrices.
This amounts to considering $\, R \, = \, P \, + \, \alpha \cdot   
{\cal I}_d\, + \, \beta \cdot  A_{sym}\, + \, \gamma \cdot
N_{ew}$, or equivalently the
 non-symmetric $\, R$-matrix :
\begin{eqnarray}
\label{rgood}
R_{z}(t, \, z) \, \, \,\, \,  = \, \,\,\, \, \, 
\left [\begin {array}{ccccccccc} 
D&0&0&0&0&0&0&0&0\\
\noalign{\medskip}0&A&0&B&0&0&0&0&0\\
\noalign{\medskip}0&0&A&0&0&0&C&0&0\\
\noalign{\medskip}0&C&0&A&0&0&0&0&0\\
\noalign{\medskip}0&0&0&0&D&0&0&0&0\\
\noalign{\medskip}0&0&0&0&0&A&0&B&0\\
\noalign{\medskip}0&0&B&0&0&0&A&0&0\\
\noalign{\medskip}0&0&0&0&0&C&0&A&0\\
\noalign{\medskip}0&0&0&0&0&0&0&0&D
\end {array}\right ] 
\end{eqnarray}
This form (\ref{rgood}) is stable by $\, {\widehat K}^{2}$.
The  $\,  \, P \, + \, \alpha \cdot 
{\cal I}_d\, + \, \beta \cdot  A_{sym}\, + \, \gamma \cdot N_{ew}\,\, $
 algebra is thus an algebra 
well-suited for the action of  $\, {\widehat K}^{2}$.
Let us consider (\ref{rgood}) for :
\begin{eqnarray}
\label{paramABCD}
A\, = \, t\cdot (1-z^6) \, , \quad \quad 
B\, = \, z^4\cdot (t^2-1) \, , \quad \quad 
C\, = \, z^2\cdot (t^2-1) \, , \quad \quad 
D \, = \, t^2-z^6
\end{eqnarray} 
The corresponding $\, R$-matrix
((\ref{rgood}), (\ref{paramABCD}))
  Baxterises in a very simple way :
\begin{eqnarray} 
\label{666}
&&\beta( t, \, z) \cdot {\widehat K}^{2} \Bigl(R_{z}(t, \, z)\Bigr)
\, \,  = \, \,  \, 
\,  \,  R_{z}(t, \,\,  t \cdot z) \\
&&{\rm where :} \qquad \qquad  \beta(t, \, z) \, = \, \,
 {\frac {\left (1-{z}^{3}\right )\left (1+{z}^{3}\right )\left ({z}^{3}{t}^{
3}-1\right )\left ({z}^{3}{t}^{3}+1\right ){t}^{2}}{\left (t+{z}^{3}\right 
)\left ({z}^{3}t+1\right )\left ({z}^{3}t-1\right )\left (t-{z}^{3}\right )
}} \nonumber 
\end{eqnarray} 
The $\, R$-matrix  (\ref{rgood}), with (\ref{paramABCD}),
is actually one of the Yang-Baxter integrable 
models of Perk and Schultz~\cite{PerkSchul,PerkSchul2}, and can be
seen as a deformation of $\, R_2(x)$. Actually, 
in the limit $\, (z, \, t)  \, = \, (1\, + Z \cdot h, \, 1\, + T \cdot
h)$, where $\, h \, \rightarrow \, 0$, one finds that 
 operator $\, A_{sym}\, $  occurs  at order one in $\, h \, $,
while  operator $\,  N_{ew}\,  $  occurs at order two in $\, h$ :
\begin{eqnarray} 
&&{{ 2\, R_z(t, \, z)} \over {z^2\cdot  (z^2+1 ) \cdot  (t^2-1)
}}\,  \, \,\, \longrightarrow \quad \quad  \Bigl( P \,  
 - \, 3 \, {{Z} \over {T}}  \cdot {\cal I}_d\Bigr) \, \, \, \, + 
 \, \,\,  \,\,   \Bigl( Z \cdot
A_{sym}\, \, + \,3 \,{\frac {Z\left (Z-T\right )}{2 \, T}} \cdot 
{\cal I}_d\Bigr) \cdot h  \nonumber \\
&&\,\quad \quad \quad  \, + \, \Bigl(  {{ Z \cdot (8\,Z-3\,
T ) } \over {2}}
  \cdot N_{ew} \,\, \,  - {{Z^2} \over {2}}  \cdot  A_{sym}\, \, 
 - \,{ {Z\, (16\,Z^2 -3\,T Z-3\,T^2 )} \over {4\, T}} \cdot 
{\cal I}_d \Bigr) \cdot h^2 \,+ \,  \cdots \nonumber 
\end{eqnarray}

Parametrization (\ref{paramABCD}) is equivalent to
the (Yang-Baxter) integrability~\cite{PerkSchul,PerkSchul2} condition :
\begin{eqnarray} 
\label{ABCD}
C_{ond}(A,\, B, \, C, \, D)
 \,\,  = \,\, \,  \, {B}^{3} D \, -{B}^{2}{C}^{2}\, +{C}^{3}D\, +BC{A}^{2}\,
 -BC{D}^{2} 
\,\, \, = \, \,\, \, 0
\end{eqnarray} 
This algebraic variety 
is an extension of the simple interpolation between the
identity matrix and $\, P$ (Yang $\, R$-matrix), which corresponds to 
$\, C\, = \, B$ and $\, D \, = \, \, A\, + \, B$.
Actually, for $\, C\, = \, B$, condition 
(\ref{ABCD}) factorizes as follows :
\begin{eqnarray} 
\label{ABCD2}
C_{ond}(A,\, B, \, B, \, D)
 \, = \, \, -B^2 \cdot \Bigl(D-(B-A)\Bigr)
 \cdot \Bigl(D-(A+B)\Bigr) \,\, \, = \, \,\, \, 0
\end{eqnarray} 

Considering the general $\, R$-matrix of the form (\ref{rgood})
({\em beyond} the integrability condition (\ref{paramABCD})),
the factorization analysis, which extracts
the gcd's at every step (see (\ref{gcd})), 
 yields a $\, 2^n$ {\em exponential 
 growth} of the calculations. The generating functions
of the degrees of the $\, M_n$'s and $G_n$'s (see (\ref{gcd}), (\ref{haha2}))
read  (the expansions of $\,
D(w) \,$ and
$\, G(w)\, $ have been obtained up to $\, w^8$) :
\begin{eqnarray}
\label{2puissancen}
D(w) \, = \, \, 
3 \cdot {\frac {3-2\,w+6\,{w}^{2}+2\,{w}^{3}-6\,{w}^{4}}{\left (1-w\right
)\left (1+w \right 
)\left (1-2\,w\right )}}\, , \quad \quad \quad
G(w) \, = \, \, 
w \cdot {\frac { 6-7\,w+18\,{w}^{2}+6\,{w}^{3}-16\,{w}^{4} }
{\left (1-w\right )
(1+w)\left (1-2\,w\right )}}
\end{eqnarray} 

In order to have a {\em polynomial growth} of the calculations
some {\em additional factorizations must occur}. Actually, for
(\ref{paramABCD}), that is when the integrability condition
(\ref{ABCD}) is satisfied,
the generating function of the degrees~\footnote{The degree, here, being the
degree in  $\, t$,  the calculations being performed for
 (\ref{paramABCD}) with $\, z\, = \, 3$. Note that there is
nothing specific with $\, z\, = \, 3$.} reads (the expansions of $\,
d(w) \,$ and
$\, g(w)\, $ have been obtained up to $\, w^8$) :
\begin{eqnarray}
\label{14}
d(w) \, = \, \,\,3 \cdot {\frac {6-2\,w+7\,{w}^{2}+7\,{w}^{3}}
{\left (1-{w}^{2}\right )\left (1-w\right )}}\, , \quad \quad \quad
g(w) \, = \, \,\,
w \cdot {\frac {\left (14-9\,w+19\,{w}^{2}+18\,{w}^{3}\right )}
{\left (1-{w}^{2}\right )\left (1-w\right )}}
\end{eqnarray}

\subsection{Integrability emerging from
 the occurrence of additional factorizations}
\label{requres}

In fact one can actually find  the integrability condition
(\ref{ABCD}), as a {\em condition corresponding to the occurrence of 
additional factorizations}. The ``bifurcation'' 
from the polynomial 
growth of the degrees to a $\, 2^n$ exponential growth of the degrees,
  takes place immediately when leaving the
integrability 
condition (\ref{ABCD}). In the $\, (A, \, B, \, C, \, D)\, $
parameter space, the integrability 
condition (\ref{ABCD}) is ``surrounded'' by 
a   $\, 2^n\, $ exponential growth. More specifically, the ``bifurcation''
 from the polynomial 
growth of the degrees to a $\, 2^n$ exponential growth of the degrees,
  takes place {\em with the first gcd},
 namely $\, G_1$.
Let us consider the factorization 
of the $\, R$-matrix corresponding to  (\ref{paramABCD}) (that is  (\ref{ABCD})),
and the one where $\, A$ has been replaced 
by  $\, A\, +\, u$, which corresponds to a
 $\, 2^n$ exponential growth. 
The first gcd's corresponding to these two situations,
namely for $\, u\, = \, 0\, $ (polynomial growth, integrable) $\, g_1$,
and  for $\, u\, \ne \, 0\, $ ($ 2^n$  exponential growth) $\, G_1(u)$,
read respectively~\footnote{
Note that the factors of $\, G_1(u)\, $ 
actually correspond to some factors 
in $\, \det(R(u))\, $ which reads $\, 
 \det(R(u))\, \, = \,
\left (-u+{z}^{3}-t+t{z}^{6}-{z}^{3}{t}^{2}\right )^{3}\cdot 
\left (-u-{z}^{3}-t+t{z}^{6}+{z}^{3}{t}^{
2}\right )^{3} \cdot \left ({t}^{2}-{z}^{6}\right )^3$.
This is a general result : the cofactors 
associated with $\, K^n\, $ or $\, t_1$ and $\, I$
are necessarily related with powers of the factors 
occurring in the determinant (or the whole determinant),
and their transformed by the group generated 
by $\, t_1$ and $\, I$.} :
\begin{eqnarray} 
&&G_1 (u)\, = \, \, 
\left (-u+{z}^{3}-t+t{z}^{6}-{z}^{3}{t}^{2}\right )^{2} \cdot 
\left (-u-{z}^{3}-t+t{z}^{6}+{z}^{3}{t}^{
2}\right )^{2} \cdot \left ({t}^{2}-{z}^{6}\right )^{2}\, = \, \, 
D^2 \cdot (B\cdot C\, -A^2)^2  \nonumber \\
&&g_1 \, = \, \,
\left ({t}^{2}-{z}^{6}\right )^{5} \cdot 
\left ({z}^{6}t^2-1\right )^{2}
\end{eqnarray} 
One remarks that :
\begin{eqnarray} 
 g_1  \, \,\, = \, \,  \, \,(t^2-z^6)
\cdot   G_1 (u=0)\, \, \,\, = \, \,  \, \,
 D \cdot   G_1(u=0)\,
\end{eqnarray} 
which amounts to saying that, restricting to 
the integrability condition (\ref{ABCD}) 
(here $\, u \, = \, 0$), an {\em additional factorization
occurs}, changing (\ref{2puissancen}), and its $\, 2^N$ exponential growth
of the calculations, into (\ref{14}) and its associated {\em polynomial
growth}. When the integrability condition (\ref{ABCD}) is not
satisfied, one has 
$\, G_1 \, = \,  \, \, D^2 \cdot (B\cdot C\, -A^2)^2$, and 
the first ``reduced'' matrix $\, M_1\, $ 
reads :
\begin{eqnarray} 
\label{reducedm1}
M_1 \,\, = \,\,
{{K(R)} \over {G_1}} \, \,= \, \, \, \,
\left [\begin {array}{ccccccccc} 
{A}^{2}-BC&0&0&0&-D\,C&0&0&0&-D\,B\\
\noalign{\medskip}0&D\,A&0&0&0&0&0&0&0\\
\noalign{\medskip}0&0&D\,A&0&0&0&0&0&0\\
\noalign{\medskip}0&0&0&D\,A&0&0&0&0&0\\
\noalign{\medskip}-D\,B&0&0&0&{A}^{2}-BC&0&0&0&-D\,C\\
\noalign{\medskip}0&0&0&0&0&D\,A&0&0&0\\
\noalign{\medskip}0&0&0&0&0&0&D\,A&0&0\\
\noalign{\medskip}0&0&0&0&0&0&0&D\,A&0\\
\noalign{\medskip}-D\,C&0&0&0&-D\,B&0&0&0&{A}^{2}-BC
\end {array}\right ]
\end{eqnarray} 
Condition (\ref{ABCD}), namely $\, 
 (B^3\, +\, C^3) \, D \, +\,  \Bigl(
({A}^{2}\, -\, BC) \, - D^2 \Bigr) \, B\, C 
 \, = \,  \,  0 $,
 is {\em nothing but the condition of 
divisibility} of $\, {A}^{2}-BC\, $ by $\, D$, which obviously enables
to extract an {\em additional factor} $\, D$ 
in all the entries of (\ref{reducedm1}).
The fact that (\ref{ABCD}) is such a  condition of 
divisibility, becomes obvious if one uses
the fact that (\ref{ABCD}) is a {\em rational surface}, and 
one actually uses this rational parameterization.
With parameterization (\ref{paramABCD})
the division of  $\, {A}^{2}-BC\, $ by $\, D$
reads :
\begin{eqnarray} 
 {{ {A}^{2}-BC\,} \over {D}} \, \,  = \,  \, 
 \, \, 1\, -\, z^6 \cdot t^2 
\end{eqnarray} 

Finding systematically 
algebraic varieties (like (\ref{ABCD}) 
but not necessarily rational) which correspond 
to {\em additional factorizations} in the factorization
scheme, and thus correspond to a smaller 
complexity of the calculations 
(smaller Arnold complexity~\cite{zeta}, smaller 
topological entropy~\cite{topo}), 
{\em can be seen as a new approach} of integrable models.
The set of all these singled-out
algebraic varieties, of non generic topological 
entropy~\cite{zeta}, should be a {\em complicated
stratified space}. A remarkable subset
is the set of algebraic varieties associated with a {\em polynomial growth} of
the calculations. A remarkable subset of the previous subset
is the set of algebraic varieties 
which are Yang-Baxter integrable.

In this {\em integrable deformations} 
framework  (\ref{intdefo}) one  notes that the $\, 9 \times 9\, $ matrices
$\, H_{n,n+1}$, considered for different values of 
their parameters $\, V, \, J, \, t$, commute all together
and commute with $\, P\, $ and $\, P\cdot  H_{n,n+1}$. 
Therefore one could imagine that the matrices $\, R_1$, and 
 $\, R_2$, could belong to
 the {\em same} integrable algebraic variety.
Is it possible to give  a complete description of {\em all} the
 {\em integrable deformations} (\ref{intdefo}), and a complete
description of the algebraic varieties associated 
with the Baxterisation of the largest integrable deformation 
of such quantum Hamiltonians ? 
These questions will be studied elsewhere.

\section{Let us Baxterise differential operators }
\label{differentialope}

The Baxterisation is the building of the spectral parameters 
 wherever they live : on an elliptic curve, on a rational surface, on
 a Jacobian variety, on an Abelian
 variety, ... The Baxterisation of $\, R$-matrices, 
or monodromy matrices, can always be performed
systematically using various 
methods (analytical methods, by formal calculations, 
by visualization of the orbits~\cite{BoMaRo95,BoMa95}, ...).
Let us show, in this section, that these results 
can even be extended to {\em infinite dimensional}
generalizations
of monodromy matrices, namely when the block matrices in the $\,
R$-matrix
are infinite dimensional linear
 operators (L-operators, local quantum Lax matrices ...).

Let us consider, here, the L-operator corresponding to
the Toda chain~\cite{Toda} :
\begin{eqnarray}
\label{Loper}
L[\lambda] \, = \, \, 
\left [\begin {array}{cc}
 \lambda\, - \, {{d} \over {dx}} \,&e^{x}\\
\noalign{\medskip} e^{-x}&0
\end {array}\right ]
\end{eqnarray}
This can be seen as a generalization of
the monodromy matrices  (\ref{t1bloc}) 
where the $m \times m$ matrices $A$,  $B$, $C$ and  $D$
become operators which can only have  {\em infinite dimensional 
representations}. Let us introduce an ``inverse''  L-operator :
\begin{eqnarray}
I(L)[\lambda'] \, = \, \, 
\left [\begin {array}{cc}
0 \,&e^{x}\\
\noalign{\medskip} e^{-x}&\lambda'\, + \, {{d} \over {dx}} \,
\end {array}\right ]
\end{eqnarray}
and let us perform the products of these two  L-operators :
\begin{eqnarray}
&&L[\lambda] \cdot I(L)[\lambda'] \, = \, \, 
\left [\begin {array}{cc}
 \lambda\, - \, {{d} \over {dx}} \,&e^{x}\\
\noalign{\medskip} e^{-x}&0
\end {array}\right ]
\cdot
 \left [\begin {array}{cc}
0 \,&e^{x}\\
\noalign{\medskip} e^{-x}&\lambda'\, + \, {{d} \over {dx}} \,
\end {array}\right ]
\, = \, \,
 \left [\begin {array}{cc}
 1 \,& N \\
\noalign{\medskip} 0 &1
\end {array}\right ] \nonumber \\
&&\quad \quad {\rm where :} \qquad \qquad  N \, = \, \, \Bigl(\lambda\, - \, {{d}
\over {dx}} \Bigr)
\cdot e^{x} \, + \,  e^{x} \cdot  \Bigl( \lambda'\, + \, {{d} \over {dx}} \Bigr)
\end{eqnarray}
or :
\begin{eqnarray}
&&I(L)[\lambda'] \cdot L[\lambda] \, = \, \, 
\left [\begin {array}{cc}
0 \,&e^{x}\\
\noalign{\medskip} e^{-x}&\lambda'\, + \, {{d} \over {dx}} \,
\end {array}\right ] \cdot \left [\begin {array}{cc}
 \lambda\, - \, {{d} \over {dx}} \,&e^{x}\\
\noalign{\medskip} e^{-x}&0
\end {array}\right ] \, = \, \, 
 \left [\begin {array}{cc}
 1 \,& 0 \\
\noalign{\medskip} N' &1
\end {array}\right ] \nonumber \\
&&\quad \quad {\rm where :} \qquad \qquad N' \, = \, \, e^{-x} \cdot
\Bigl(\lambda\, - \, {{d} \over {dx}}\Bigr) \, + \, 
\Bigl( \lambda'\, + \, {{d} \over {dx}} \Bigr) \cdot  e^{-x}
\end{eqnarray}
It is a straightforward calculation to see that
the derivative operator $\, {{d} \over {dx}} \, $ 
disappears in the off-diagonal $\, N\, $ and $\, N'\, $ operators, and that $\, N\, $ 
and $\, N'\, $ reduce to the operators of multiplication by the function
 $\, (\lambda' \, + \, \lambda\, -1) \cdot   e^{x}$, and 
 $\, (\lambda' \, + \, \lambda\, -1) \cdot   e^{-x}$, respectively.
The operators $\, N$  and $\, N'\, $ reduce to the null operator
for  $\, \lambda' \, = \, 1 \, - \,  \, \lambda$.
Therefore one has the following result :
\begin{eqnarray}
L[\lambda] \cdot I(L)[ 1 \, - \,  \, \lambda] \, = \, \, 
I(L)[ 1 \, - \,  \, \lambda]  \cdot L[\lambda] \, = \, \, 
 \left [\begin {array}{cc}
 1 \,& 0 \\
\noalign{\medskip} 0 &1
\end {array}\right ]
\end{eqnarray}
As it should the left ``inverse'' coincides with 
the right  ``inverse''. Let us use such inverse and combine it 
with  the partial transposition $\, t_1$, previously
described. Let us now perform the partial transposition $\, t_1\, $ 
 which amounts to permuting the two off-diagonal operators :
\begin{eqnarray}
\label{truc1}
t_1(L)[\lambda] \, = \, \, 
\left [\begin {array}{cc}
 \lambda\, - \, {{d} \over {dx}} \,&e^{-x}\\
\noalign{\medskip} e^{x}&0
\end {array}\right ]
\end{eqnarray}
One easily finds, as a consequence of the nullity
of  two operators $\, N_1\, $ and $\, N_2$, 
when  $\,  \lambda'\, = \, \,-1\, -\lambda\, $ :
\begin{eqnarray}
N_1 \, = \, \Bigl( \lambda\, - \, {{d} \over {dx}}\Bigr) \cdot e^{-x} \, + \,  e^{-x}
 \cdot \Bigl( \lambda' \, + \, {{d} \over {dx}}\Bigr)
\qquad {\rm and :} \qquad 
N_2 \, = \,  e^{x} \cdot \Bigl( \lambda\, - \, {{d} \over {dx}}\Bigr)  \, + \,  
  \Bigl( \lambda' \, + \, {{d} \over {dx}}\Bigr)  \cdot e^{x} \nonumber
\end{eqnarray}
that the inverse 
of (\ref{truc1}) reads :
\begin{eqnarray}
\label{truc3}
I(t_1(L))[\lambda'] \, = \, \, 
\left [\begin {array}{cc}
0 \,&e^{-x}\\
\noalign{\medskip} e^{x}&\lambda'\, + \, {{d} \over {dx}} \,
\end {array}\right ] \qquad \quad {\rm where :} \quad  \qquad 
\lambda'\, = \, \,-1\, -\lambda 
\end{eqnarray}
Let us perform, again, the partial transposition $\, t_1\, $ 
on the previous  L-operator  :
\begin{eqnarray}
\label{truc}
t_1(I(t_1(L)))[\lambda'] \, = \, \, 
\left [\begin {array}{cc}
0 \,&e^{x}\\
\noalign{\medskip} e^{-x}&\lambda'\, + \, {{d} \over {dx}} \,
\end {array}\right ]
\end{eqnarray}

The inverse of (\ref{truc}) reads :
\begin{eqnarray}
I(t_1(I(t_1(L))))[\lambda''] \, = \, \, 
\left [\begin {array}{cc}
 \lambda'' \, - \, {{d} \over {dx}} \,&e^{x}\\
\noalign{\medskip} e^{-x}&0
\end {array}\right ]
\end{eqnarray}
with $\, \lambda'' \, = \, -\lambda'\, +1$.
Actually :
\begin{eqnarray}
t_1(I(t_1(L)))[\lambda'] \cdot I(t_1(I(t_1(L))))[\lambda'']
\, = \, \, 
\left [\begin {array}{cc}
0 \,&e^{x}\\
\noalign{\medskip} e^{-x}&\lambda'\, + \, {{d} \over {dx}} \,
\end {array}\right ]
\cdot \left [\begin {array}{cc}
 \lambda'' \, - \, {{d} \over {dx}} \,&e^{x}\\
\noalign{\medskip} e^{-x}&0
\end {array}\right ]\, = \, \, 
 \left [\begin {array}{cc}
 1 \,& 0 \\
\noalign{\medskip} N_3 &1
\end {array}\right ] \nonumber
\end{eqnarray}
\begin{eqnarray}
I(t_1(I(t_1(L))))[\lambda'']\cdot t_1(I(t_1(L)))[\lambda'] \, = \,
\, \left [\begin {array}{cc}
 \lambda'' \, - \, {{d} \over {dx}} \,&e^{x}\\
\noalign{\medskip} e^{-x}&0
\end {array}\right ] 
\cdot \left [\begin {array}{cc}
0 \,&e^{x}\\
 \noalign{\medskip} e^{-x}&\lambda'\, + \, {{d} \over {dx}} \,
\end {array}\right ] \, = \, \, 
 \left [\begin {array}{cc}
 1 \,& N_4 \\
\noalign{\medskip} 0 &1
\end {array}\right ]
\end{eqnarray}
where :
\begin{eqnarray}
&&N_3 \, = \, \, e^{-x} \cdot \Bigl( \lambda'' \, - \, {{d} \over {dx}}
\Bigr) \, + \,  \Bigl( \lambda' \, + \, {{d} \over {dx}}
\Bigr) \cdot  e^{-x} \, = \, \, 
\Bigl( \lambda''\, + \,  \lambda' -1 \Bigr) \cdot e^{-x} 
\nonumber \\
&&\qquad {\rm and :} \qquad 
N_4 \, = \, \,  \Bigl( \lambda'' \, - \, {{d} \over {dx}}
\Bigr) \cdot  e^{x}\, +  e^{x} \cdot \Bigl( \lambda' \, + \, {{d} \over {dx}}
\Bigr)\, = \, \, \Bigl( \lambda''\, + \,  \lambda' -1 \Bigr) \cdot e^{x}
\end{eqnarray}
Combining this result, and the previous result (\ref{truc3}),
one gets that  $\, I(t_1(I(t_1(L))))\, $ {\em  has exactly the same form
as} $\, L$, {\em  but with} $\, \lambda \, $ {\em changed into}  $\, \lambda \, +2$ :
\begin{eqnarray}
 I(t_1(I(t_1(L))))[\lambda] \, = \, \, L[\lambda \, +2] \, = \, \, 
\left [\begin {array}{cc}
 (\lambda \, +2)\, - \, {{d} \over {dx}} \,&e^{x}\\
\noalign{\medskip} e^{-x}&0
\end {array}\right ]
\end{eqnarray}
$\, I\, t_1\, I\, t_1 \, $ is an {\em infinite order} 
transformation acting on the L-operator (\ref{Loper}).
This provides a first, and very simple, example of 
{\em rational} Baxterisation of the  L-operator (\ref{Loper}).
The iteration of $\, I\, t_1\, I\, t_1 \, $ 
yields a ``trajectory'', in the space of  the L-operator (\ref{Loper}),
corresponding to a {\em straight line} : 
$\, \lambda\, \rightarrow \, \lambda \, +\, 2$.
It seems necessary, here, to perform $\, (I\, t_1)^2$, 
instead of $\, I\, t_1$, to preserve 
the form of the L-operator. In fact one can only perform 
 $\, I\, t_1 $ if one recognizes that (\ref{truc3})
is like $L(\lambda\,+1)$, up to a transformation $\,  c \cdot s\, $
which commutes with the inversion $I$ and 
with the partial transposition $t_1$ :
\begin{eqnarray}
\left [\begin {array}{cc}0 \,&e^{-x}\\
\noalign{\medskip} e^{x}& -1-\lambda\, + \, {{d} \over {dx}} \,
\end {array}\right ] \, = \, c \Bigl( s \Bigl(\left [\begin {array}{cc}
 \lambda\,+1\,  - \, {{d} \over {dx}} \,&e^{x}\\
\noalign{\medskip} e^{-x}&0
\end {array}\right ]
\Bigr)\Bigr) 
\end{eqnarray}
where $\, c\, $ and $\, s \, $ are
 the two following transformations ($U$, $\, V$,
$\, W$, $\, T$ are any operator) :
\begin{eqnarray}
\label{cets}
c \Bigl( \left [\begin {array}{cc}
U \,&V\\
\noalign{\medskip} W& T\,
\end {array}\right ] \Bigr) \, = \, \, \left [\begin {array}{cc}
T \,& W\\
\noalign{\medskip} V& U\,
\end {array}\right ] \qquad {\rm and :}\qquad 
s \Bigl( \left [\begin {array}{cc}
U \,&V\\
\noalign{\medskip} W& T\,
\end {array}\right ] \Bigr) \, = \, \, \left [\begin {array}{cc}
-U \,&V\\
\noalign{\medskip} W& -T\,
\end {array}\right ] 
\end{eqnarray}
Therefore  $\, I\, t_1 $ is associated with the shift 
$\, \lambda \, \rightarrow \, \lambda + \, 1$.
One sees that $\, \lambda \, $ actually {\em plays the role of the spectral
parameter} for the Toda L-operator.

Many other examples, corresponding to
differential operators
and L-operators, or local quantum Lax matrices
associated with the $\, XXX$ quantum Hamiltonian,
yield similar calculations :  for instance, a very simple
 example of L-operator, 
associated with the discrete self-trapping model
 (see for
instance~\cite{Sklyanin})
can easily be Baxterised, the calculations being 
extremely similar to the one sketched in this section.

\subsection{Let us Baxterise higher derivatives }

Let us consider a (slightly) more complicated 
example of Baxterisation of differential operators.
In the following we will denote
by $p$ the derivative operator 
$\, {{d} \over {dx}} \,$.
More generally, let us consider 
the following   L-operator :
\begin{eqnarray}
\label{Lopergen}
L_N \, = \, \, 
\left [\begin {array}{cc}
  {\cal A}^{(0)}(x)\,  \,&B(x)\\
\noalign{\medskip} C(x)& 0
\end {array}\right ]
\end{eqnarray}where $\, {\cal A}^{(0)}(x)\, $ denotes the differential operator 
$\,  A_0(x)\, + \, A_1(x)\cdot p \,  + \cdots 
\, + \, A_m(x)\cdot p^m\, + \cdots \, +\,A_N(x)\cdot p^N$
and where the two functions $\, B(x)$ and $\, C(x)$ are related by 
$\, B(x) \cdot C(x) \, = \, 1$.
Let us first perform the partial transposition
$\, t_1$ :
\begin{eqnarray}
t_1(L_N) \, = \, \, 
\left [\begin {array}{cc}
 A_0(x)\, + \, A_1(x)\cdot p \,  + \cdots 
\, + \, A_m(x)\cdot p^m\, + \cdots \, +\,A_N(x)\cdot p^N  \,&C(x)\\
\noalign{\medskip} B(x)& 0
\end {array}\right ]
\end{eqnarray}

Let us try to find the inverse of this  L-operator. 
Let us denote  $\, {\cal A}^{(1)}(x)\, $
another similar
 differential operator $\, \tilde{A}_0(x)\, + \, \tilde{A}_1(x)\cdot p \,  + \cdots 
\, + \, \tilde{A}_m(x)\cdot p^m\, + \cdots \, +\,\tilde{A}_N(x)\cdot p^N $.
It is straightforward to see
that $\, t_1(L_N) \cdot I(t_1(L_N)) \, $ is the identity
 matrix when $\, I(t_1(L_N))\, $ reads :
\begin{eqnarray}
\label{ILoper}
I(t_1(L_N)) \, = \, \, 
\left [\begin {array}{cc}
 0 \,&\, C(x)\\
\noalign{\medskip} \, B(x)&  -\, {\cal A}^{(1)}(x)
\end {array}\right ]
\end{eqnarray}
with :
\begin{eqnarray}
\label{superrel}
{\cal A}^{(0)}(x) \cdot C(x)\, \, = \, \,  \, \, \, C(x) 
\cdot  {\cal A}^{(1)}(x) 
\end{eqnarray}
The fact that the left inverse 
and the right inverse identify 
is just a consequence of the fact that
(\ref{superrel}) is
equivalent to
$\, B(x) \cdot {\cal A}^{(0)}(x)\, = \, \,  \, 
  {\cal A}^{(1)}(x)\cdot B(x)\, $ when $\, B(x) \cdot C(x) \, = \, 1$.
One easily sees that, up to the previous $\, c$ and $\, s$ transformations,
 the infinite order transformation $\, I\, t_1\, $ reads :
\begin{eqnarray}
I ( t_1 \Bigl( \left [\begin {array}{cc}
 {\cal A}^{(0)}(x)   \,&B(x)\\
\noalign{\medskip} C(x)& 0
\end {array}\right ]
\Bigr) \, ) \, \, = \, \,  \,  \, \,
c  \Bigl( s \Bigl(  \left [\begin {array}{cc}
 {\cal A}^{(1)}(x)   \,&B(x)\\
\noalign{\medskip} C(x)& 0
\end {array}\right ]
\Bigr)\Bigr)
\end{eqnarray}
where  $\, {\cal A}^{(1)}(x)\, $ is deduced from 
 $\, {\cal A}^{(0)}(x)\, $
by (\ref{superrel}). From now on let us note
 $\, \tilde{A}_N$, $\, A_N$, $C$ the functions
$\tilde{A}_N(x)$, $\, A_N(x)$ and $\, C(x)$. We will denote $C'$, $C''$ and
$C^{(N)}$
the first, second and $\, N$-th derivatives of $\, C(x)$.
For $\, N=3\, $ the transformation
$\, {\cal A}^{(0)}(x)\,\rightarrow \, {\cal A}^{(1)}(x)\, $    reads :
\begin{eqnarray}
\label{tildeA}
&& C \cdot \tilde{A}_3 \, = \, C \cdot A_3
\nonumber \\
&&   C \cdot\tilde{A}_2 \, = \, C \cdot A_2 \, + 3 \cdot C' \cdot A_3
\nonumber \\
&&    C \cdot\tilde{A}_1 \, = \, C \cdot A_1 \, 
+ 2 \cdot C' \cdot A_2 + 3 \cdot C'' \cdot A_3
\nonumber \\
&&   C \cdot\tilde{A}_0 \, = \, C \cdot A_0 \,
 +  \cdot C' \cdot A_1 +  \, C'' \cdot A_2
 + \, C^{(3)} \cdot A_3
\end{eqnarray}
Let us denote $\, {\it S_1}\, = \, C'/C$, $\, {\it S_2}\, = \, C''/C$
and $\,  {\it S_3} \, = \, C^{(3)}/C$.
One can straightforwardly associate to  transformation (\ref{tildeA})
a $\, 4 \times 4\, $ matrix :
\begin{eqnarray}
\label{MC}
M \, = \, \, \left [\begin {array}{cccc}
 1&{\it S_1}&{\it S_2}&{\it S_3} \\
\noalign{\medskip}0&1&2\,{\it S_1}&3\,{\it S_2}\\
\noalign{\medskip}0&0&1&3\,{\it S_1}\\
\noalign{\medskip}0&0&0&1
\end {array}\right ]
\end{eqnarray}
The Baxterisation process corresponds to the iteration of this matrix.
Note that this iteration yields a simple 
{\em group structure} :
\begin{eqnarray}
\label{itera}
M^n \, = \, \, 
\left [\begin {array}{cccc}
 1&n\cdot {\it S_1}&n\cdot {\it S_2}+\, n\cdot (n-1)
 \cdot {{\it S_1}}^{2}& n \cdot {\it S_3}
+ 3 \cdot n \cdot (n-1) \cdot 
{\it S_1}\,{\it S_2}+ n\cdot (n-1) \cdot (n-2) \cdot {{\it S_1}}^{3}\\
\noalign{\medskip}0&1& 2 \cdot n\cdot {\it S_1}& 3 \cdot n\cdot {\it
S_2}\,
+\,3 \cdot n\cdot (n-1) \cdot {{\it S_1}}^{2}\\
\noalign{\medskip}0&0&1& 3 \cdot n\cdot {\it S_1}\\
\noalign{\medskip}0&0&0&1
\end {array}\right ]
\end{eqnarray}

If, instead of seeing  $\, I(t_1(L_N))\, $ as an operator
similar to $\, L_N$, 
up to the previous $\, c$ and $\, s$ transformations  (\ref{cets}),
one sees $\, I(t_1(L_N))\, $ as an operator
similar to $\, L_N$, 
up to transformation $\, c$ {\em only},
one associates to $\, I \, t_1\, $ 
the $\, 4 \times 4\, $ matrix $\, -M$, instead of 
$\, M$, yielding to a $(-1)^n \, $ factor in front of (\ref{itera}).
If, similarly to the first Toda L-operator
example, 
one just considers the iteration of $\, (I \, t_1)^2$, 
one does not have  a $(-1)^n \, $ factor problem, but  (\ref{itera})
is only valid for $\, n\, $ {\em even}. Note that 
the ``time reversal'' $\, I \cdot t_1 \, \rightarrow t_1 \cdot I\, $
corresponds to the same transformation as (\ref{tildeA}),
but where the function $\, C(x)\, $ is changed into 
the function $\, B(x)\, = 1/C(x)$. When $\, C(x)\, $ is
an exponential $\, C(x) \, = \, e^{r \cdot x}\, $
the entries in the  matrix $\, M\, $ are not functions of $\, x\, $
but  numbers. For $\, N\, = \, 1$ and $\, C(x) \, = \, e^{ x}\, $
one recovers the previous Toda result associated to 
the $\, 2 \times 2\, $ (shift) matrix :
\begin{eqnarray}
M \, = \, \, 
\left [\begin {array}{cc} 1&1\\
\noalign{\medskip}0&1\end {array}
\right ]
\end{eqnarray}
To sum up, transformation $\widehat{K}^2\, $ is associated with 
 matrix $\, M^2\,$ (with $\,M\,$ given by  (\ref{MC})). We are thus
 reduced to the analysis of the iteration of $\,M\,$ given
by  (\ref{MC}). It is clear from (\ref{itera}) that the
orbits of $\widehat{K}^2\, $  in the 
``huge'' functional space ($\,A_0(x)$, $\,A_1(x)$,  $\,
A_2(x)$,  $A_3(x)$), 
correspond to a {\em rational curve}. This rational curve corresponds to 
the rational parametrization (Baxterisation)
of  (\ref{itera}) which consists in considering the integer
$\,n\, $ as a real (or complex) number.

{\bf Remark :} It is worth noticing that all these 
Baxterisations of $\, L$-operators have been performed
{\em without using any Yang-Baxter} ($R\,  L\, L \, = \, L\, L \, R$)
{\em hypothesis}. These calculations are similar to the 
calculations on the sixteen vertex model
where the iteration of $\, {\widehat K}$ yields a foliation of the whole 
fifteen dimensional space of the model in elliptic curves, {\em beyond}
the Yang-Baxter integrable cases~\cite{prl2}.

\subsection{ Some exercises of Baxterisation}

Of course all these calculations can be generalized to 
more general   L-operators. We let the reader 
 perform a few exercises of Baxterisation.

$\bullet$ Let us Baxterise the L-operators :
\begin{eqnarray}
L \, = \, \, \, 
\left [\begin {array}{cc}
 u\, +y\cdot {{d} \over {dy}}\, -l& \quad {{d} \over {dy}}\\
\noalign{\medskip}-{y}^{2}\cdot {{d} \over {dy}}\, +2\,l\, y& \quad u\, -y\cdot
{{d} \over {dy}}\, +\, l
\end {array}\right ]
\end{eqnarray}
where one recognizes~\cite{Boouwk} the realization of the finite 
dimensional Lie algebra $\, \widehat{sl}(2)\, $ in terms of
differential
operators : 
\begin{eqnarray}
 e \,  = \, \, {{d} \over {dy}}, \,\quad \quad  \, h \, = \, -2\, y\cdot
{{d} \over {dy}}\, +\, 2\, l , \, \quad\quad  \, f \, = \, \, -{y}^{2}\cdot {{d}
\over {dy}}\, +2\,l\, y
\end{eqnarray}
What is the ``status'' of the $\, u$ and $\, l$ parameters from a
Baxterisation point of view : gauge parameters, spectral parameters ... ? 

$\bullet$ Show that the Baxterisation
of the R-matrix~\cite{Ruiz}
\begin{eqnarray}
\label{119}
R \, = \, \, 
\left [\begin {array}{cccc} 1&0&\quad 0&0\\
\noalign{\medskip}0&u&\quad 1-u&0\\
\noalign{\medskip}0&0&\quad 1&0\\
\noalign{\medskip}1-u&0&\quad 0&u
\end {array}\right ]
\end{eqnarray}
yields a {\em finite order} group :
the inverse $\, {\widehat I}\, $ amounts to changing
$\, u \, $ into $\, 1/u$, and 
transformation $\, \widehat{K}^2\, = $
$\, t_1 \cdot \widehat{I} \cdot t_1 \cdot \widehat{I} \, $
leaves the R-matrix (\ref{119}) invariant :  $\, \widehat{K}^2(R)\, = \, R$.

$\bullet$ Show that the Baxterisation
of the R-matrix~\cite{Resh}
\begin{eqnarray}
R(q) \, = \, \, \left [\begin {array}{cccc} 
\sqrt {q}&0&0&0\\
\noalign{\medskip}0&1& (q-1)/\sqrt {q}&0\\
\noalign{\medskip}0&0&1&0\\
\noalign{\medskip}0&0&0&\sqrt {q}
\end {array}\right ]
\qquad {\rm yields :} \qquad 
R(q, \, n) \, = \, 
\left [\begin {array}{cccc} \sqrt {q}&0&\quad 0&0\\
\noalign{\medskip}0&1&\quad \left (q-1\right 
){q}^{n-1/2}&0\\\noalign{\medskip}0&0&\quad 1&0\\
\noalign{\medskip}0&0&\quad 0&\sqrt {q}
\end {array}\right ]
\end{eqnarray}
Hint : the inverse of $\, R(q)\, $ is  $\, R(1/q)\, $
but  $\, \widehat{K}^{2\, n} (R(q))\, = \, \, R(q, n)$ .

$\bullet\, $ Recalling that the quantum group $\, SL_q(2)\, $
is a Hopf algebra with the antipode $\, S\, $ 
given explicitly by~\cite{Kosm} : 
\begin{eqnarray}
S(T) \, = \, \, \,\,
S \left [\begin {array}{cc} A&B\\
\noalign{\medskip}C&D
\end {array}\right ]
\, = \, \, \,\,
\left [\begin {array}{cc} 
D&-B/q\\
\noalign{\medskip}-\, q\, C&A\end {array}
\right ]
\end{eqnarray}
show that the antipode $\, S\, $ 
plays exactly the role of the inverse in the Baxterisation procedure.
One will use the relations :
\begin{eqnarray}
C \, D \, - q\, D\, C\, = \, \, 0\, , \, \, \, \,\, \, \,\,
q\, B \, A \, - \, A\, B\, = \, \, 0\, , \, \, \, \,\, \, \,\,
A\, D\, -\, D\, A \, = \, \, (q-{{1} \over {q}}) \cdot B\, C\, , \, \,\,\, \, \,\,
\,  B\, C \, = \, \, C\, B
\end{eqnarray}
Hint : calculate the product $\, T \cdot S(T) \,$ and introduce the
quantum determinant 
$\,  \det_q(T)\, = \,\, \,\, A\, D \, - \, q \, B\, C$. This gives  :
\begin{eqnarray}
T \cdot S(T) \, = \, \, \,\,
\left [\begin {array}{cc}
 A&B\\
\noalign{\medskip}C&D
\end {array}\right ]
\cdot
\left [\begin {array}{cc} 
D&-B/q\\
\noalign{\medskip} -\, q\, C&A
\end {array}
\right ]\,\, \,= \, \, \, \,
 \det_q(T) \cdot 
\left [\begin {array}{cc} 1&0\\
\noalign{\medskip}0&1
\end {array}\right ]
\end{eqnarray}

Introducing transformation $\, t_1\, $ :
\begin{eqnarray}
t_1(T) \, \, = \, \,\, t_1 \left [\begin {array}{cc} A&B\\
\noalign{\medskip}C&D
\end {array}\right ]
 \, \, = \, \,\,
\left [\begin {array}{cc} A&C\\
\noalign{\medskip}B&D
\end {array}\right ]
\end{eqnarray}
let us consider the infinite dihedral group generated
 by these two involutions $\, S\, $ and $\, t_1$, or just the iteration
of transformation $\, t_1 \cdot S$. What is the ``quantum group meaning''
of this infinite dihedral group?

$\bullet\, $ Let us recall the L-operator corresponding to the
Liouville model on a lattice~\cite{Kosm} :
\begin{eqnarray}
L\, = \, \, 
\left [\begin {array}{cc} 
\sqrt {1+{e}^{2\,Q+ih}}\, {e}^{P}&{e}^{Q}\\
\noalign{\medskip}
{e}^{Q}&{e}^{-P}\sqrt {1+{e}^{2\,Q+ih}}
\end {array}\right ]
\end{eqnarray}
where $\, P\, $ and $\, Q$ verify the Heisenberg commutation relation
$\, [P, \, Q]\, = \, \, i\, h$, or the Hermann-Weyl
relation $\,e^P \, e^Q \, = \, \, e^{i\, h} \, e^Q \, e^P$.
Try to Baxterise it, or remark that this problem reduces
to the previous Hopf algebra, with a quantum determinant equal to $\,
1$.

$\bullet\, $ 
 Let us recall the L-operator corresponding to the
Sine-Gordon model~\cite{Faddeev} :
\begin{eqnarray}
L\, = \, \, 
\left [\begin {array}{cc}
 u^{-}&1/4\,m\left (v^{-}/\lambda\, -\lambda\,v^{+}\right )\\
\noalign{\medskip}1/4\,m \, \left (\lambda\,v^{-}
-\, v^{+}/\lambda\right )&u^{+}
\end {array}\right ]
\end{eqnarray}
where $\, u^{\pm} \,$ and $\,  v^{\pm} \,$ are the Weyl operators :
$\, u^{\pm} \, = \, e^{\pm i \beta/4\, p}\, $
and $\,  v^{\pm} \, = \, e^{\pm i \beta/2\, q}$.
The spectral parameter $\, \lambda\, $ is explicit here. 
Show that the inverse transformation  $\, {\widehat I} $, and the
transformation $\, {\widehat K}^2 $, yield very simple 
multiplicative transformations on the  spectral parameter $\, \lambda$.
In the classical limit this  L-operator
becomes:
\begin{eqnarray}
L\, = \, \, -i \cdot 
\left [\begin {array}{cc} 
-2\,i\gamma\,\pi&1/4\,m\left (\lambda\,{e}^{-1/2\,i\phi}\, 
-{e}^{1/2\,i\phi}/\lambda\right )\\
\noalign{\medskip}1/4\,m\left (\lambda\,
{e}^{1/2\,i\phi}-\, {e}^{-1/2\,i\phi}/\lambda\right )&2\,i\gamma\,\pi
\end {array}\right ]
\end{eqnarray}
In this classical limit what means the inversion relation $\,
{\widehat I}$ ?

$\bullet$ Is it possible to Baxterise the universal
R-matrix of $\, U_q(sl_2)\, $ which corresponds to the
 formal series~\cite{Felder} :
\begin{eqnarray}
R \, \, = \, \, \, \sum^{\infty}_{k=0} \, q^{k\, (k-1)/2} \,
{{(q-q^{-1})^k} \over {[k]!}}\, q^{1/2\, H \otimes H} \, E^k \otimes F^k
\end{eqnarray}

\section{Conclusion}
The Baxterisation procedure is very efficient, and powerful, and is
{\em not restricted} to any Bethe Ansatz framework,
or any particular ``mathematical object'' : one can actually Baxterise
$\, R$-matrices, monodromy matrices, $\, L$-operators, 
quantum Hamiltonians, ...
in order to see the spectral parameter, and the integrability, become
crystal clear. Typically, all the  calculations sketched here,
can be performed for $\, R$ belonging to a quite general algebra
$\, R \, = \, \, \sum_i c_i \cdot A_i$, the 
generators $\,  A_i \, $ being {\em not necessarily 
associated with semi-simple Lie
algebras} : they can, for instance, be 
elements of a {\em Bose-Mesner algebra}~\cite{Jaeger}
associated with {\em distance regular graphs} ... The Baxterisation 
procedure corresponds to very simple 
calculations, namely performing the inverse of a matrix, 
or of some element of an algebra, and performing permutations of
the entries of a matrix, and combining these
two transformations to get a (generically) {\em infinite
order} (birational, polynomial) transformation  
one studies as a {\em discrete dynamical system}.
The Baxterisation procedure is a very powerful tool enabling
to find, or simply analyse, Yang-Baxter integrable models.
It is probably the quickest, and most powerful, way to get the solution
(parametrization) of the Yang-Baxter equations, {\em even if this 
parametrization is extremely complicated}. Actually, the
 problem of the parametrization 
of the Yang-Baxter equations is often a quite difficult, or subtle,
one. Recalling the 
free-fermion condition for the asymmetric eight vertex model,
it is not always clear to see if a variable is a spectral parameter,
an invariant (like the modulus of elliptic functions), 
or simply a gauge variable~\cite{Kr81,BaSt85abc} (see
also~\cite{Wadati}). A variable can 
be seen as a spectral parameter for a row-to-row transfer matrix, and
as an invariant for the column-to-column  transfer matrix, looking ``like a
gauge variable'', but corresponding, in fact, to 
symmetries like (\ref{35}) ...
The Baxterisation procedure is a very efficient method to clarify 
these ``subtleties''.

It is striking to note that the  Baxterisation procedure
actually provides results {\em beyond the Yang-Baxter integrable
framework}. The canonical elliptic parameterization of the sixteen
vertex model is such a good example~\cite{prl2}.  What does the {\em integrability 
of  discrete symmetries} of the parameter space of the model
mean {\em outside} the  Yang-Baxter integrable
framework~\cite{MeAnMaRo94} ? 
Could it be possible that it could help to solve the
model  {\em in the absence} of Yang-Baxter integrability ?
Is integrability restricted to Yang-Baxter integrability ?
What is integrability? In our introduction we have recalled that
the Yang-Baxter structure is a {\em sufficient} condition 
for the commutation of $\, q^N \times q^N\, $ 
transfer matrices (for any $\, N$), which is a fundamental property for 
the integrability of the model. However, if one {\em just} wants 
to calculate the partition function of the model (largest eigenvalue),
{\em one does not need such commutation property in  the whole}  $\, q^N \, $
{\em dimensional space} : if the transfer matrix can be block diagonalized,
the commutation in some block, including  the 
eigenvector corresponding to the largest eigenvalue,
is {\em sufficient to calculate} exactly the partition function 
per site. The exact calculation of the partition function
on the so-called ``disorder solutions''~\cite{disorder} is such an example : 
the disorder solution ``calculability'' is {\em not a Yang-Baxter
integrability}~\cite{disorderb}. Other examples exist in the literature 
called ``quasi-integrability'' : they correspond to Hamiltonians
for which one can {\em only} find exactly  the ground state and the corresponding 
eigenvalue. 

Along this line let us also recall the ideas developed by  V. Jones on {\em planar
algebras}~\cite{Jones} where he considers {\em local relations}, similar 
to the  $\, ABC = CBA\, $ Yang-Baxter relations, that could be sufficient
to calculate {\em global} objects like generating functions 
equivalent to partition functions.
He introduces, for instance, {\em deformations} of Yang-Baxter relations,
like  $\, ABC \, = \, CBA \, + \, S$, where $\, S\, $ is
``something'' with ``at most'' two $\, R$ matrices 
(see~\cite{Jones}). In such a framework
one does not have a commutation of transfer matrices anymore.
On could easily imagine to Baxterise the $\, ABC \, = \, CBA \, + \, S
\, $ relation in a similar way it can be done for the 
$\, ABC = CBA\, $ Yang-Baxter relations~\cite{bmv2,bmv2b}.
One could thus imagine the following situation : a parameterization
of the deformed Yang-Baxter relation in terms of elliptic curves,
no Yang-Baxter integrability stricto sensu, 
but, may be, a possibility to
calculate exactly the partition function per site !

Leaving these speculative ``dreams'' let us just
underline the fact that, beyond the Yang-Baxter 
framework, the Baxterisation procedure has 
already provided a very large number of new 
exact results for lattice models in
 statistical mechanics~\cite{HaMaOiVe87,MaRo94}
and, more generally, for {\em discrete
 dynamical systems}~\cite{Zittartz,McGuire}.

\vskip 1cm \noindent {\bf Acknowledgment}: One of us (JMM)
would like to thank 
R. J. Baxter for so many discussions 
and  for so many warm and friendly encouragements, in the 
last two decades. One of us (JMM)
would also like to thank P. A. Pearce for his great hospitality 
in the Mathematics and Statistics Dept. of Melbourne University
where part of this work has been completed. The authors would like to
thank J-C. Angl\`es d'Auriac for many discussions on the 
Baxterisation of quantum Hamiltonians.

\section{Appendix A : Polynomial representations of $\, \lambda \,
\rightarrow \, M \cdot  \lambda$}

For $\, N\, = \,  \, 7, \,  \, \, 11$, 
one has a completely similar structure than the one depicted in
section (\ref{shift}), for the polynomial
representation
of the  multiplication of the shift by $\, N$, namely 
 $\, (J_x, \, J_y, \, J_z) \,
\rightarrow \,  (J^{(N)}_x, \, J^{(N)}_y, \, J^{(N)}_z)$ where :
\begin{eqnarray}
\label{mult7}
&&J^{(N)}_x \, = \, \,J_x \cdot P^{(N)}_x(J_x, \, J_y, \, J_z) 
 \nonumber \\
&&J^{(N)}_y \, = \, \,J_y \cdot P^{(N)}_y(J_x, \, J_y, \, J_z) 
\, = \, \,J_y \cdot P^{(N)}_x(J_y, \, J_z, \, J_x) 
 \nonumber \\
&&J^{(N)}_z \, = \, \,J_z \cdot P^{(N)}_z(J_x, \, J_y, \, J_z) 
\, = \, \,J_z \cdot P^{(N)}_x(J_z, \, J_x, \, J_y) 
\end{eqnarray}

For $\, N\, = \, 7$ polynomial $\, 
P^{(7)}_x(J_x, \, J_y, \, J_z)\, $ reads :
\begin{eqnarray}
&&P^{(7)}_x(J_x, \, J_y, \, J_z) \, = \, \, -52 \, J_y^4 \, J_z^{22}
\, J_x^{22}
-12 \, J_y^2 \, J_z^{24} \, J_x^{22}-495 \, J_y^8 \, J_z^{16} \, J_x^{24}+
260 \, J_y^8 \, J_z^{18} \, J_x^{22}+220 \, J_y^6 \, J_z^{18} \,J_x^{24}
\nonumber \\
&&+188 \, J_y^6 \, J_z^{20} \, J_x^{22}+1680 \, J_y^{10} \, J_z^{20} \,
J_x^{18}
-42 \, J_y^{20} \, J_z^{24} \, J_x^4
+3192 \, J_y^{12} \, J_z^{22} \, J_x^{14}-516 \, J_y^{22} \, J_z^{20} \, J_x^6 \nonumber \\
&&-5936 \, J_y^{12}
 \, J_z^{18} \, J_x^{18}+700 \, J_y^{12} \, J_z^{20} \, J_x^{16}
-364 \, J_y^6 \, J_z^{24} \, J_x^{18}-66 \, J_y^4 \, J_z^{20}
 \, J_x^{24}+12 \, J_y^2 \, J_z^{22} \, J_x^{24}-2520 \, J_y^{14} \,J_z^{22} \, J_x^{12}
\nonumber \\
&&-2184 \, J_y^{10} \, J_z^{22} \, J_x^{16}+752 \, J_y^{10} \, J_z^{18}
\, J_x^{20}
+3404 \, J_y^{12} \,J_z^{16} \, J_x^{20}
-6536 \, J_y^{14} \, J_z^{14} \, J_x^{20}
+5704 \, J_y^{14} \, J_z^{18} \, J_x^{16}-J_y^{24} \, J_x^{24} \nonumber \\
&&+3864 \, J_y^{14} \, J_z^{16} \, J_x^{18}+1800 \, J_y^{14} \,
J_z^{24} \, J_x^{10}
-44 \, J_y^6 \, J_z^{22} \, J_x^{20}
-J_z^{24} \, J_x^{24}+7 \, J_y^{24} \, J_z^{24}+756 \, J_y^8
 \, J_z^{22} \, J_x^{18}\nonumber \\
&&-1260 \, J_y^{12}\, J_z^{24} \, J_x^{12}
+118 \, J_y^4 \, J_z^{24} \, J_x^{20}
+168 \, J_y^{10} \, J_z^{24} \, J_x^{14}+441 \, J_y^8 \, J_z^{24}
\, J_x^{16}
-962 \, J_y^8 \, J_z^{20} \, J_x^{20}-28 \, J_y^{24} \, J_z^{22} \, J_x^2\nonumber \\
&&-1311\, J_y^{24} \, J_z^{16} \, J_x^8
-42 \, J_y^{24} \, J_z^{20} \, J_x^4+3756 \, J_y^{20} \, J_z^{16} \, J_x^{12} \nonumber \\
&&+1800 \, J_y^{24} \, J_z^{14} \, J_x^{10}+484 \, J_y^{24} \, J_z^{18} \, J_x^6-
1260 \, J_y^{24} \, J_z^{12} \, J_x^{12}+756 \, J_y^{22} \, J_z^8 \, J_x^{18}
-12 \, J_y^{24} \, J_z^2 \, J_x^{22}-28 \, J_y^{22} \, J_z^{24} \, J_x^2\nonumber \\
&&-2544 \, J_y^{20} \, J_z^{18} \, J_x^{10}
-9322 \, J_y^{16} \, J_z^{16} \, J_x^{16}+3756 \, J_y^{16} \, J_z^{20}\, J_x^{12}
+12 \, J_y^{22} \, J_z^2 \, J_x^{24} \nonumber \\
&&-3928 \, J_y^{14} \, J_z^{20} \, J_x^{14}+484 \, J_y^{18} \,J_z^{24} \, J_x^6
+444 \, J_y^{22} \, J_z^{18} \, J_x^8
+792 \, J_y^{10} \, J_z^{14} \, J_x^{24}+744 \, J_y^{22} \, J_z^{16} \, 
J_x^{10}-2544 \, J_y^{18} \, J_z^{20} \, J_x^{10}\nonumber \\
&&+444 \, J_y^{18} \,J_z^{22} \, J_x^8
+168 \, J_y^{24} \, J_z^{10} \, J_x^{14}+1734 \, J_y^{20} \, J_z^{20} \, J_x^8
-52 \, J_y^{22} \, J_x^{22} \, J_z^4+48 \, J_y^{18} \, J_z^{18} \, J_x^{12}+3864
 \, J_y^{16} \, J_z^{14} \, J_x^{18} \nonumber \\
&&+568 \, J_y^{16} \, J_z^{18} \, J_x^{14} +3404 \, J_y^{16} \, J_z^{12} \, J_x^{20}+744 \, J_y
^{16} \, J_z^{22} \, J_x^{10}-5936 \, J_y^{18} \, J_z^{12} \, J_x^{18}
+752 \, J_y^{18} \, J_z^{10} \, J_x^{20}\nonumber \\
&&+5704 \, J_y^18 \, J_z^{14} \, J_x^{16}-516 \, J_y^{20} \, J_z^{22}
\, J_x^6\,
-495 \, J_y^{16} \, J_z^8 \, J_x^{24}+220 \, J_y^{18} \, J_z^6 \,J_x^{24}
+196 \, J_y^{22} \, J_z^{22} \, J_x^4
+824 \, J_y^{12} \, J_z^{14} \, J_x^{22} \nonumber \\
&&+260 \, J_y^{18} \, J_z^8 \, J_x^{22}-1208 \, J_y^{16} \, J_z^{10}
\, J_x^{22}
-924 \, J_y^{12} \, J_z^{12} \, J_x^{24}+792 \, J_y^{14} \, J_z^{10}
\, J_x^{24}
-1208 \, J_y^{10} \, J_z^{16} \, J_x^{22}\nonumber \\
&&+1680 \, J_y^{20} \, J_z^{10} \, J_x^{18}
-2184 \, J_y^{22} \, J_z^{10} \, J_x^{16}+188 \, J_y^{20} \, J_z^6 \,J_x^{22}
-962 \, J_y^{20} \, J_z^8 \, J_x^{20}
+824 \, J_y^{14} \, J_z^{12} \, J_x^{22}
\nonumber \\
&&-3928 \, J_y^{20} \, J_z^{14} \, J_x^{14}+568 \, J_y^{18} \, J_z^{16} \,J_x^{14}
+700 \, J_y^{20} \, J_z^{12} \, J_x^{16}-66 \, J_y^{20} \, J_x^{24} \,J_z^4
+118 \, J_y^{24} \, J_x^{20} \, J_z^4 +441 \, J_y^{24} \, J_z^8 \, J_x^{16}\nonumber \\
&&-2520 \, J_y^{22} \, J_z^{14} \, J_x^{12}-1311 \, J_y^{16} \, 
J_z^{24} \, J_x^8+3192 \, J_y^{22} \, J_z^{12} \, J_x^{14}-44 \,
J_y^{22} \, J_z^6 \, J_x^{20}-364 \, J_y^{24} \, J_z^6 \, 
J_x^{18}
\end{eqnarray}
The modulus  (\ref{Mod}) is invariant by (\ref{mult7})
corresponding to the multiplication of the
shift by seven.

The multiplication of the shift by eleven
has a polynomial representation $\, (J_x, \, J_y, \, J_z) \,
\rightarrow \,  (J^{(11)}_x, \, J^{(11)}_y, \, J^{(11)}_z) $
satisfying (\ref{mult7}) for $\, N \, =\, 11$, 
where  $\, P^{(11)}_x(J_x, \, J_y, \, J_z) \, $ is a homogeneous
polynomial of degree $\, 120 \, $ sum of $\, 496\, $ 
monomial expressions:
\begin{eqnarray}
&&P^{(11)}_x(J_x, \, J_y, \, J_z) \, = \, \, -30045015\, J_y^{40}\,
J_z^{20}\, J_x^{60}+2035800\, J_y^{46}\, J_z^{14}\, J_x^{60}\, +142506\, J_y^{50}\,
J_z^{10}\, J_x^{60}\, -27405\, J_y^8\, J_z^{52}\, J_x^{60}\nonumber \\
&&\quad \quad \quad+54627300\, J_y^{38}\, J_z^{22}\, J_x^{60}
-435\, J_z^{56}\, J_y^4\, J_x^{60}+54627300\, J_y^{22}\, J_z^{38}\, J_x^{60} \,
+\,  \cdots 
\end{eqnarray}
The actual expression of $\, P^{(11)}_x(J_x, \, J_y, \, J_z) \, $
will be given elsewhere.

When $\, N\, $ is not a prime number one has slightly modified
results: one does not have 
(\ref{mult7}) anymore. 
Actually, the polynomial
representation
of the  multiplication of the shift by six 
can be obtained in many different ways, namely substituting
$\, J^{(2)}$ in $\, J^{(3)}$, or 
$\, J^{(3)}$ in $\, J^{(2)}$, or 
by various eliminations  between various
 biquadratic curves $\, \Gamma_i$. The result reads
 $\, (J_x, \, J_y, \, J_z) \,\rightarrow \,  (J^{(6)}_x, \, J^{(6)}_y,
\, J^{(6)}_z)$ where :
\begin{eqnarray}
\label{mult6}
&&J^{(6)}_x \, = \, \,J^{(2)}_x \cdot {\tilde P}^{(6)}_x(J_x, \, J_y, \, J_z) 
 \nonumber \\
&&J^{(6)}_y \, = \, \,J^{(2)}_y \cdot {\tilde P}^{(6)}_y(J_x, \, J_y, \, J_z) 
\, = \, \,J^{(2)}_y \cdot {\tilde P}^{(6)}_x(J_y, \, J_z, \, J_x) 
 \nonumber \\
&&J^{(6)}_z \, = \, \,J^{(2)}_z \cdot {\tilde P}^{(6)}_z(J_x, \, J_y, \, J_z) 
\, = \, \,J^{(2)}_z \cdot {\tilde P}^{(6)}_x(J_z, \, J_x, \, J_y) 
\end{eqnarray}
such that $\,  J^{(6)}_y(J_x, \, J_y, \, J_z) 
\, = \, \,  J^{(6)}_x(J_y, \, J_z, \, J_x)\, $ 
and $\,  J^{(6)}_z(J_x, \, J_y, \, J_z) 
\, = \, \,J^{(6)}_x(J_z, \, J_x, \, J_y)$. The $\, {\tilde P}^{(6)}$'s
 and the  $\, J^{(6)}$'s
are functions of $\, J_x^2$,  $\, J_y^2$ and $\, J_z^2$. 
One also has $\, J^{(6)}_x(J_x, \, J_y, \, J_z) 
\, = \, \, J^{(6)}_x(J_x, \, J_z, \, J_y)$, 
$\,  J^{(6)}_y(J_x, \, J_y, \, J_z) 
\, = \, \, J^{(6)}_y(J_z, \, J_y, \, J_x)$, 
$\,  J^{(6)}_z(J_x, \, J_y, \, J_z) 
\, = \, \, J^{(6)}_z(J_y, \, J_x, \, J_z)$
and $\, {\tilde P}^{(6)}_x(J_x, \, J_y, \, J_z) 
\, = \, \,{\tilde P}^{(6)}_x(J_x, \, J_z, \, J_y)$ , 
$\, {\tilde P}^{(6)}_y(J_x, \, J_y, \, J_z) 
\, = \, \, {\tilde P}^{(6)}_y(J_z, \, J_y, \, J_x)$ , 
$\, {\tilde P}^{(6)}_z(J_x, \, J_y, \, J_z) 
\, = \, \, {\tilde P}^{(6)}_z(J_y, \, J_x, \, J_z)$. 

All these expressions can thus be deduced 
from $\, {\tilde P}^{(6)}_x(J_x, \, J_y, \, J_z) $ :
\begin{eqnarray}
&&{\tilde P}^{(6)}_x(J_x, \, J_y, \, J_z) \, = \, \, 
70\, J_z^{16}\, J_x^{8}\, J_y^{8}-250\, J_z^{8}\, J_x^{16}\,
J_y^{8}-8\, J_z^{16}\, J_x^{14}
\, J_y^{2}+8\, J_z^{14}\, J_x^{16}\, J_y^{2}\,+184\, J_z^{10}\,
J_x^{16}\, J_y^{6} \nonumber \\
&&+28\, J_z^{16}\, J_x^{12}\, J_y^{4}+40\, J_z^{14}\, J_x
^{14}\, J_y^{4}-68\, J_z^{12}\, J_x^{16}\, J_y^{4}
-56\, J_z^{16}\, J_x^{10}\, J_y^{6}-56\, J_z^{14}\, J_x^{12}\, J_y^{6}
-72\, J_z^{12}\, J_x^{14}\, J_y^{6}+J_z^{16}\, J_x^{16}\nonumber \\
&&+228\, J_z^{12}\, J_x^{12}\, J_y^{8}\, +40\, J_z^{10}\, J_x^{14}\, J_y^{8}
+70\, J_y^{16}\, J_z^{8}\, J_x^{8}-56\, J_y^{10}\, J_z^{16}
\, J_x^{6}+88\, J_y^{10}\, J_z^{14}\, J_x^{8}+28\, J_y^{12}\,
J_z^{16}\, J_x^{4}
+56\, J_y^{12}\, J_z^{14}\, J_x^{6}\nonumber \\
&&-40
\, J_y^{14}\, J_z^{14}\, J_x^{4}+56\, J_y^{14}\, J_z^{12}\,
J_x^{6}+88\, J_y^{14}
\, J_z^{10}\, J_x^{8}-8\, J_y^{16}\, J_z^{14}\, J_x^{2}+28\,
J_y^{16}\, J_z^{12}\, J_x^{4}
-56\, J_y^{16}\, J_z
^{10}\, J_x^{6}\nonumber \\
&&+40\, J_y^{10}\, J_x^{14}\, J_z^{8}+184\, J_y^{10}\,J_x^{16}\, J_z^{6}
+228\, J_y^{12}\, J_x^{12}\, J_z^{8}-72\, J_y^{12}\, J_x^{14}\, J_z^{6}
-68\, J_y^{12}\, J_x^{16}\, J_z
^{4}-88\, J_y^{14}\, J_x^{10}\, J_z^{8}\nonumber \\
&&-56\, J_y^{14}\, J_x^{12}\,
J_z^{6}+40\, J_y^{14}\, J_x^{14}\, J_z^{4}+8\, J_y^{14}\, J_x^{16}\,
J_z^{2}-56\, J_y^{16}\, J_x^{10}\, J_z^{6}
+28\, J_y^{16}\, J_x^{12}\, J_z^{4}-88\, J_z^{14}\, J_x^{10}\,
J_y^{8}\,
-316\, J_y^{12}\, J_z^{12}\,
J_x^{8}\nonumber \\
&&-8\, J_y^{16}\, J_x^{14}\, J_z^{2}
+144\, J_z^{12}\, J_x^{10}\, J_y^{10}\,
 -400\, J_z^{10}\, J_x^{12}\,J_y^{10}+144\, J_z^{10}
\, J_x^{10}\, J_y^{12}+J_y^{16}\, J_z^{16}+J_y^{16}\, J_x^{16}\, 
-8\, J_y^{14}\, J_z^{16}\, J_x^{2}\nonumber 
\end{eqnarray}
Note that $\, J^{(6)}_x \,$
is a homogeneous polynomial expression
of degree $\, 36$.

Similarly, the polynomial
representation
of the  multiplication of the shift by nine
 reads
 $\, (J_x, \, J_y, \, J_z) \,
\rightarrow \,  (J^{(9)}_x, \, J^{(9)}_y, \, J^{(9)}_z) $ 
where :
\begin{eqnarray}
\label{mult9}
&&J^{(9)}_x \, = \, \,J^{(3)}_x \cdot {\tilde P}^{(9)}_x(J_x, \, J_y, \, J_z) 
 \nonumber \\
&&J^{(9)}_y \, = \, \,J^{(3)}_y \cdot {\tilde P}^{(9)}_y(J_x, \, J_y, \, J_z) 
\, = \, \,J^{(3)}_y \cdot P^{(9)}_x(J_y, \, J_z, \, J_x) 
 \nonumber \\
&&J^{(9)}_z \, = \, \,J^{(3)}_z \cdot {\tilde P}^{(9)}_z(J_x, \, J_y, \, J_z) 
\, = \, \,J^{(3)}_z \cdot {\tilde P}^{(9)}_x(J_z, \, J_x, \, J_y) 
\end{eqnarray}
The expression of  $\, {\tilde P}^{(9)}_x(J_x, \, J_y, \, J_z) \, $ will be given
elsewhere. It is, again, a function of
 $\, J_x^2$, $\, J_y^2$ and $\, J_z^2$. It is the sum of 
$\, 190\, $ monomial expressions of degree $\, 72$, $\, J^{(9)}_x \,$
being a homogeneous polynomial expression
of degree $\, 81$.

\section{Appendix B : Finite order conditions}

It can be shown that the points of the Baxter model on the algebraic varieties :
\begin{eqnarray}
V^{(3)}(J_x,\, J_y, \, J_z)\,\, = \, \,\, J_x \, J_y\, 
+\, J_z \, J_x\, +J_z\, J_y\, , \qquad 
\widehat{V}_z^{(3)}(J_x,\, J_y, \, J_z)\, = \, \,-J_x \, J_y\, 
+\, J_z \, J_x\, +J_z\, J_y \nonumber
\end{eqnarray}
are actually such that $\, 
K^{6}(R) \, = \, \zeta \cdot R$.
One has the following factorization property :
\begin{eqnarray}
\label{union}
&&V^{(3)}(J^{(2)}_x,\, J^{(2)}_y, \, J^{(2)}_z)\, = \, \\
&&\,\, V^{(3)}(J_x,\, J_y, \, J_z) \cdot \widehat{V}_x^{(3)}(J_x,\, J_y,
\, J_z) \cdot \widehat{V}_y^{(3)}(J_x,\, J_y, \, J_z)
\cdot \widehat{V}_z^{(3)}(J_x,\, J_y, \, J_z) \, = \, 0  
\end{eqnarray}
where $\,  \widehat{V}_x^{(3)}(J_x,\, J_y, \, J_z) \, =  \, 
\widehat{V}_z^{(3)}(J_y,\, J_z, \, J_x) \, $ and $\,  
\widehat{V}_y^{(3)}(J_x,\, J_y, \, J_z) \, =  \, 
\widehat{V}_z^{(3)}(J_z,\, J_x, \, J_y)$. One has the relation :
\begin{eqnarray}
\label{rel3}
\widehat{V}_z^{(3)}(J^{(2)}_x,\, J^{(2)}_y, \, J^{(2)}_z)\,- \, 
P_z^{(3)}(J_x,\, J_y, \, J_z)\,\, = \,\, \, 0
\end{eqnarray}
Note that the points of the Baxter model
on the algebraic varieties
$\, P_x^{(3)}(J_x,\, J_y, \, J_z)\,\, = \,\,  0$, 
or $\, P_y^{(3)}(J_x,\, J_y, \, J_z)\,\, = \,\,  0$, 
or $\, P_z^{(3)}(J_x,\, J_y, \, J_z)\,\, = \,\,  0$, 
are such that $\, K^{12}(R) \, = \, \zeta \cdot R$.
Relation (\ref{rel3}) is in agreement with the fact that 
$\, (J_x,\, J_y, \, J_z)\, \rightarrow \, (J^{(2)}_x,\, J^{(2)}_y, \,
J^{(2)}_z)\, $ is a representation of the 
shift doubling. The points of order 
six (namely $K^{6}(R) \, = \, \zeta \cdot R$)
correspond to (\ref{union}),  
their image by the shift doubling $\, (J_x,\, J_y, \, J_z)\,
\rightarrow \, (J^{(2)}_x,\, J^{(2)}_y, \,
J^{(2)}_z)\, $ giving $\, P_x^{(3)}(J_x,\, J_y, \, J_z) \cdot  
P_y^{(3)}(J_x,\, J_y, \, J_z) \cdot  P_z^{(3)}(J_x,\, J_y, \, J_z)\,\,
= \,\,  0$,  together, of course, with (\ref{union}). 

Let us note that the points of the
algebraic variety  $\, \widehat{V}_z^{(3)}(J_x,\, J_y, \, J_z)\, =
\,0\, $ are (projectively) of order {\em three} for the Baxter model : $\, 
K^{3}(R) \, = \, \zeta \cdot R$. The points of the Baxter model on the variety :
\begin{eqnarray}
\label{orderfive}
&&V_y^{(5)}(J_x,\, J_y, \, J_z)\, = \, \, J_z^2 \, J_x \, J_y^3\, 
-2 \, J_z^2 \, J_x^2 \, J_y^2\, 
+J_z^2 \, J_x^3 \, J_y+J_z \, J_x^2 \, J_y^3-J_z \, J_x^3 \, J_y^2
-J_z^3 \, J_y^3-J_z^3 \, J_x \, J_y^2\nonumber \\
&&\qquad \qquad +J_z^3 \, J_x^2 \, J_y+J_z^3 \,
J_x^3-J_x^3 \, J_y^3 \,\, \, = \, \,\, \, 0
\end{eqnarray}
are of order five $\, 
K^5(R) \, = \, \zeta \cdot R$. The points
 of the Baxter model on the algebraic variety :
\begin{eqnarray}
\label{orderten}
&&\widehat{V}_y^{(5)}(J_x,\, J_y, \, J_z)\,\, = \,\,\, V^{(5)}_y(J_y,\,
J_x, \, J_z)\,\,=\, \,
 \, J_z^2 \, J_x^3 \, J_y-2 \, J_z^2 \,
J_x^2 \, J_y^2\, 
+J_z^2 \, J_x \, J_y^3\, +J_z \, J_x^3 \, J_y^2 \nonumber \\
&&\quad \quad \quad \quad \quad -J_z \, J_x^2 \, J_y^3 \, -J_z^3 \, J_x^3
 -J_z^3 \, J_x^2 \, J_y\, +J_z^3 \, J_x \, J_y^2\, 
+J_z^3 \, J_y^3\, -J_x^3 \, J_y^3 \,\, \, = \,\,\, \, 0
\end{eqnarray}
are of order ten : $\, 
K^{10}(R) \, = \, \zeta \cdot R$. The  point of the Baxter model on the variety 
$\, P_x^{(5)}(J_x,\, J_y, \, J_z)\, = \, 0$, 
$\,\,  P_y^{(5)}(J_x,\, J_y, \, J_z)\, = \, 0$,
or $\, P_z^{(5)}(J_x,\, J_y, \, J_z)\, = \, 0$,
are of order {\em twenty} : $\, 
K^{20}(R) \, = \, \zeta \cdot R$. Let us note that:
\begin{eqnarray}
\label{rel5}
V_y^{(5)}(J^{(2)}_x,\, J^{(2)}_y, \, J^{(2)}_z)\,+ \, 
P_y^{(5)}(J_x,\, J_y, \, J_z)\, = \, 0\, , \quad 
\widehat{V}_y^{(5)}(J^{(2)}_x,\, J^{(2)}_y, \, J^{(2)}_z)\,+ \, 
P_x^{(5)}(J_x,\, J_y, \, J_z)\, = \, 0  
\end{eqnarray}
which is in agreement with the fact that 
$\, (J_x,\, J_y, \, J_z)\, \rightarrow \, (J^{(2)}_x,\, J^{(2)}_y, \,
J^{(2)}_z)\, $
represents the shift doubling.

{\bf Remark :} Relation (\ref{rel5}) 
is in  agreement with  the shift doubling, however one seems to have
an apparent contradiction with previous relations. The points 
of (\ref{orderfive}), namely 
$\, V_y^{(5)}(J_x,\, J_y, \, J_z)\,= \, \, 0\, $
being of order five,
one expects that their image by the shift doubling  will give points
of order ten,
 like (\ref{orderten}),
and not points of order twenty, like 
$\, P_y^{(5)}(J_x,\, J_y, \, J_z)\, = \, 0$.
In fact, this algebraic variety is an order five
algebraic variety {\em only when restricted to the 
Baxter model}. For the sixteen vertex model
one can actually show that  $\, V_y^{(5)}(J_x,\, J_y, \, J_z)\,= \, \,
0$ and $\, \widehat{V}_y^{(5)}(J_x,\, J_y, \, J_z)\,
=\, \, 0$, are (codimension-one) algebraic varieties 
of order {\em ten}, the algebraic varieties 
of order five being higher codimension algebraic varieties.

\end{document}